\documentclass[11pt,a4paper]{article}
\pdfoutput=1
\usepackage{jheppub}
\usepackage{graphicx}
\usepackage{lipsum}
\usepackage[all]{hypcap}
\allowdisplaybreaks

\def\Eq#1{eq.~(\ref{#1})}

\def\Sec#1{sect.~\ref{#1}}

\newcommand{\be}{\begin{equation}}
\newcommand{\ee}{\end{equation}}
\newcommand{\bea}{\begin{eqnarray}}
\newcommand{\eea}{\end{eqnarray}}
\newcommand{\beq}{\begin{equation}}
\newcommand{\eeq}{\end{equation}}
\newcommand{\comment}[1]{}

\begin{document}

{\raggedleft CERN-TH-2017-219 \\}

\title{Clockwork / Linear Dilaton: Structure and Phenomenology}

\affiliation[a]{Theoretical Physics Department, CERN, Geneva, Switzerland}
\affiliation[b]{Department of Physics, Ben-Gurion University, Beer-Sheva 8410501, Israel}
\affiliation[c]{INFN, Sezione di Genova, Via Dodecaneso 33, 16146 Genova, Italy}
\affiliation[d]{INFN, Sezione di Trieste, SISSA, Via Bonomea 256, 34136 Trieste, Italy
 \vspace{2mm}}
 
\author[a]{Gian F. Giudice,}
\author[a,b]{Yevgeny Kats,}
\author[a]{Matthew McCullough,}
\author[a,c]{Riccardo Torre,}
\author[a,d]{\\and Alfredo Urbano}

\emailAdd{gian.giudice@cern.ch}
\emailAdd{yevgeny.kats@cern.ch}
\emailAdd{matthew.mccullough@cern.ch}
\emailAdd{alfredo.leonardo.urbano@cern.ch}
\emailAdd{riccardo.torre@cern.ch}

\abstract{The linear dilaton geometry in five dimensions, rediscovered recently in the continuum limit of the clockwork model, may offer a solution to the hierarchy problem which is qualitatively different from other extra-dimensional scenarios and leads to distinctive signatures at the LHC. We discuss the structure of the theory, in particular aspects of naturalness and UV completion, and then explore its phenomenology, suggesting novel strategies for experimental searches. In particular, we propose to analyze the diphoton and dilepton invariant mass spectra in Fourier space in order to identify an approximately periodic structure of resonant peaks. Among other signals, we highlight displaced decays from resonantly-produced long-lived states  and high-multiplicity final states from cascade decays of excited gravitons.}

\maketitle

\section{Introduction}

It has been shown recently~\cite{Giudice:2016yja} that the clockwork mechanism, introduced in refs.~\cite{Choi:2014rja,Choi:2015fiu,Kaplan:2015fuy} to reconcile super-Planckian field excursions with renormalizable quantum field theories, is a much broader tool with many possible applications and generalisations, some of which have been explored in refs.~\cite{Giudice:2016yja,Kehagias:2016kzt,Farina:2016tgd,Ahmed:2016viu,Hambye:2016qkf,Craig:2017cda,vonGersdorff:2017iym,Giudice:2017suc,Coy:2017yex,Ben-Dayan:2017rvr,Hong:2017tel,Im:2017eju,Park:2017yrn,Lee:2017fin,Marzola:2017lbt,Carena:2017qhd,Ibanez:2017vfl,Kim:2017mtc,Agrawal:2017cmd,Ibarra:2017tju,Patel:2017pct,Choi:2017ncj}. A particularly interesting result is the observation that discrete clockworks have a non-trivial continuum limit that singles out a five-dimensional theory with a special geometry. This geometry coincides with the one obtained in theories with a five-dimensional dilaton which acquires a background profile linearly varying with the extra-dimensional coordinate. This theory can address the Higgs naturalness problem in setups where the Standard Model (SM) lives on a brane embedded in a truncated version of the five-dimensional linear-dilaton space. Earlier, the same setup was proposed in ref.~\cite{Antoniadis:2001sw} motivated by the seven-dimensional gravitational dual~\cite{Aharony:1998ub,Giveon:1999px} of Little String Theory~\cite{Berkooz:1997cq,Seiberg:1997zk}, which is a six-dimensional strongly-coupled non-local theory that arises on a stack of NS5 branes. Additionally, several recent studies have examined in more detail how the linear dilaton setup can be embedded in supergravity~\cite{Kehagias:2017grx,Antoniadis:2017wyh}.

The clockwork interpretation of the linear dilaton theory has helped in elucidating its relation with Large Extra Dimension (LED)~\cite{ArkaniHamed:1998rs,Antoniadis:1998ig,ArkaniHamed:1998nn} and Randall-Sundrum (RS)~\cite{Randall:1999ee} theories, thus providing a coherent map of approaches to the hierarchy problem in extra dimensions. Because of the double interpretation of the same theory either as a linear dilaton setup emerging from an effective description of non-critical string theory, including duals of Little String Theory, or as the continuum version of a clockwork model, we will refer to this theory as Clockwork / Linear Dilaton (CW/LD).

So far, CW/LD has received very little experimental attention, in spite of its attractive and distinguishing features. Nonetheless, some phenomenological aspects of CW/LD have already been discussed in the literature. The distinctive KK graviton spectrum, with a mass gap followed by a narrowly spaced spectrum of modes, and some of its associated collider signatures have been pointed out in ref.~\cite{Antoniadis:2011qw}. A more detailed study of the KK graviton phenomenology, including in particular the case of a small mass gap, has been undertaken in ref.~\cite{Baryakhtar:2012wj}. Finally, the KK dilaton / radion collider signatures have been studied in ref.~\cite{Cox:2012ee}.

In the current paper we aim at providing a comprehensive picture of the collider phenomenology of CW/LD, extending previous studies, deriving new constraints from present LHC data, and suggesting new characteristic signatures, in the hope of motivating dedicated experimental searches. 

We start by reviewing in section~\ref{sec1} the structure of CW/LD and the consequences of its UV embedding in string theory for the low-energy parameters. We also discuss the effect of adding cosmological constant terms to the effective theory, arguing that the bulk theory should be supersymmetric to avoid destabilising the setup. Then, we
explain the salient features of CW/LD for collider applications. The theory describes a tower of massive spin-two particles which, depending on the point of view, can be interpreted either as the Kaluza-Klein (KK) excitations of the five-dimensional graviton or as the continuum version of the clockwork gears. Their mass spectrum and couplings are completely fixed in terms of only two parameters: the fundamental gravity scale $M_5$ and the mass $k$ which characterizes the geometry of CW/LD. We encounter also a tower of spin-zero particles obtained from the combination of the single radion state with the KK excitations of the dilaton. In CW/LD the same scalar field both induces the non-trivial geometry and stabilises the extra dimension (thus playing the role of the Goldberger-Wise field~\cite{Goldberger:1999uk} in RS). The mass spectrum and couplings of the scalar modes are not fully determined by $M_5$ and $k$ alone, but also depend on the brane-localised stabilising potential and on a possible Higgs-curvature coupling. However, many features of the scalar phenomenology are independent of these details.

The rest of the paper is devoted to the collider phenomenology of CW/LD. In section~\ref{sec21} we study the effect of $s$-channel and $t$-channel exchange of KK gravitons in diphoton, dilepton, and dijet distributions at high invariant mass. An interesting feature of CW/LD is that these processes can be reliably computed within the effective theory, unlike LED where these processes are dominated by incalculable UV contributions. We also analyse the resonant production of new states decaying into diphoton and dilepton final states. All signatures discussed in section~\ref{sec21} correspond to already existing searches performed by the LHC collaborations and results can be adapted to the case of CW/LD using studies of continuum distributions (previously done for LED) or resonant production (previously done for RS).

In section~\ref{sec22} we explore new strategies that can be used at the LHC to discover or constrain CW/LD. The near-periodicity of invariant mass distributions with characteristic separations in the 1-5\% range (at the edge of experimental resolution) has prompted us to suggest a data analysis based on a Fourier transform, similarly to what is routinely encountered in other fields, for example in analyses of CMB temperature fluctuations. We show that such an analysis is competitive with other searches as a discovery mode, as well as being effective for extracting model parameters. 

A distinguishing feature of CW/LD with respect to LED or RS is the mass gap of the KK tower, followed by a near-continuum of modes. We suggest that such a turn-on of the spectrum may be observable even when it occurs at low invariant mass, where the individual resonances are difficult to see due to the experimental resolution. Such searches would require the use of trigger-level analysis / data scouting or ISR-based triggers.

An important new result presented in this paper is the calculation of the decay chains of graviton or scalar excited modes into lighter KK modes. We find that such cascades are the dominant decay mode for most of the scalar KK tower, and in certain parameter regions also a significant decay mode for part of the graviton KK tower. We study the properties of the high-multiplicity final states that arise from such decays. We also explore the possibility of displaced vertices originating from resonant particle production, which is another signature characteristic of CW/LD. 

Finally we collect in the appendices a compendium of formul{\ae} for the production cross sections and decay rates of the new particles in CW/LD. This should be a useful resource for those interested in experimental or phenomenological simulations of CW/LD in collider environments. Other appendices are dedicated to additional details and discussions concerning the structure of the theory.

\section{Properties of the model}
\label{sec1}

\subsection{Basic setup}
\label{sec:setup}

We consider a 5D space in which
the extra dimension is a circle, with the circumference parameterized by a coordinate $y$ in the range $-\pi R \leq y \leq \pi R$. The SM lives on a brane (TeV brane) at $y = y_{\rm T} = 0$, another brane (Planck brane) is at $y = y_{\rm P}=\pi R$, and a $Z_2$ orbifold symmetry identifies $y \leftrightarrow -y$. The full action in the Einstein frame is
\beq\label{eq:Einstein}
\mathcal{S}=\mathcal{S}_{\text{bulk}}+\mathcal{S}_{\mathcal{B}}+\mathcal{S}_{\text{GHY}}+\mathcal{S}_{\text{SM}}\,,
\eeq
\begin{align}
 \mathcal{S}_{\text{bulk}}&=M_{5}^{3}\int d^{4}x\int_{0}^{\pi R}dy\sqrt{-g}\left[\mathcal{R}-\frac{1}{3}g^{MN}\partial_{M}S\partial_{N}S-V(S)\right]\,,\label{actionbulk}\\[2mm]
 \mathcal{S}_{\mathcal{B}}&=-2\sum_{i={\rm T},{\rm P}}\int d^{4}x\int_{0}^{\pi R}dy \sqrt{-\gamma}\,\delta(y-y_{i})\lambda_{i}(S)\,,\label{actionboundary}\\[2mm]
 \mathcal{S}_{\text{GHY}}&=4M_{5}^{3} \sum_{i={\rm T},{\rm P}}\int d^{4}x\int_{0}^{\pi R}dy \sqrt{-\gamma}\,\mathcal{K}^{i}\delta(y-y_{i}) \nonumber \\
&=2M_{5}^{3} \int d^{4}x\int_{0}^{\pi R}dy\, \partial_{y}\left(\frac{1}{\sqrt{g_{55}}}\partial_{y}\sqrt{-\gamma}\right)
\,,\label{actionGHY}\\[2mm]
 \mathcal{S}_{\text{SM}}&=2\int d^{4}x\int_{0}^{\pi R}dy \sqrt{-\gamma}\, e^{-S/3}\mathcal{L}_{\text{SM}}\delta(y-y_{\rm T})\,,\label{actionSM}
\end{align}
where
\beq
V(S)=- 4 k^2 e^{-2 S/3}
\eeq
is a bulk potential and
\begin{eqnarray}
\lambda_{\rm T}(S) &=& e^{-\frac{S}{3}}\, M_5^3\left[
-4k + \frac{\mu_{\rm T}}{2}(S - S_{\rm T})^2
\right]~,\label{eq:Smart1}\\
\lambda_{\rm P}(S) &=& e^{-\frac{S}{3}}\, M_5^3\left[
+4k + \frac{\mu_{\rm P}}{2}(S - S_{\rm P})^2
\right] \label{eq:Smart2}
\end{eqnarray}
parameterise the scalar potentials at the TeV and Planck branes near their local minima at $S = S_{\rm T,\rm P}$.
Here $S$ is the dilaton field, $M_5$ is the five-dimensional reduced Planck mass, $k$ is a mass parameter, $\mathcal{R}$ is the 5D Ricci scalar, $g$ is the determinant of the 5D metric, $\gamma=g/g_{55}$ is the determinant of the induced metric at the boundaries, $\mu_{i}$ are two masses that determine the strength of the dilaton boundary potentials, $\mathcal{K}^{i}$ are the extrinsic curvatures of the two boundaries, which determine the Gibbons-Hawking-York (GHY) term~\cite{York:1972sj,Gibbons:1976ue}, and $\mathcal{L}_{\text{SM}}$ is the SM Lagrangian.\footnote{This is one possible choice for the dilaton coupling to the SM fields, however in general the coupling is model-dependent.  For instance, one could have taken a coupling $e^{S}\mathcal{L}_{\text{SM}}$ in the Jordan frame, which would result in a different Einstein frame action.}

Einstein's equations and the dilaton equation of motion (which are presented in detail in appendix~\ref{app:background}) are solved by
\beq
ds^2 = e^{\frac43 k |y|}\left(\eta_{\mu\nu}dx^\mu dx^\nu + dy^2\right) ,\qquad
S(y) = 2 k |y| \, .
\label{eq:ScalarBackground}
\eeq
With the above metric, the four-dimensional reduced Planck mass, $M_P \equiv 1/\sqrt{8\pi G} \approx 2.4 \times 10^{18}$~GeV, is given by
\beq
M_P^2 = \frac{M_5^3}{k}\left(e^{2\pi kR} - 1\right) \, .
\label{MPl}
\eeq
The fundamental scale, $M_5$, which is the scale at which the theory must be UV-completed, is assumed to be not much higher than the electroweak scale, while the exponentially greater scale $M_P$ is an illusion, created by the exponential factor in~\eqref{MPl}. To account for the hierarchy, one needs
\beq
kR \simeq \frac{1}{\pi}\ln\left(\frac{M_P}{M_5}\sqrt{\frac{k}{M_5}}\right) \approx 10 +\frac{1}{2\pi} \ln \left( \frac{k}{\rm TeV}\right) -
\frac{3}{2\pi} \ln \left( \frac{M_5}{10~{\rm TeV}}\right) \, .
\label{kR}
\eeq
In the parameter range of interest, the logarithmic dependence on $k$ and $M_5$ is very mild.

A natural explanation of the hierarchy in this setup relies on the possibility of having mild differences between the values of $S$ on the two branes. While na\"{i}vely the brane potentials in eqs.~\eqref{eq:Smart1}--\eqref{eq:Smart2} naturally allow for this, one may argue that since the dilaton realises dilatations non-linearly one expects it would enter the action as an analytic function of $\phi \sim e^S$.  The boundary potentials in eqs.~\eqref{eq:Smart1}--\eqref{eq:Smart2} would correspond to non-analytic functions, since $S \sim \log \phi$.  However, it has been argued that such logarithmic terms can actually arise in string-theoretic setups~\cite{Antoniadis:2001sw} and we will assume this to be the case.

To understand the origin of the hierarchy in different extra-dimensional setups it is useful to introduce the proper size of the extra dimension $L_5$, given by 
\beq
L_5 \equiv \int_{-\pi R}^{\pi R} dy\,  \sqrt{g_{55}}\, ,
\eeq
and the warp factor $w$, which can be defined through the ratio of the total spacetime volumes on the two different branes
\beq
w \equiv \left[ \frac{\int d^5 x \, \sqrt{-g}\, \delta (y-\pi R)/ \sqrt{g_{55}}}{\int d^5 x \, \sqrt{-g}\, \delta (y-0)/ \sqrt{g_{55}}} \right]^{1/4}
\, ,
\eeq
where the extra $\sqrt{g_{55}}$ comes from the definition of the covariant $\delta$-function. Both $L_5$ and $w$ are purely geometrical quantities that characterise the extra-dimensional compactified space and their definitions are explicitly invariant under coordinate reparametrization. The quantity $L_5$ corresponds to the proper length of the compactified space (in the special case of a single extra dimension). The warp factor $w$ encodes the information of how much each of the four spacetime dimensions is stretched between one end of the compactified space and the opposite end.

In the case of a single flat extra dimension of radius $R_{\rm LED}$,
\beq
L_5 = 2\pi R_{\rm LED} \, , \qquad w=1 \,, \qquad
M_P^2 = L_5 \, M_5^3  \qquad {\rm (LED)}.
\label{l5led}
\eeq
We recognize in eq.~(\ref{l5led}) the familiar LED result that the largeness of $M_P^2/M_5^2$ comes entirely from the effect of a large extra-dimensional volume ($L_5$). 

For RS one finds
\beq
L_5 = 2\pi R_{\rm RS} \, , \qquad w=e^{k_{\rm RS}\pi R_{\rm RS}}\, , \qquad M_P^2= \frac{w^2-1}{\ln w^2}\, L_5 \, M_5^3 \qquad {\rm (RS)},
\label{l5rs}
\eeq
where $k_{\rm RS}$ is the AdS inverse radius and $\pi R_{\rm RS}$ is the location of the brane (in the coordinate system where $g_{55}=1$). In RS the proper length $L_5$ is a number of order one, in natural units, while the hierarchy $M_P/M_5$ comes entirely from the warp factor $w$. Notice also that RS reproduces LED when $k_{\rm RS} \to 0$ and $R_{\rm RS}\to R_{\rm LED}$.

\begin{figure}[t]
\begin{center}
\includegraphics[width=0.47\textwidth]{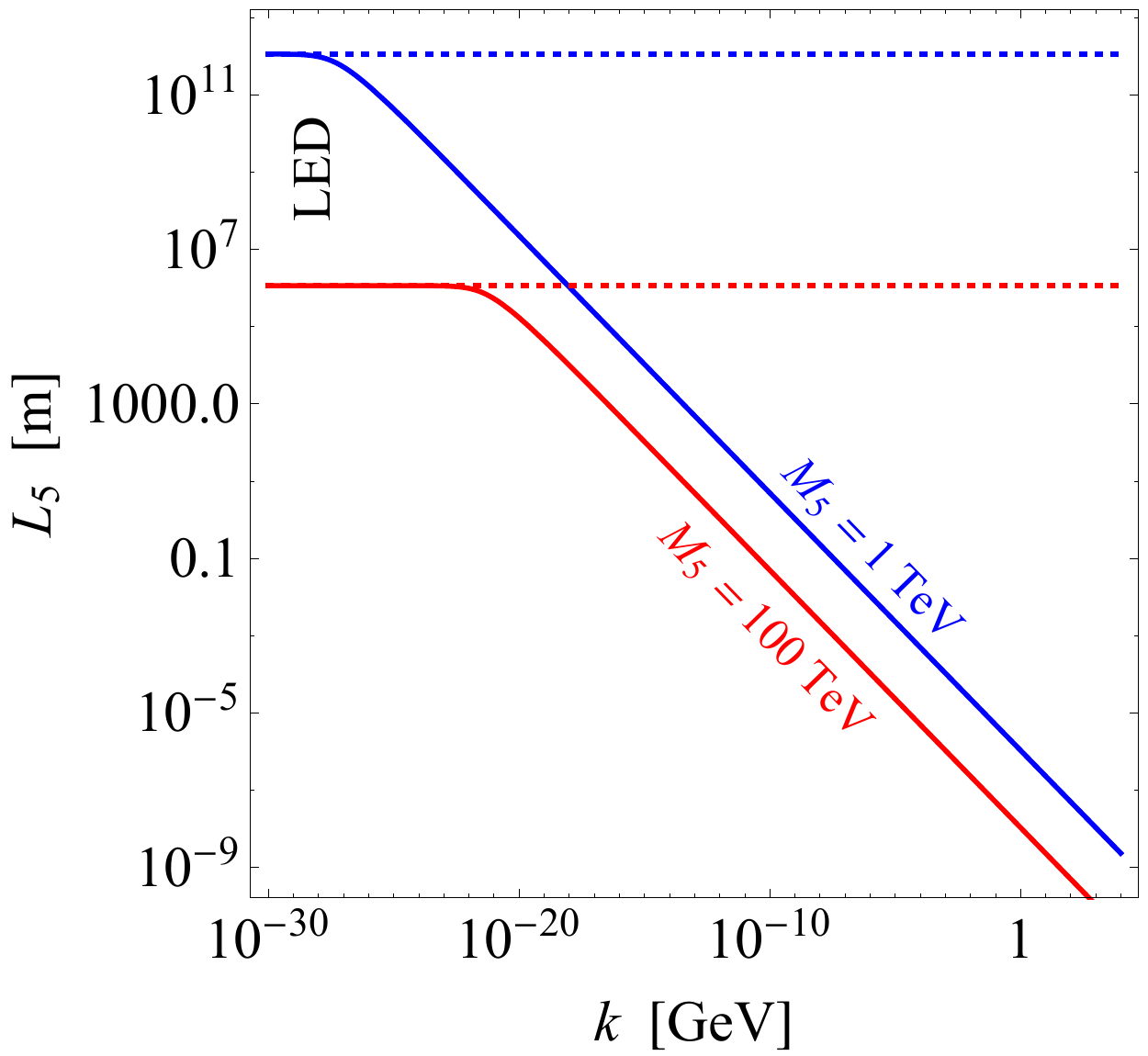}\qquad
\includegraphics[width=0.46\textwidth]{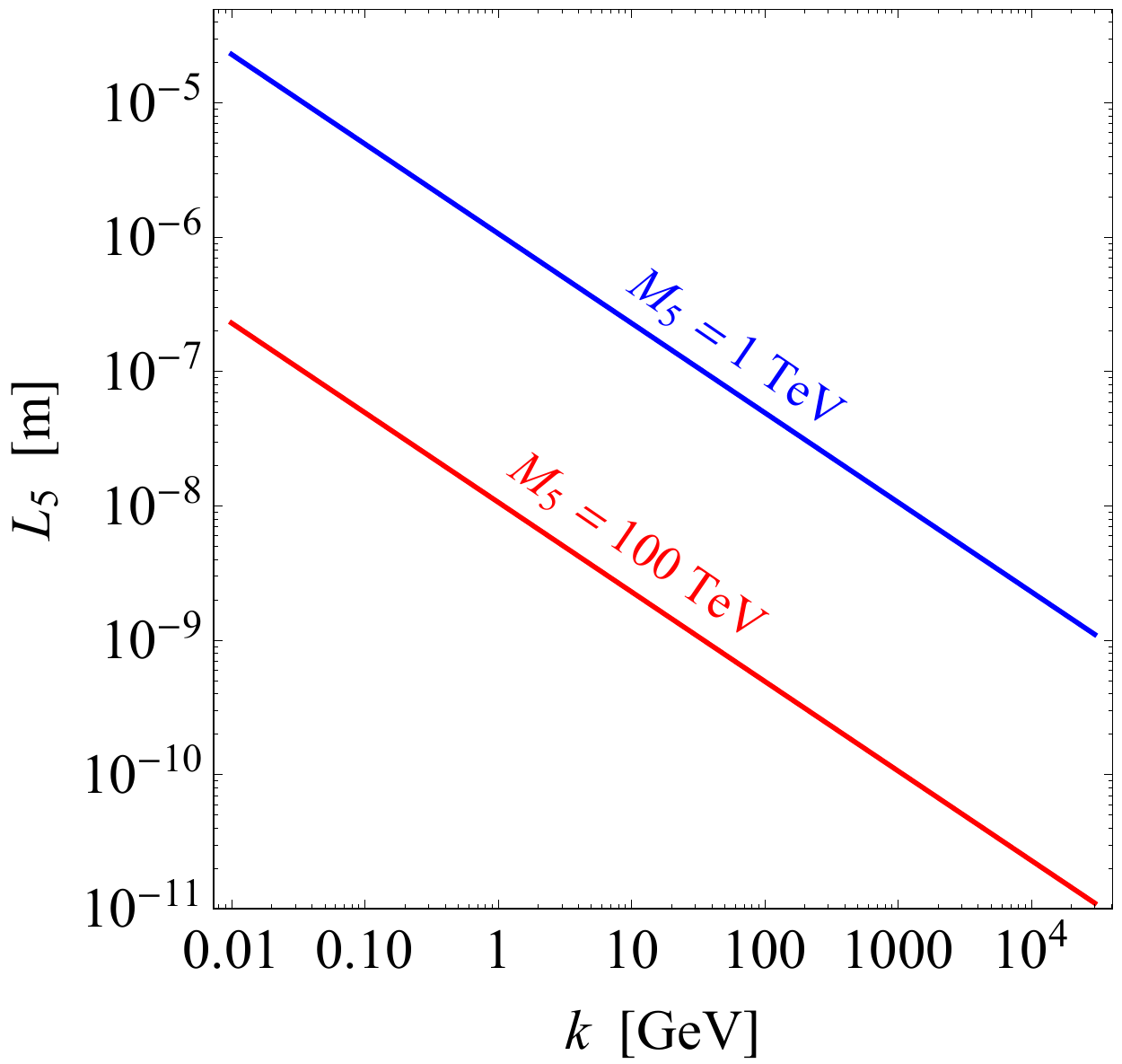}
\end{center}
\caption{The required physical size of the extra dimension as a function of $k$ for $M_5 = 1$~TeV (blue) and $100$~TeV (red). The horizontal dotted lines indicate the LED limit. The plot on the right zooms in on the phenomenologically most relevant range of $k$.}
\label{L5}
\end{figure}

For CW/LD, we obtain
\beq
L_5 = \frac{3}{k} \left(e^{\frac23 k\pi R} -1\right)   , ~~~ w=e^{\frac23 k\pi R}  , ~~~ M_P^2 = \frac{w^2+w+1}{3}\, L_5 \, M_5^3
  ~~~ {\rm (CW/LD)}.
\label{l5rs}
\eeq
In this case $R$ does not measure the proper size of the extra dimension, which  is much larger than its natural value $1/M_5$, as illustrated in figure~\ref{L5}, but not as extreme as in LED with one extra dimension. As a result, in CW/LD the hierarchy $M_P/M_5$ is explained by a combination of volume (as in LED) and warping (as in RS). While in RS the warp factor depends exponentially on the proper size, the warp factor of CW/LD is linearly proportional to $L_5$, so that $M_P \simeq k (L_5M_5/3)^{3/2}$.
However, $M_P$ is still exponentially sensitive to the parameters $k$ and $R$ which, as shown in section~\ref{sec:KK-masses-couplings}, determine the physical mass spectrum of the graviton excitations.

The difference between RS and CW/LD lies in the geometry of the corresponding compactified spaces. To appreciate this difference it is useful to work in both cases in the coordinate basis where the line element is $ds^2 =A^2(z)dx^2+dz^2$, such that the function $A(z)$ measures the warping of four-dimensional spacetime as a function of the physical distance along the extra dimension. For RS one finds $A(z)=e^{k_{\rm RS} z}$ and, because of the steep exponential behavior, an order-one separation between the branes is sufficient to obtain the large warp factor needed to explain the hierarchy. For CW/LD one finds $A(z)=2kz/3$ and, because of the slower linear dependence on the coordinate $z$, an exponentially large separation between the branes is needed to obtain the required warping factor. Nonetheless, the stabilization of the compactification radius is naturally obtained with order-one parameters, and the KK masses and interaction scales are set by the typical size of $k$, $R$, and $M_5$ and not by the size of $L_5$, which corresponds to a much larger distance.

Unlike the case of flat geometry, where a single extra dimension would need to have a Solar System size and is therefore excluded, CW/LD makes the possibility of a single relatively large extra dimension viable again. This is illustrated in figure~\ref{L5}, which shows the transition between the regime of roughly constant $kR$ (and $L_5 \propto k^{-2/3}$) at fixed $M_P$ to the regime at small $k$, in which the proper size of the extra dimension is frozen at $2\pi R$ and CW/LD turns into LED. This feature of CW/LD is important because the phenomenology of a single extra dimension differs significantly from that of multiple extra dimensions, as we will discuss. (A similar possibility arises in the low-curvature RS model, as pointed out and analyzed in refs.~\cite{Giudice:2004mg,Franceschini:2011wr}.)

A remarkable feature of CW/LD is that the same field $S$ that determines the spacetime geometry also stabilizes the compactification radius by fixing the factor $k\pi R$ in terms of the boundary conditions of $S$ on the branes, set by the brane potentials for the dilaton field appearing in $\mathcal{S}_{\mathcal{B}}$~\cite{Cox:2012ee,Im:2017eju}.  In this way the model automatically leads to radion stabilization. This is to be contrasted with the case of RS.  The original RS model suffers from a stabilization problem, especially pressing because a massless radion with TeV-scale interactions is experimentally ruled out. Luckily, a simple solution in RS is readily found by adding a bulk scalar field with a small mass term, which does not significantly perturb the AdS metric, but generates a stabilizing potential for the radion~\cite{Goldberger:1999uk}. CW/LD has this feature already built in in its structure. A more detailed discussion of the stabilisation is given in appendix~\ref{app:stabilization}.

\subsection{Stringy origins of the linear dilaton field theory}
Since the linear dilaton action is rather peculiar from a field theory perspective, it is instructive to consider the UV motivation for this setup.  Let us briefly review the stringy setting for the field theory we are considering, beginning with the worldsheet CFT in string theory.\footnote{For more details see \textit{e.g.}~refs.~\cite{Green:2012oqa,Green:2012pqa,Polchinski:1998rq,Polchinski:1998rr,Tong:2009np} and the recent papers~\cite{Kehagias:2017grx,Antoniadis:2017wyh}.}  We will only consider the spacetime metric and dilaton in the Polyakov action, omitting for now the antisymmetric two-form Ramond-Ramond (RR) field and the superpartner fields that are also present in the massless spectrum.\footnote{We do not consider the bosonic string theory due to its inherent problems with tachyons.}  The worldsheet action is
\be
\mathcal{S} = \frac{1}{2 \alpha'} \int d^2 \sigma \sqrt{-h} \left( g_{MN} (X) \partial_\alpha X^M \partial_\beta X^N h^{\alpha\beta}  + \frac{\alpha'}{4 \pi} S(X) \mathcal{R}^{(2)} +...\right) \, .
\label{eq:cft}
\ee
Here $X^M$ are the target space coordinates of an as-yet unfixed spacetime dimension $D$, $\alpha'$ is the string tension, and $h$ is the worldsheet metric.  To determine if the action is Weyl-invariant at the quantum level we may consider the $\beta$-functions that describe the renormalization of the massless fields, which are in superstring theory
\bea
\beta_{MN} (g) &  = & \alpha' \mathcal{R}_{MN} - 2 \alpha' \nabla_M \nabla_N S  +... + \mathcal{O} (\alpha'^2) \;, \\
\beta (S) &  = & \frac{10-D}{3} +\frac{\alpha'}{2} \nabla^2 S+ \alpha' \nabla_M S \nabla^M S+... + \mathcal{O} (\alpha'^2) \;.
\eea
In matching to the effective theory, these $\beta$-functions arise as the equations of motion for massless fields in the target spacetime. Thus the relevant action is given by~\cite{Fradkin:1984pq,Lovelace:1983yv,Callan:1985ia}\footnote{Note that this effective action may be written in numerous forms, all related by field redefinitions.  We will use the convention of ref.~\cite{Veneziano:1991ek}.}
\be
\mathcal{S} = \frac{M_D^{D-2}}{2} \int d^D x \sqrt{-g} \, e^{S} \left[\frac{D-10}{3 \alpha'} + \mathcal{R}  + \partial_M S \partial^M S +...+ \mathcal{O} (\alpha') \right]\,.
\label{eq:EFT}
\ee
This is analogous to the 5D linear dilaton action in the Jordan frame
\be
\mathcal{S} = \frac{M_5^{3}}{2} \int d^5 x \sqrt{-g}\, e^{S} \left(\mathcal{R}  + \partial_M S \partial^M S + 4 k^2 \right)\,,
\label{eq:Jordan}
\ee
from which, with a Weyl transformation $g \to e^{-2 S/3} g$, one obtains the Einstein frame bulk action \eqref{actionbulk}. From this matching we see that $k^2$ is in fact related to the string tension. The action \eqref{eq:EFT} is also often called the string frame action.  It describes the low energy effective theory for the dilaton and graviton below the scale of string excitations.

It is known that the string effective action is classically scale-invariant.\footnote{Discussions of this point are found in refs.~\cite{Witten:1985xb,Dine:1985kv}, however we will essentially follow the discussion of ref.~\cite{Green:2012pqa} Sec.~13.2, albeit in a different basis.}  To see this symmetry realised linearly we may make a Weyl transformation $g \to e^{- 2 S/(D-2)} g$ and write the dilaton field as $\phi = e^{- 2 S/(D-2)}$.  The effective action becomes
\be
\mathcal{S} = \frac{M_D^{D-2}}{2} \int d^D x \sqrt{-g} \left[\phi\, \frac{D-10}{3 \alpha'} + \mathcal{R}  -\frac{D-2}{4} \frac{\partial_M \phi \partial^M \phi}{ \phi^2} +...+ \mathcal{O} (\alpha') \right] .
\label{eq:weird}
\ee
In this basis we see that under the transformation $g_{MN} \to \lambda^2 g_{MN}$, $\phi \to \phi/\lambda^2$, the action is rescaled $\mathcal{S} \to \lambda^{D-2} \mathcal{S}$.  Thus, since the overall coefficient factors out of the classical equations of motion, they are invariant under this transformation.  Alternatively, in the standard Jordan-frame action \Eq{eq:Jordan} the transformation is realised non-linearly as $S\to S+\kappa$, where $\kappa$ is a constant, which again only rescales the total action.

This scale invariance is only classical, and will not be respected quantum mechanically as the normalisation of the action is physical.  Nonetheless, if the action only contains the dilaton and metric there are selection rules on how the quantum corrections enter.  They may be determined by considering $\hbar$ to transform as $\hbar \to \hbar e^{\kappa}$ (essentially, one can think of $e^{-S}$ as \emph{being} $\hbar$).  As fractional powers of $\hbar$ will not arise in the perturbative expansion, we see that the only additional scalar potential terms in the bulk action will be proportional to the dimensionless quantity $\delta V \propto (\hbar e^{-S})^n$, where $n$ is an integer.  Essentially the perturbative series is an expansion in $e^{-S}$, since this plays the role of a coupling constant~\cite{Dine:1985kv}.  This argument shows that the classical scale invariance can persist perturbatively if only the dilaton and metric are present.  However, non-perturbative effects, or the presence of additional fields with couplings that do not follow the classical scale invariance, may spoil any selection rules. 

An additional structural aspect motivated by the UV picture is that, since we are considering a superstring origin for this 5D action, the full effective action should also be supersymmetric.  Due to the supersymmetry of the effective action, the classical scale invariance of the action in fact survives quantum corrections.  This means that additional terms, such as a cosmological constant, are not expected to arise perturbatively in the bulk action.

We note that in superstring theory for $D\neq 10$ neither the dilaton Weyl anomaly nor the dilaton potential vanish.  However, 
one may consider the linear dilaton background
\be
\partial_M S \partial^M S = \frac{D-10}{3 \alpha'}  \,,\quad g_{MN} = \eta_{MN} \,,
\ee
for which the Weyl anomalies vanish, at the expense of a curved metric in the Einstein frame.  Since this background allows for a vanishing $\beta$-function, it describes a worldsheet CFT.  Such theories are known as non-critical string theories, in that the critical number of dimensions has not been chosen, however they still describe string worldsheet CFTs~\cite{Myers:1987fv,Veneziano:1991ek}.  This describes the basic features of the non-critical string UV motivation for the linear dilaton field theory.

In addition to the non-critical string motivation, the linear dilaton theory in 7D has been shown to arise in the Little String Theory (LST) limit of critical superstring theory with a stack of NS5 branes.  This limit corresponds to vanishing string coupling~\cite{Aharony:1998ub}.  It has been argued that LST-like theories are dual to backgrounds which asymptote to string theory in the linear dilaton background (see Sec.~3 of ref.~\cite{Aharony:1999ks}).
However, in this case one has two extra dimensions.  The linear dilaton form of the effective action persists to 5D when two of the additional dimensions are compactified~\cite{Antoniadis:2001sw}.
In this framework, $M_5$ and $k$ are determined by the string scale $M_s$, the number of the NS5 branes $N$, and the volume of the six compactified dimensions $V_6$ as $M_5 \sim M_s^3 V_6^{1/3}/N^{1/6}$ and $k \sim M_s/\sqrt N$~\cite{Antoniadis:2011qw}. Thus we have $M_5/k \sim M_s^2 (N V_6)^{1/3}$, hence the ratio depends on the UV parameters.

In this work we will not consider the phenomenology of the additional RR two-form field, nor the additional states required by supersymmetry.   All of these states would likely have the usual LD spectrum, with a mass gap and densely packed states.  Furthermore, the superpartners, such as the dilatino and gravitino, may be charged under remnants of the $R$-symmetries that may make the lightest states stable.   These neutral fermions will only be pair-produced in colliders and thus their greatest effect will likely be to contribute to the decay channels of the KK gravitons.  Furthermore, we do not include the effects of genuine string excitations which, if entering below the cutoff $M_5$, could lead to additional signatures.

\subsection{Impact of cosmological constant terms}
Having considered the stringy origins of the action, including the bulk supersymmetry, we can also take a viewpoint that is agnostic of the UV completion, and consider the action from a purely non-supersymmetric field theory perspective.  There are two terms that enter Einstein's equations in the form of a cosmological constant (CC): one from the bulk and one from the branes.  These terms may be written in the Jordan frame action as
\be
S_{{\rm CC}} = \int d^D x \sqrt{-g}\, \Lambda^D\, e^{\frac{2}{D-2} S}  ~~,~~ S_{\mathcal{B}{\rm CC}} = \int d^{D-1} x \sqrt{-\gamma}\, \Lambda_{\mathcal{B}}^{D-1}\, e^{\frac{2 (D-1)}{D-2} S} \;,
\ee
where $\gamma$ is the determinant of the induced metric. These terms violate classical scale invariance, which is why they did not arise in the string effective action at tree level.  However, from a purely field theory perspective classical scale invariance is impotent unless the UV theory and non-perturbative corrections also respect it. If we are being agnostic as to the UV, including considering non-supersymmetric theories, then we cannot rely on such arguments to forbid these cosmological constant terms.  Continuing in this vein, let us determine how large these terms can be before they significantly modify the solution.

Since the inclusion of the cosmological constant terms generally leads to Einstein's equations that cannot in general be solved analytically,\footnote{For its novelty, in appendix~\ref{app:exact} we include an isolated exact solution that satisfies a different set of boundary conditions, and is thus not of interest here.} we will instead perform a perturbative analysis valid for small cosmological constants, which we will parameterise as $\varepsilon$ (bulk) and $\varepsilon_\mathcal{B}$ (brane).  We consider a bulk potential which, in Einstein frame, is given by
\be
 \mathcal{S}_{\rm bulk} =  \int d^{5}x  \sqrt{-g}\, \frac{M_{5}^{3}}{2}\Bigg[\mathcal{R}(g) -\frac{1}{3} \partial_M S \partial^M S + 4 k^2 e^{-2 S/3} + \varepsilon k^2 \Bigg]
\label{ancora}
\ee
and a boundary potential given by\footnote{The junction conditions, which follow immediately from integrating the equations of motion following \Eq{ancora} at the branes give, for a given solution to the bulk equations of motion, conditions on the absolute value and gradient of the brane potentials $\lambda_i(S)$ and $d\lambda_i(S)/dS$. Thus there are an infinite class of brane potentials that can satisfy these constraints.  We present only those that correspond to the usual linear dilaton brane potential and a cosmological constant on the branes.  The quadratic brane terms proportional to $\mu_{\rm T, \rm P}$ are irrelevant for this discussion.}
\bea
&& \mathcal{S}_\mathcal{B}  =   \int d^{4}x\, dy  \sqrt{-\gamma} \, 4 k M_{5}^{3} \bigg\{ \left[ (1+\varepsilon_\mathcal{B}) e^{-S/3} + \frac{3\, \varepsilon}{32} - \frac{3\, \varepsilon_\mathcal{B}}{4}    \right]\delta (y)  \\
&  -&\left[ \left(1-\frac{3\, \varepsilon}{40} (\kappa^{-2}-\kappa^3)+\varepsilon_\mathcal{B} \kappa^{3} \right) e^{-S/3}  
+ \frac{3\, \varepsilon}{20} \left(\kappa^{-1}-\frac{3}{8} \kappa^{4}   \right)- \frac{3\, \varepsilon_\mathcal{B} }{4} \kappa^{4}   \right] \delta (y-\pi R) \bigg\} \nonumber \,,
\eea
where $\kappa = e^{-2 k \pi R/3}$.  The specific form of the terms at the boundaries has been chosen to give a vanishing 4D cosmological constant, thus that particular fine-tuning, present in all extra-dimensional models, has already been performed.   This means that any further tuning of $\varepsilon$ or $\varepsilon_\mathcal{B}$ is now \emph{in addition} to the tuning for a vanishing 4D CC.

Since $\kappa$ is exponentially small in our scenario,
the limit $\varepsilon_\mathcal{B} \to 0$ corresponds to the solution whenever the bulk CC is non-vanishing, which is one possibility of interest here, and the limit $\varepsilon \to 0$ corresponds to only having a CC on the $0$-brane, up to tiny corrections on the $\pi R$-brane, which is the second possibility of interest.  We present both solutions together for convenience.

Working in the conformally flat metric $ds^2 = e^{2 \sigma(y)} (\eta^{\mu\nu} dx_\mu dx_\nu + dy^2)$, we find that the bulk Einstein's equations and $S$ equation of motion
\begin{eqnarray}
-3\left[
3(\sigma^{\prime})^2 + \sigma^{\prime\prime}
\right]\eta_{\mu\nu} &=& e^{2\sigma}V(S)\eta_{\mu\nu} \,, \\
-4\sigma^{\prime\prime} &=& \frac{1}{3}\left[
e^{2\sigma}V(S) + (S^{\prime})^2
\right] , \\
e^{-2\sigma}\left(
3S^{\prime}\sigma^{\prime} + S^{\prime\prime}\right)  &=&  \frac{3}{2}V^{\prime}(S) \,,
\end{eqnarray}
the boundary conditions
\be
S(0) = 0 ~~,~~  \sigma(0) = 0 ~~,
\ee
and the junction conditions are all satisfied for the following dilaton and metric profiles
\bea
S & = & 2 k y + \frac{3 \varepsilon}{32} \left(1 - 2 k y - e^{- 2 k y} \right)+ \frac{\varepsilon_\mathcal{B}}{4} \left(5 - 2 k y - 5 e^{- 2 k y} \right) ~~,  \\
\sigma & = & \frac{2}{3} k y + \frac{\varepsilon}{80} \left( 6 e^{4 k y/3} - 5 (1+k y) - e^{-2 k y} \right) + \frac{\varepsilon_\mathcal{B}}{6} \left(1 - k y - e^{- 2 k y}  \right) ~~.
\label{eq:sol}
\eea

The physical impact of these modifications can be expressed in numerous ways. One way is to derive from the expression of the 4D Planck mass the value of $kR$, which characterizes the physical mass spectrum of the KK modes.
We find that, at leading order in $\varepsilon$ and $\varepsilon_\mathcal{B}$ and for the same input values of $M_5$ and $k$, the usual CW/LD value of $R$ given by eq.~(\ref{kR}) is modified as follows
\be
\frac{R_{\varepsilon , \varepsilon_\mathcal{B}}}{R} = 1 + \varepsilon_\mathcal{B} - \frac{3\, \varepsilon}{32} \left( \frac{18}{25  \pi k R} e^{4  \pi k R/3} -1 \right) .
\ee

Let us consider the effect of $\varepsilon$.  For the linear dilaton solution in the conformally flat coordinates we have $k R \sim 10$, thus $e^{4 \pi k R/3} \sim 10^{18}$.  This means that unless $|\varepsilon| \lesssim 10^{-16}$, the value of $kR$ and, consequently, the mass spectrum will be significantly different from the usual CW/LD prediction. Similarly,
 the metric will have deviated significantly from the linear dilaton form in moving from the IR brane to the UV brane.  Of course, nothing radical will happen, with the metric flowing smoothly from the linear dilaton form at $|\varepsilon| \ll 10^{-16}$, to a dS or AdS-like solution.  We see this reflected in the reduced (comoving) radius required to achieve the desired hierarchy of Planck scale.  In the AdS case the mass spectrum and wavefunctions would correspond to $k_{\rm RS} \sim k\sqrt{\varepsilon/12} $, at larger $\varepsilon$.  Thus one needs to require an extremely small value of $\varepsilon$ to retain the linear dilaton solution.  From a field theory perspective this is an enormous tuning, however by considering the superstring motivation the bulk theory is supersymmetric and this protects against the generation of $\Lambda_D$ if the action started out with classical scale invariance, which is the case for the string effective action.  The requirement is actually a little stronger than requiring solely supersymmetry, since the vacuum expectation value of the superpotential must also vanish to avoid supersymmetric AdS solutions.  It has recently been shown that the CW/LD background is indeed consistent with a bulk supersymmetry~\cite{Kehagias:2017grx,Antoniadis:2017wyh}.  Supersymmetry must be broken on the SM brane, however locality still protects against the generation of supersymmetry breaking in the bulk action (see \textit{e.g.}~ref.~\cite{ArkaniHamed:1999dz} for related discussions).  Thus, given that the motivation for the linear dilaton setup is coming from string theory, the consequent bulk supersymmetry naturally protects against the generation of $\varepsilon$.

Now consider $\varepsilon_\mathcal{B}$.  The story for this parameter is very different.  The reason is that supersymmetry breaking must exist at least at the TeV scale on the SM brane, and thus there is no symmetry, including supersymmetry, that can protect against the generation of $\varepsilon_\mathcal{B} \sim M_5/k$.  Studying the perturbation to the metric we see that the metric retains the linear dilaton form for $\varepsilon_\mathcal{B}$ not too large. Similarly, the comoving radius required is only significantly modified for $\varepsilon_\mathcal{B}\sim1$.  Thus, although we have no control over the size of the $\varepsilon_\mathcal{B}$ brane cosmological constant term on the TeV brane, the setup is robust to its presence as long as $\varepsilon_\mathcal{B}$ is not too large, which in turn does not require significant fine-tuning unless $k \ll M_5$.

To summarise, bulk supersymmetry protects against the presence of a bulk cosmological constant that would otherwise radically alter the setup.\footnote{This discussion has focussed on the presence of cosmological constant terms, however in the absence of supersymmetry the dilaton potential itself is also not protected from dangerous radiative corrections, the presence of which may lead to similarly significant corrections to the form of the bulk geometry.}  On the other hand, the lack of supersymmetry on the SM brane does not imperil the solution as the brane cosmological constant does not feed into a bulk cosmological constant, due to locality, and it does not generate a significant deviation from the linear dilaton solution unless $\varepsilon_\mathcal{B}$ is $\mathcal{O}(1)$.  Thus, if the UV motivations are taken seriously, in particular the bulk supersymmetry, then the theory does not exhibit significant tuning beyond that required for a vanishing 4D cosmological constant.  On the other hand, if one were agnostic as to the UV structure one could only conclude that the necessity for $|\varepsilon| \ll 10^{-16}$ corresponds to an extreme additional tuning.

At this stage it is worth commenting on the deconstruction of the continuum model, which gives the discrete clockwork graviton model presented in Sec.~2.5 of ref.~\cite{Giudice:2016yja}.  While this is a perfectly valid, diffeomorphism invariant, multi-graviton model, since it follows from a deconstruction of a solution of Einstein's equations accompanied by the dilaton, the question of the cosmological constant for the continuum model is equally relevant for the discrete.  There is nothing to forbid a cosmological constant at each site.  Translation invariance enforces that all cosmological constants must be equal, just as $\varepsilon$ is position-independent in the continuum, however no symmetry protects the overall magnitude of each term and the cosmological constant could imperil the clockwork wavefunctions.  As with the continuum model, the only plausible solution to this problem is supersymmetry.  Thus, if one wants the discrete setup to be safe from large quadratic corrections, one really requires an entire 4D copy of \emph{Supergravity}, with the associated gravitino at each site.  Thus the discrete model of ref.~\cite{Giudice:2016yja} alone is not sufficient to address the hierarchy problem.

\subsection{$k$ \textit{vs.}\ $M_5$}

In CW/LD, $M_5$ and $k$ are two independent input parameters. As the collider phenomenology changes significantly by varying the ratio $k/M_5$, we think it is interesting to explore a wide range of this ratio and to study the corresponding experimental features, some of which are novel and  characteristic. However, we first want to discuss in this section which values of $k/M_5$ are reasonable. 

The CW/LD action in eq.~({\ref{ancora}) has an enhanced symmetry (a shift in $S$) for $k \to 0$. This suggests that small $k/M_5$ is technically natural. On the other hand,
in the string theory setup, both $M_5$ and $k$ might be expected to arise from the single dimensionful string tension parameter $\alpha'$, although their relation is not uniquely determined. In the little string theory limit a hierarchy of $\mathcal{O}(N^{1/3})$ is possible, where $N$ is the number of NS5-branes in the theory.

While the form of the bulk action is motivated by string theory in the UV, the boundary action has no such motivation.  In particular, we have no guidance as to the parameters $S_{\rm T}$, $S_{\rm P}$, which determine the form of the dilaton solution to the equations of motion.  We may thus write $\sigma_{y=0} = \sigma_0$ and $S_{y=0} = S_0$, in which case $2 k \pi R = |S_{\rm T}-S_{\rm P}|$ and $S_0 = S_{\rm T}$.  On the equations of motion the overall factors can be absorbed into a redefinition of the 5D Planck scale $M_{5,{\rm eff}} = M_5 e^{\sigma_0} $, $k_{\rm eff} = k e^{\sigma_0-S_0/3}$, thus the only effect of taking general boundary conditions on these constants is to modify the relationship between the observables $M_{5,{\rm eff}}$ and $k_{\rm eff}$.  As a result, since we do not have guidance on the scale of these parameters from the UV, in this paper we are instead open minded as to considering relatively small $k/M_5$, especially because this possibility gives rise to a variety of interesting collider phenomena.  It should also be kept in mind, however, that to have the linear dilaton solution with $k\ll M_5$ does require tuning of the brane CC at that order.

In this work, instead of using lower bounds on $k$ based on theoretical considerations, we will be guided by the phenomenological limits from beam dump experiments, supernova emission, and nucleosynthesis. For $M_5$ in the domain of interest to present and future collider experiments, the lower limits on $k$ are in the 10~MeV / 1~GeV range~\cite{Baryakhtar:2012wj}, taking into account that astrophysical and cosmological bounds are subject to large numerical uncertainties.

\subsection{KK mode mass spectrum and couplings}
\label{sec:KK-masses-couplings}

The KK gravitons have masses
\beq
m_0 = 0\,,\qquad
m_n^2 = k^2 + \frac{n^2}{R^2}\,,\qquad
n = 1, 2, 3, \ldots
\label{KKG-masses}
\eeq
and couple to the SM stress-energy tensor $T^{\mu\nu}$ as
\beq
{\cal L} \supset -\frac{1}{\Lambda_{G}^{(n)}}\, \tilde h^{(n)}_{\mu\nu}\,T^{\mu\nu} \,,
\eeq
\beq
\Lambda_G^{(0)2} = M_P^2\,,\qquad
\Lambda_G^{(n)2} = M_5^3\pi R\left(1+\frac{k^2R^2}{n^2}\right) = M_5^3\pi R\left(1-\frac{k^2}{m_n^2}\right)^{-1} \, .
\label{KKgraviton-couplings}
\eeq
The zero mode is the usual massless graviton, while the rest of the KK modes appear after a mass gap of order $k$ and their couplings to the SM are not suppressed by $M_P$. A unique property of this scenario is that the KK modes form a narrowly-spaced spectrum above the mass map. At the beginning of the spectrum, the relative mass splitting is
\beq
\frac{m_2 - m_1}{m_1} \simeq \frac{3}{2\,(kR)^2} \approx 1.5\% \,,
\eeq
{\it i.e.}, comparable to the diphoton and dielectron invariant mass resolutions in ATLAS and CMS, which are typically around $1\%$~\cite{ATLAS-CONF-2015-081,ATLAS-CONF-2016-059,Aaboud:2017yyg,Khachatryan:2016yec,Aaboud:2017buh,CMS-PAS-EXO-16-031}. The splittings then increase, as shown in figure~\ref{dm}, reaching 
a maximum value
\beq
\left( \frac{m_{n+1} - m_n}{m_n}\right)_{\rm max} \simeq \frac{1}{2kR}\approx 5\% \qquad {\rm for}~n \approx kR
\, .
\eeq
Eventually they start decreasing, becoming asymptotically
\beq
\frac{m_{n+1} - m_n}{m_n} \simeq \frac{1}{n} \qquad {\rm for}~n \gg kR \, ,
\eeq
thus dropping below the experimental resolution at $n \sim 100$.

\begin{figure}[t]
\begin{center}
\includegraphics[width=0.62\textwidth]{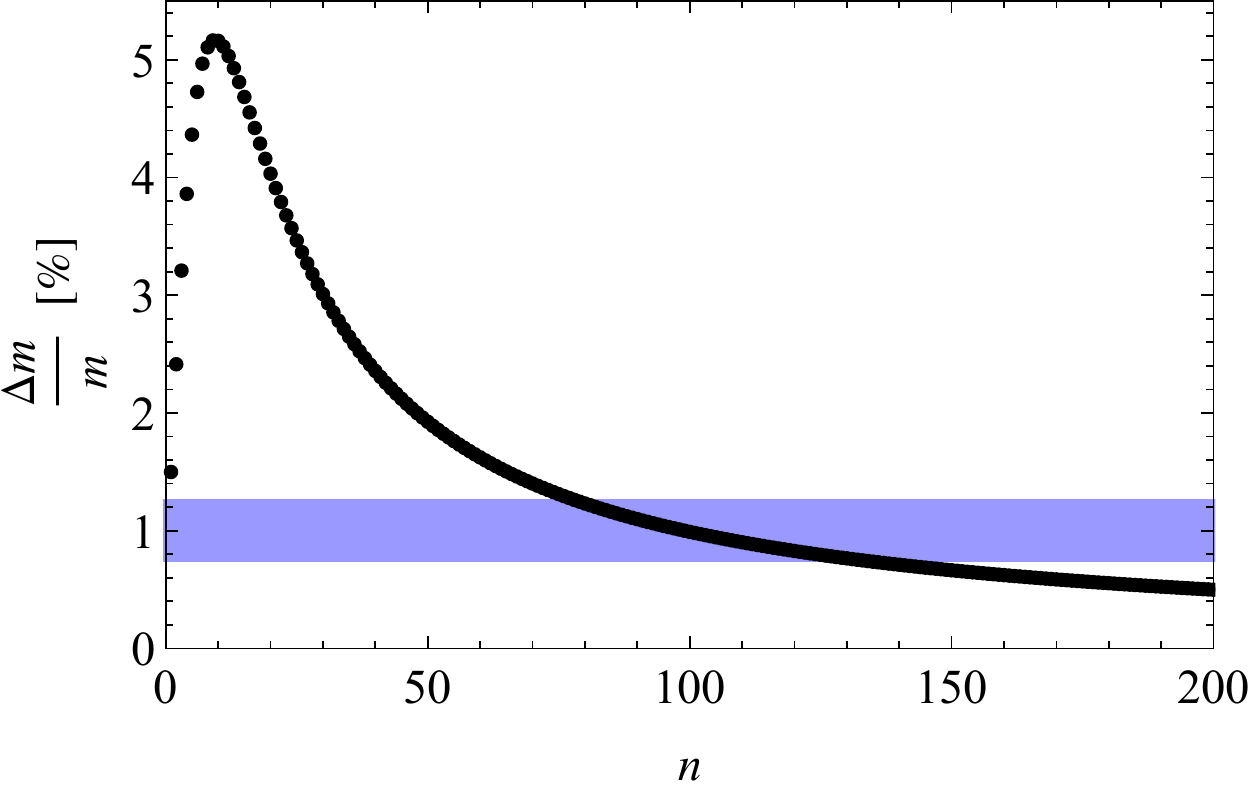}
\end{center}
\caption{KK graviton mass splittings as a function of the mode number. The behavior is shown for $M_5 = 10$~TeV, $k = 200$~GeV, however the result does not strongly depend on these parameters. The blue band indicates the typical ATLAS and CMS resolutions in the diphoton and dielectron channels ($\approx 1\%$).}
\label{dm}
\end{figure}

The physics of the KK dilatons (including the radion degree of freedom) depends on the details of the brane potentials that stabilize the extra dimension~\cite{Kofman:2004tk,Cox:2012ee}. We first focus on the limit of rigid boundary conditions for the dilaton field, obtained when the mass parameters $\mu_{\rm T,P}$ in the brane-localized potential in eq.~(\ref{actionboundary}) are infinitely large. The general case is presented in detail in appendix~\ref{app:ScalarSpectrum}. The dominant features of the phenomenology of the model as a whole are largely independent of the details of the brane-localized potential. 

For rigid boundary conditions, the KK dilatons have masses
\beq
m_0 = \sqrt{\frac89}\;k\,,\qquad
m_n^2 = k^2 + \frac{n^2}{R^2}\,,\qquad
n = 1, 2, 3, \ldots
\label{KKdilaton-masses}
\eeq
and couple to SM particles via the trace of $T^{\mu\nu}$ as\footnote{These expressions follow from eqs.\ (3), (4) and (39) of ref.~\cite{Kofman:2004tk}, and we have confirmed them. However, while the expression for $\Lambda_\Phi^{(n)2}$ agrees with the result given in eq.~(41) of ref.~\cite{Kofman:2004tk} in the limit $n \ll kR$ taken there, our expression for $\Lambda_\Phi^{(0)2}$ is \emph{larger} than theirs by a factor of $2$. Our expressions for both $\Lambda_\Phi^{(0)2}$ and $\Lambda_\Phi^{(n)2}$ (in the same limit) agree with those in eqs.~(4.4)--(4.5) of ref.~\cite{Cox:2012ee} after taking into account that the four- and five-dimensional Planck masses, $M_{Pl}$ and $M$, of ref.~\cite{Cox:2012ee} are related to our $M_P$ and $M_5$ as $M_{Pl}^2 = M_P^2/2$, $M^3 = M_5^3/2$. We thank the authors of ref.~\cite{Cox:2012ee} for clarifying this to us.\label{KKdilaton-factors-2}}
\beq
{\cal L} \supset -\frac{1}{\Lambda_{\Phi}^{(n)}}\, \phi_n T^\mu_\mu \,, \qquad
\Lambda_{\Phi}^{(0)\,2} = \frac{18 M_5^3}{k}\left(1-e^{-\frac23 \pi k R}\right)\,,
\label{KKdilaton-couplings-0}
\eeq
\beq
\Lambda_{\Phi}^{(n)\,2} = \frac34 M_5^3 \pi R \left(10 + \frac{k^2R^2}{n^2} + \frac{9\, n^2}{k^2R^2}\right)
=\frac{27}4 M_5^3 \pi R\, \frac{m_n^2}{k^2}\left( 1-\frac{8\, k^2}{9\, m_n^2}\right) \left( 1-\frac{k^2}{m_n^2}\right)^{-1} \, .
\label{KKdilaton-couplings-n}
\eeq
For large $n$, we find $\Lambda_{\Phi}^{(n)\,2}/\Lambda_G^{(n)\,2}= (27/4)(m_n^2/k^2)$, which shows a suppression of the interaction scale of dilatons with respect to gravitons, when compared at equal masses.

\begin{figure}[t]
\begin{center}
\includegraphics[width=0.48\textwidth]{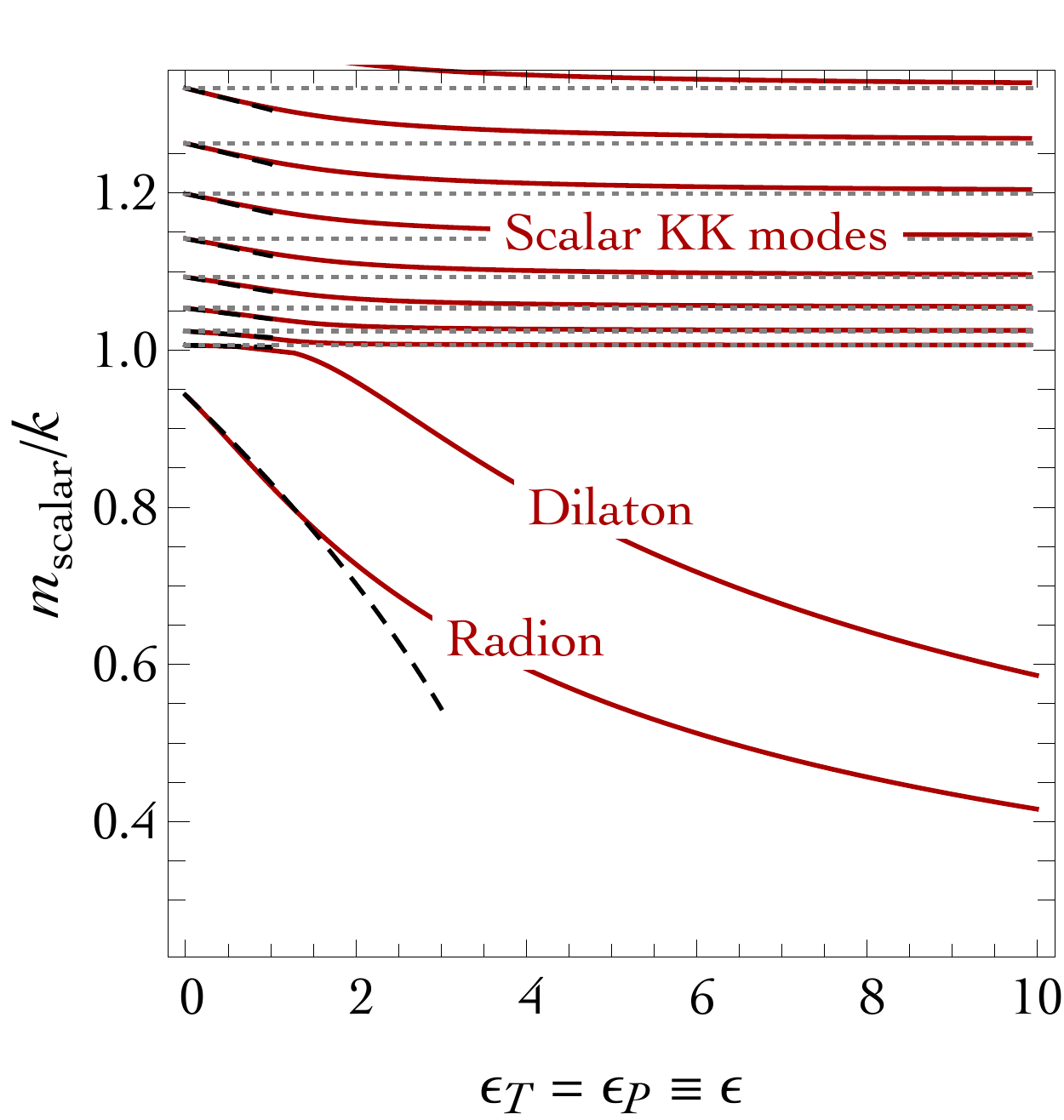}
\end{center}
\caption{Scalar KK mode masses as a function of the boundary conditions on the dilaton field, where $\epsilon = 0$ corresponds to rigid boundary conditions (see appendix~\ref{app:ScalarSpectrum} for details). The horizontal dotted gray lines correspond to the spectrum $m_n^2 = k^2 + n^2/R^2$. The dashed black lines correspond to the analytical approximations discussed in appendix~\ref{app:ScalarSpectrum}, eqs.~(\ref{eq:MassRadion})--(\ref{eq:MassKK}).}
\label{scalars_vs_BC}
\end{figure}

If the boundary conditions are relaxed by lowering the mass parameters $\mu_{\rm T,P}$ in the brane potential, the $n = 0$ and $n = 1$ modes can become significantly lighter, and both are massless in the $\mu_{\rm T,P} \to0$ limit (unstabilized limit). At the same time, all the higher KK dilaton modes shift down by one mode (as in ``Hilbert's Hotel'') so that the massive mode spectrum in the unstabilized limit ($\mu_{\rm T,P}=0$) is again described by the expression for $m_n^2$ in eq.~\eqref{KKdilaton-masses} with $n$ starting from $1$. This is shown in figure~\ref{scalars_vs_BC}. The coupling strengths to $T_\mu^\mu$ get modified as well. In the unstabilized limit, modes with $m_n \gg k$ have 
\beq
\Lambda_{\Phi}^{(n)2} \simeq \frac34 M_5^3 \pi R\, \frac{m_n^4}{k^4}\, \epsilon^2 \,,
\eeq
where the dimensionless parameter $\epsilon$, which vanishes in the rigid limit ($\mu_{\rm T,P} \to \infty$), is defined in appendix~\ref{app:ScalarSpectrum}. By comparing this with eq.~\eqref{KKdilaton-couplings-n}, one can see that the interactions with $T_\mu^\mu$ get significantly suppressed as one goes away from the rigid limit. In addition, there appear couplings to ${\cal L}_{\rm SM}$, the SM Lagrangian,
\beq
{\cal L} \supset -\frac{1}{\Lambda_{\varphi}^{(n)}}\, \phi_n {\cal L}_{\rm SM} \,,
\label{dilaton-LSM-coupling}
\eeq
where in the unstabilized limit for $m_n \gg k$
\beq
\Lambda_{\varphi}^{(n)\,2} \simeq 3 M_5^3 \pi R \,.
\eeq

\subsection{KK mode decays: final states, branching fractions, lifetimes}

\begin{table}[t]
\begin{center}
\begin{tabular}{c|c|c|c|c|c|c|c}
$gg$ & $\sum_i q_i\bar q_i$ & $W^+W^-$ & $ZZ$ & $hh$ & $\gamma\gamma$ & $\sum_i\ell_i^+\ell_i^-$ & $\sum_i\nu_i\bar{\nu}_i$ \\\hline
34\% & 38\% & 9.2\% & 4.6\% & 0.35\% & 4.2\% & 6.4\% & 3.2\%
\end{tabular}
\end{center}
\caption{Relative branching fractions of KK gravitons to SM particles, ignoring phase space effects.}
\label{BRs}
\end{table}

Since the KK gravitons couple to the SM via $T^{\mu\nu}$, the relative branching fractions into the various SM particles are the same as in any 5D model with the SM on a brane. These are shown in table~\ref{BRs} for KK gravitons that are much heavier than the SM particles. The detailed expressions, including phase space effects, which must be taken into account for lighter KK gravitons, are given in appendix~\ref{app:grav-SM-decays}. The total decay rate of a mode-$n$ KK graviton into SM particles in this limit is
\beq
\Gamma_{G_n \to {\rm SM}} = \frac{283}{960\pi}\,\frac{m_n^3}{\Lambda_G^{(n)2}}
= \frac{283}{960\pi^2}\frac{m_n^3}{R M_5^3}\left( 1-\frac{k^2}{m_n^2}\right) \, .
\eeq
In the absence of other types of decays, the resulting lifetime is
\beq
c\tau_n \approx 6.6\times 10^{-8}\,\mbox{m}\left(\frac{M_5}{10~\mbox{TeV}}\right)^3\left(\frac{1~\mbox{GeV}}{k}\right)\left(\frac{100~\mbox{GeV}}{m_n}\right)^3\left(\frac{kR}{10}\right) \left( 1-\frac{k^2}{m_n^2}\right)^{-1} \, .
\label{ctauap}
\eeq
We see that the KK graviton decays can be prompt, but it is also possible, especially if $M_5 \gtrsim 10$~TeV, that KK modes below a certain mass will be displaced, or even stable on detector scale.
The last factor in eq.~(\ref{ctauap}) gives a further enhancement to the lifetimes of the lightest modes besides the $m_n^{-3}$ factor. For instance, for the first mode we find $(1-k^2/m_1^2)^{-1} \approx k^2 R^2 \approx 100$. As a result, it is possible that, within the same theory at a given $M_5$ and $k$, some KK gravitons decay promptly, while others lead to displaced vertices.

Importantly, we find that decays of KK gravitons to SM particles are not the full story. Graviton self-interactions allow a heavy KK graviton to decay to a pair of lighter KK gravitons.\footnote{Numerical results for graviton-to-graviton decays were also considered for RS models in \cite{Davoudiasl:2001uj}.} The method of calculation and the detailed expressions for the rates of these decays are presented in appendix~\ref{app:grav-grav-decays}. Their total rate, in the limit $n \gg kR \gg 1$, is given by
\beq
\Gamma_{G_n \to \sum G_\ell G_m} \simeq 
\frac{595}{3 \times 2^{14}\, \pi^2}\,
\frac{m_n^{7/2}}{k^{1/2}RM_5^3} \, ,
\label{approx-KKG-KKG}
\eeq
while for smaller values of $n$ it is reduced relative to this expression in the fashion shown in figure~\ref{KKG-KKG-correction}. This rate can be quite sizable, and even dominate over decays to SM particles for large $n$:
\beq
\frac{\Gamma_{G_n \to \sum G_\ell G_m}}{\Gamma_{G_n \to {\rm SM}}} 
\approx 4.1\times 10^{-2}\sqrt{\frac{m_n}{k}} \;.
\eeq
This has significant effects on the phenomenology. First, decays to SM particles are diluted, as shown for the example of the diphoton channel in figure~\ref{BRs-w-KKG-KKG} (left). Second, signatures due to decays to lighter KK gravitons can be important because the branching fraction for such decays can be large, as shown in figure~\ref{BRs-w-KKG-KKG} (right).

\begin{figure}[t]
\begin{center}
\includegraphics[width=0.7\textwidth]{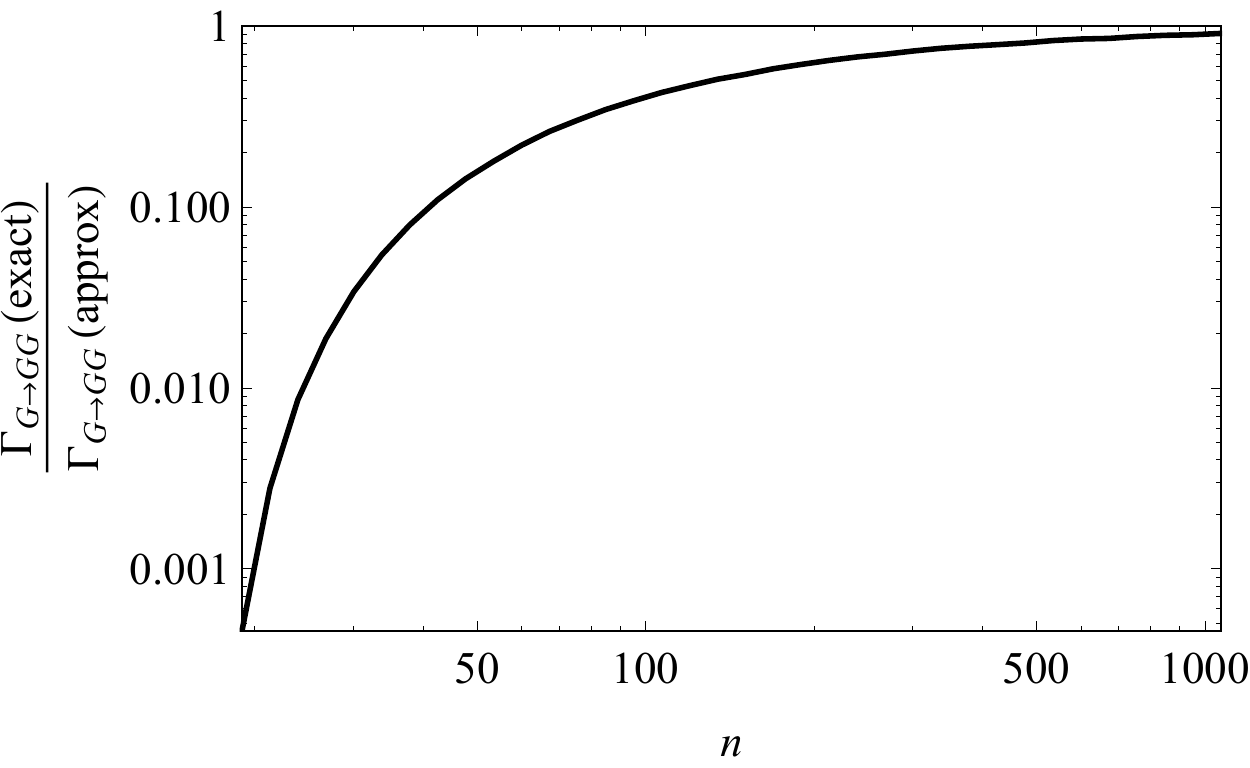}
\end{center}
\caption{Exact partial width of a mode-$n$ KK graviton to pairs of lighter KK gravitons relative to the asymptotic expression given in eq.~\eqref{approx-KKG-KKG}. Here we have taken $M_5 = 10$~TeV, $k = 10$~GeV, however the dependence on these parameters is weak.}
\label{KKG-KKG-correction}
\end{figure}

\begin{figure}
\begin{center}
\includegraphics[width=0.48\textwidth]{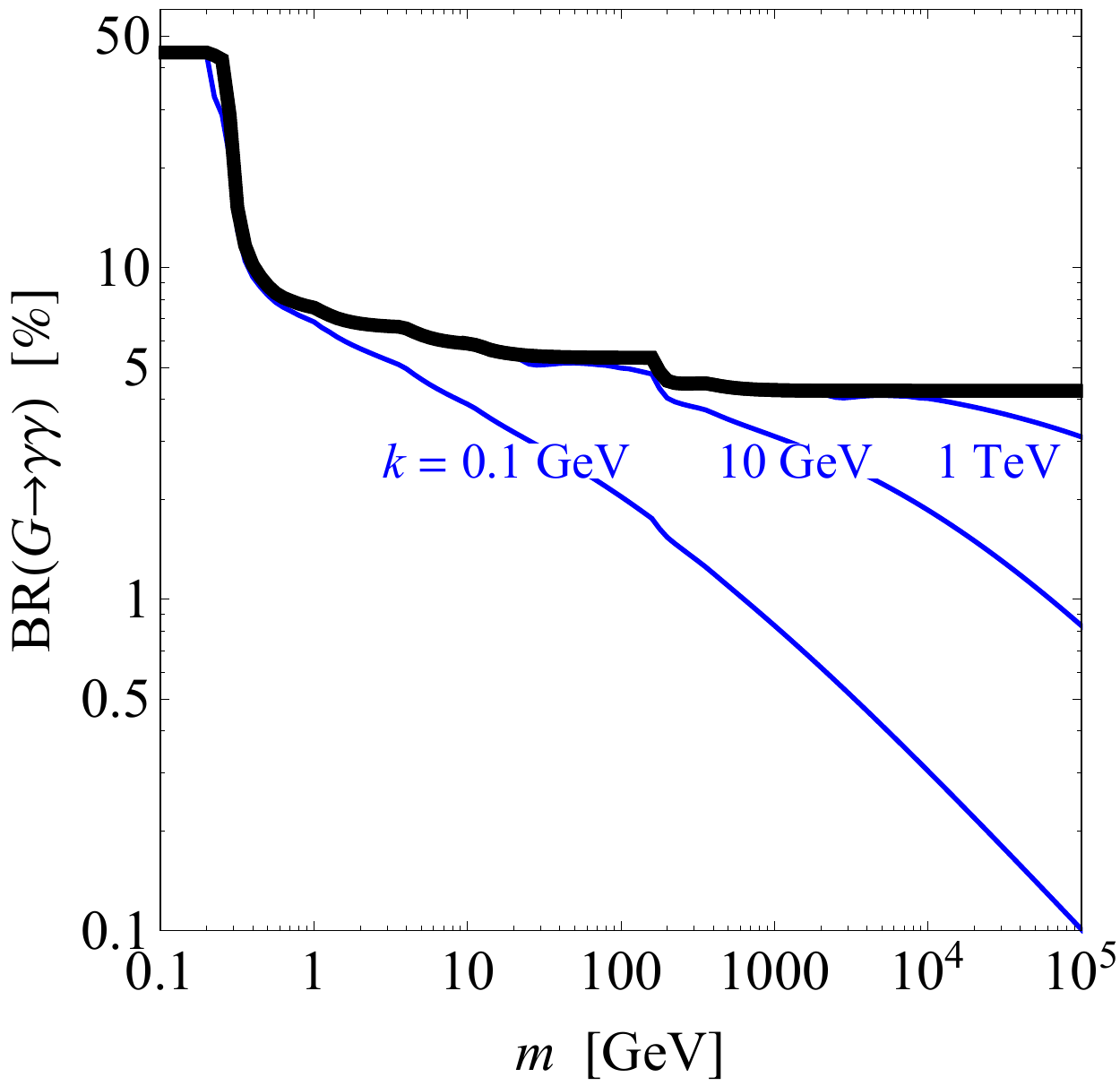}\quad
\includegraphics[width=0.48\textwidth]{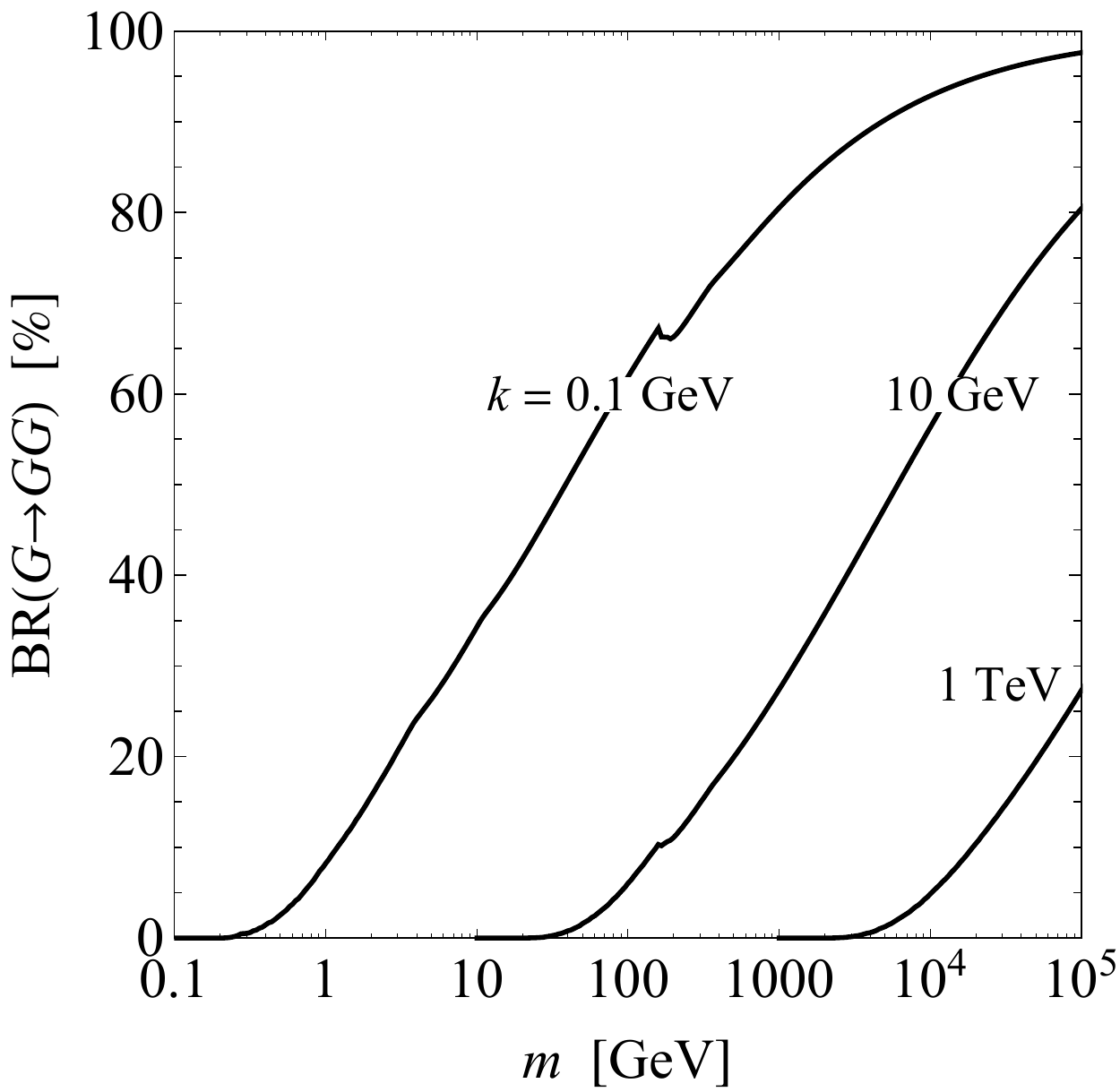}
\end{center}
\caption{KK graviton branching fractions to photons (left) and to lighter KK gravitons (right) for $k = 0.1$, 10 and 1000~GeV, as a function of the KK graviton mass. In the left plot, the thick black curve shows the result that would be obtained without accounting for decays to lighter KK modes.}
\label{BRs-w-KKG-KKG}
\end{figure}

Even though the contribution to the total width ($\Gamma_n$) from decays to lighter KK gravitons can be significant, note from~\eqref{approx-KKG-KKG} that all modes within the range of validity of the theory ($m_n \lesssim M_5$) have $\Gamma_n \ll m_n$. Even the stricter condition $\Gamma_n < m_n - m_{n-1}$, which allows us to treat the KK excitations as individual resonances, is fulfilled as long as
\beq
m_n \lesssim 6.8 \left(\frac{k}{M_5}\right)^{1/7}M_5 \,.
\eeq
This is satisfied by all modes with $m_n \lesssim M_5$ if
\beq
k \gtrsim 1.5\times 10^{-6}\, M_5 \approx 15~\mbox{MeV}\left(\frac{M_5}{10~\mbox{TeV}}\right) .
\eeq
Note also that the extra contribution to the decay rate does not preclude displaced decays, since only high-$n$ modes are affected (see figure~\ref{KKG-KKG-correction}).

A KK graviton can also decay to a KK scalar and a KK graviton, or to a pair of KK scalars.\footnote{Cascade decays of gravitons to KK scalars were also considered for RS models in \cite{Dillon:2016tqp}.} The branching fractions for such decays are usually small, as shown in figure~\ref{BRs-KKG-to-KKS}.

\begin{figure}
\begin{center}
\includegraphics[width=0.47\textwidth]{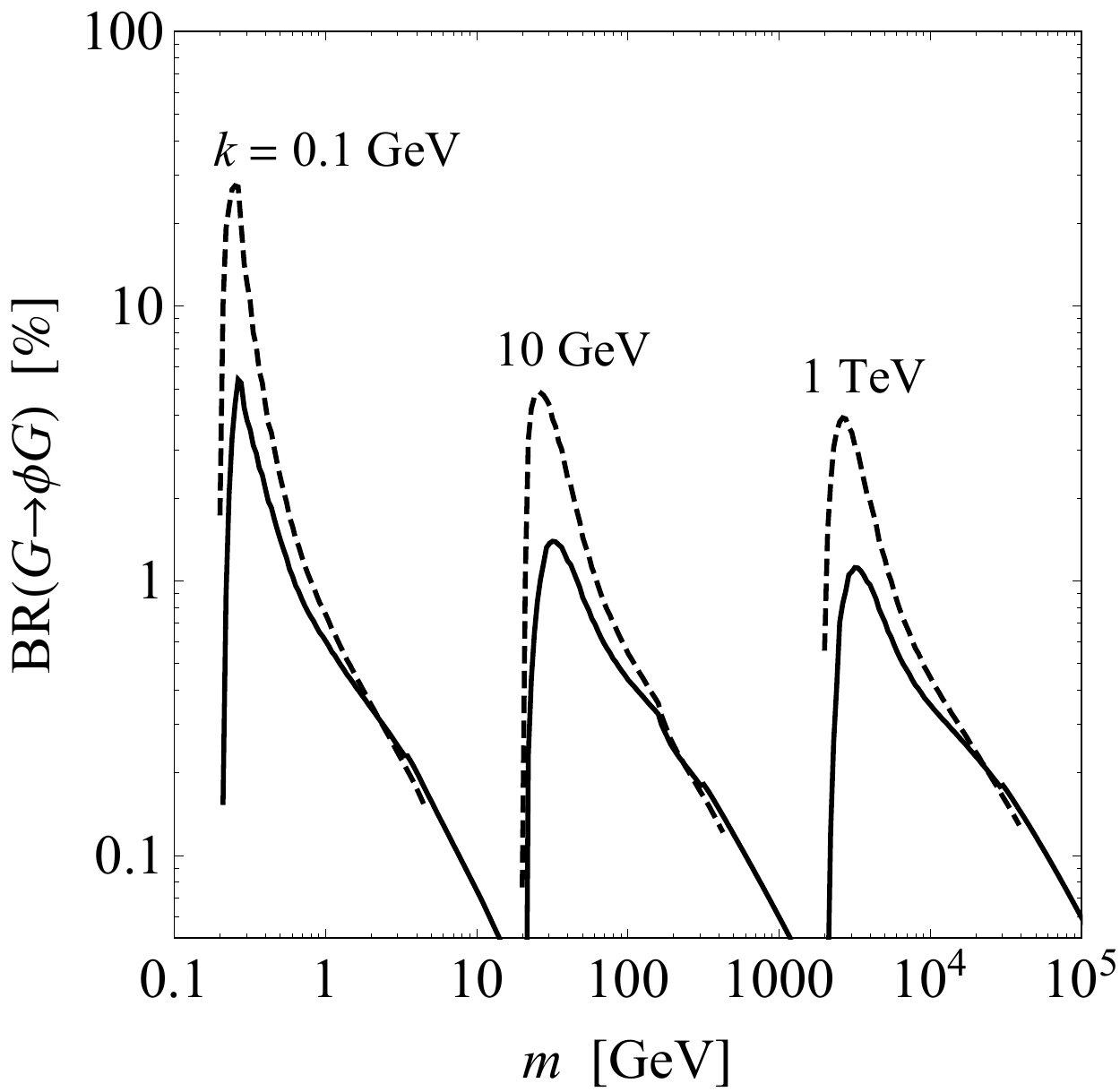}\quad
\includegraphics[width=0.49\textwidth]{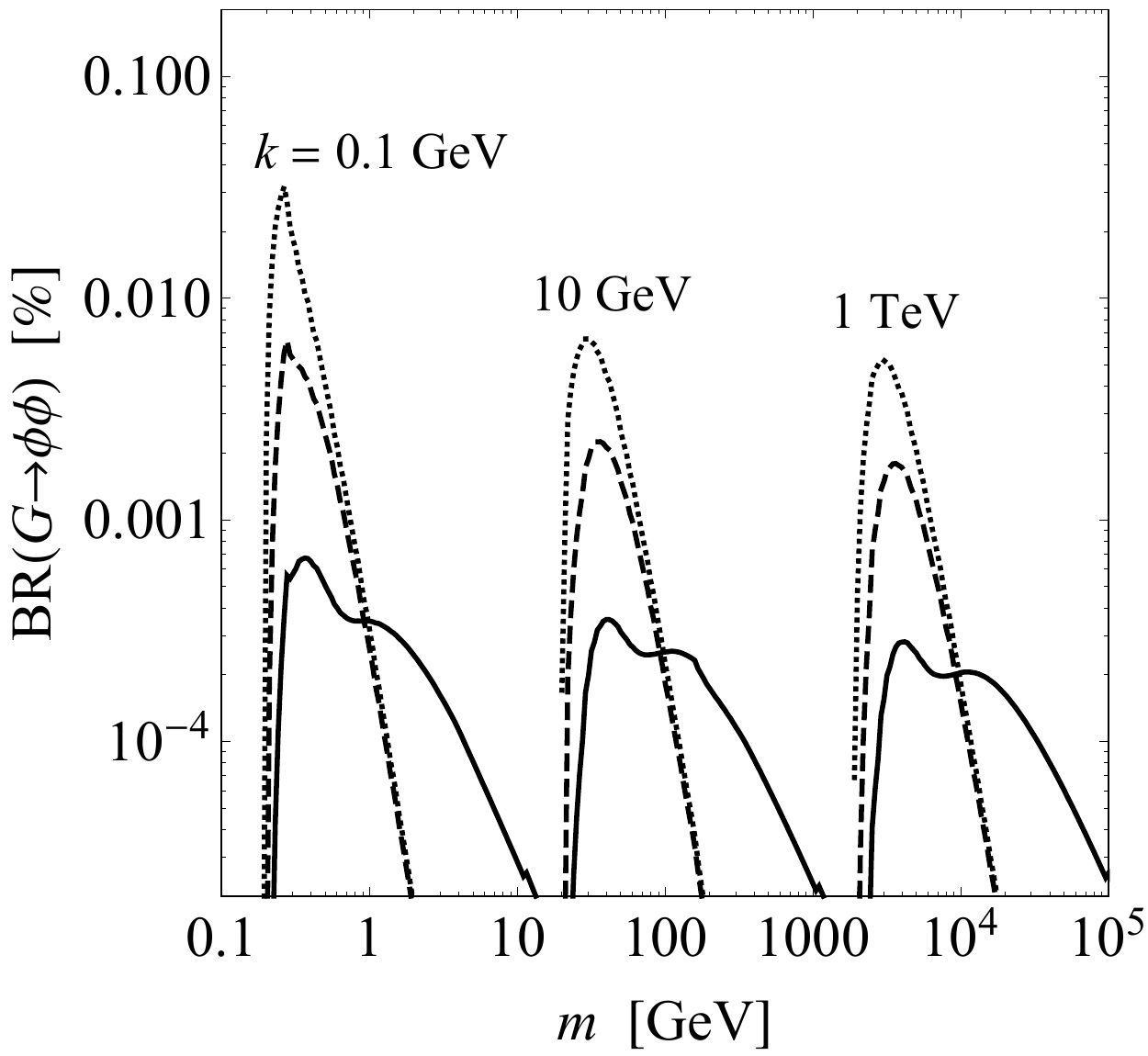}
\end{center}
\caption{KK graviton branching fractions (in \%) for decays to a KK scalar + KK graviton (left) or a pair of KK scalars (right) for $k = 0.1$, 10 and 1000~GeV, as a function of the KK graviton mass, assuming rigid boundary conditions for the dilaton field. We show separately contributions involving two scalar zero modes (dotted), one scalar zero mode (dashed) and no zero modes (solid).}
\label{BRs-KKG-to-KKS}
\end{figure}

Let us now consider the decays of KK scalars. As we have already mentioned, their couplings to the SM are model dependent. The decay rates have been computed in ref.~\cite{Cox:2012ee} and we present the full expressions in appendix~\ref{app:scalar-SM-decays}. For rigid boundary conditions (which implies couplings to $T_\mu^\mu$ only), assuming the Higgs-curvature interaction (which we will discuss below) to be absent, and neglecting the SM particle masses relative to the KK scalar mass, the total partial width to SM particles is dominated by $W^+W^-$, $ZZ$ and $hh$, and is given by
\beq
\Gamma_{\phi_n \to {\rm SM}} 
\simeq \frac{m_n^3}{8\pi\Lambda_\Phi^{(n)2}} = \frac{m_n k^2}{54\pi^2RM_5^3}\left( \frac{1-\frac{k^2}{m_n^2}}{1-\frac{8k^2}{9m_n^2}}\right) \, ,
\label{KKdilaton-width}
\eeq
where the last expression applies for $n \neq 0$.
The corresponding lifetime is
\beq
c\tau_n \approx 1.1\,\mbox{mm} \times \left(\frac{M_5}{10~\mbox{TeV}}\right)^3 \left(\frac{1~\mbox{GeV}}{k}\right)^3 \left(\frac{1~\mbox{TeV}}{m_n}\right)\left(\frac{kR}{10}\right) \left( \frac{1-\frac{8k^2}{9m_n^2}}{1-\frac{k^2}{m_n^2}}\right) \, .
\label{uffa}
\eeq
For large $n$, the last factor in eq.~(\ref{uffa}) is equal to one, while for $n=1$ it is equal to $(kR/3)^2\approx 11$, showing a smaller enhancement than in the graviton case. However, there is a higher overall tendency for the decays to be displaced than in the KK graviton case, at least for the low modes, where KK tower cascades are irrelevant. The decay rate into SM particles is even smaller below the $W^+W^-$ threshold, where $gg$ becomes the dominant SM decay channel, with the rate approximately given by eq.~\eqref{KKdilaton-width} times $49\,(\alpha_s/2\pi)^2$. In that regime
\beq
c\tau_n \approx 66\,\mbox{cm} \times \left(\frac{M_5}{10~\mbox{TeV}}\right)^3 \left(\frac{1~\mbox{GeV}}{k}\right)^3 \left(\frac{100~\mbox{GeV}}{m_n}\right)\left(\frac{kR}{10}\right)\left( \frac{1-\frac{8k^2}{9m_n^2}}{1-\frac{k^2}{m_n^2}}\right)\, .
\eeq

Relaxed boundary conditions, which allow KK scalars to decay via a direct coupling to ${\cal L}_{\rm SM}$, eq.~\eqref{dilaton-LSM-coupling}, rather than only through $T_\mu^\mu$, generate sizable branching ratios to $gg$ and $\gamma\gamma$ relative to the total rate to pairs of SM particles, of ${\cal O}(30\%)$ and ${\cal O}(3\%)$, respectively. The partial widths to $W^+W^-$, $ZZ$, $hh$ are modified by ${\cal O}(1)$ factors, so the total SM rates of the heavy modes stay in the same ballpark. The light modes, on the other hand, can become significantly shorter lived.

\begin{figure}[t]
\begin{center}
\includegraphics[width=0.55\textwidth]{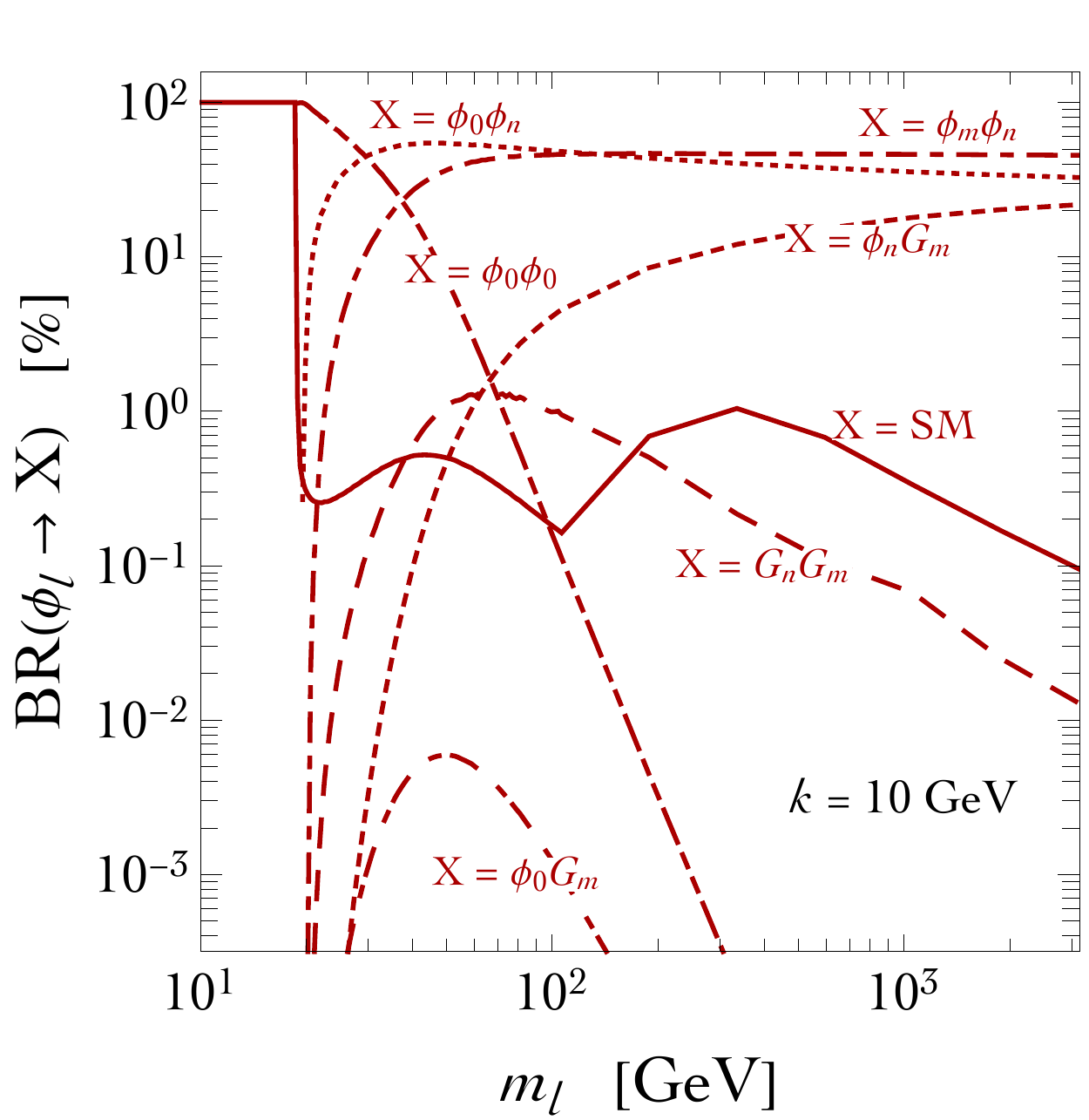}
\end{center}
\caption{Branching ratios of the scalar KK modes $\phi_l \to X$ for $k = 10$~GeV, rigid boundary conditions and vanishing curvature-Higgs interaction from eq.~\eqref{xi}.}
\label{scalar_BRs_10}
\end{figure}

Because the couplings of the scalar KK modes to the SM are somewhat suppressed, cascade decays within the KK tower play an important role. We compute these decays in appendix~\ref{app:trilinear-decays}. We find that the KK scalar decay rates, with the exception of the first few modes, are dominated by the cascades, as shown in figure~\ref{scalar_BRs_10}.

Figure~\ref{lifetime-map} demonstrates how the KK graviton and KK scalar lifetimes vary throughout the KK mode spectrum for different values of $M_5$ and $k$. Interestingly, even for a single choice of the parameters $M_5$ and $k$, the model typically contains particles with a very wide range of lifetimes. For fixed $M_5$ and $k$, the variation of the KK mode lifetimes with $m_n$ is more evident for gravitons than for scalars.

\begin{figure}[t]
\begin{center}
\vspace{6mm}
\includegraphics[width=0.48\textwidth]{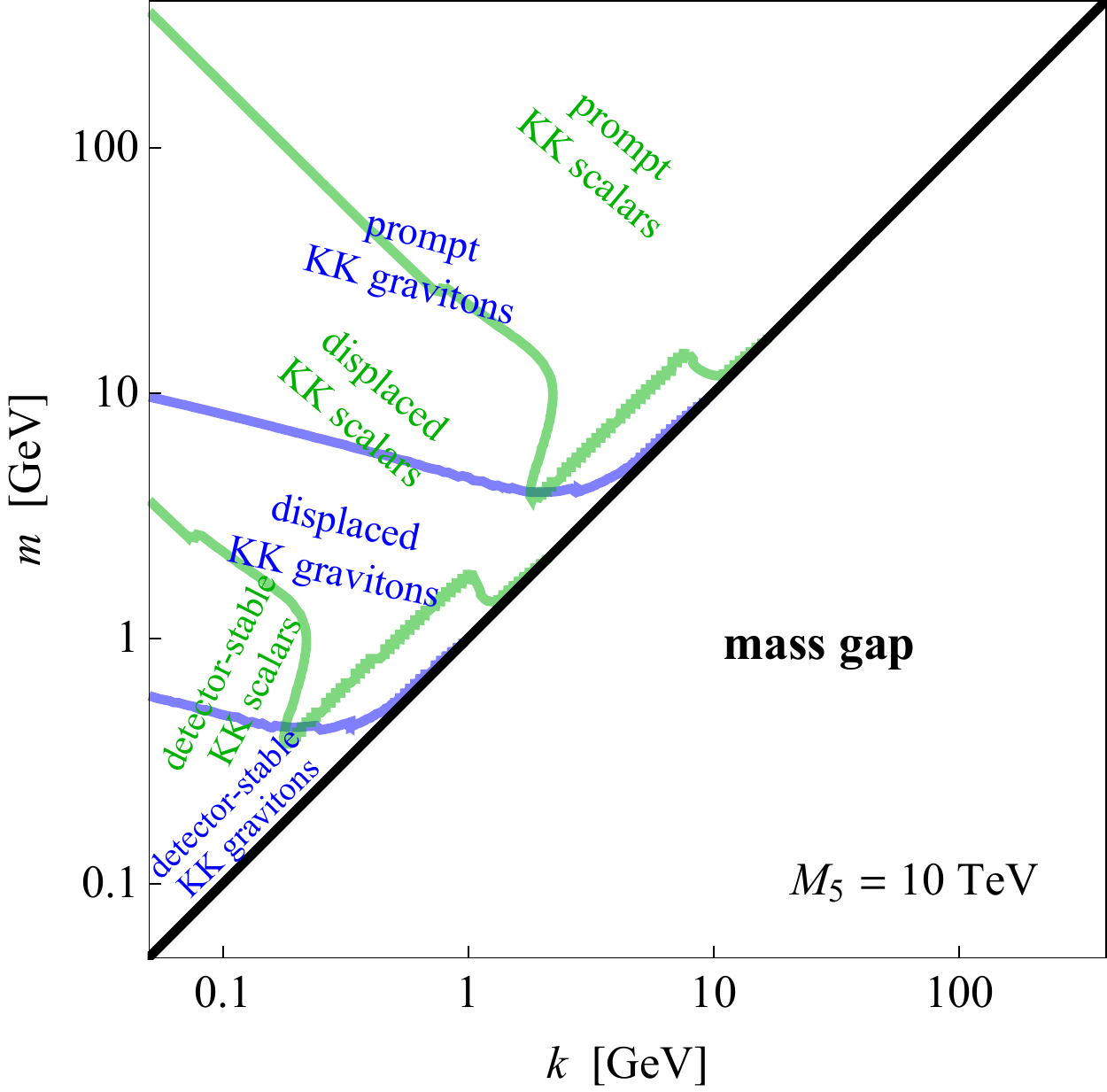}\quad\;
\includegraphics[width=0.48\textwidth]{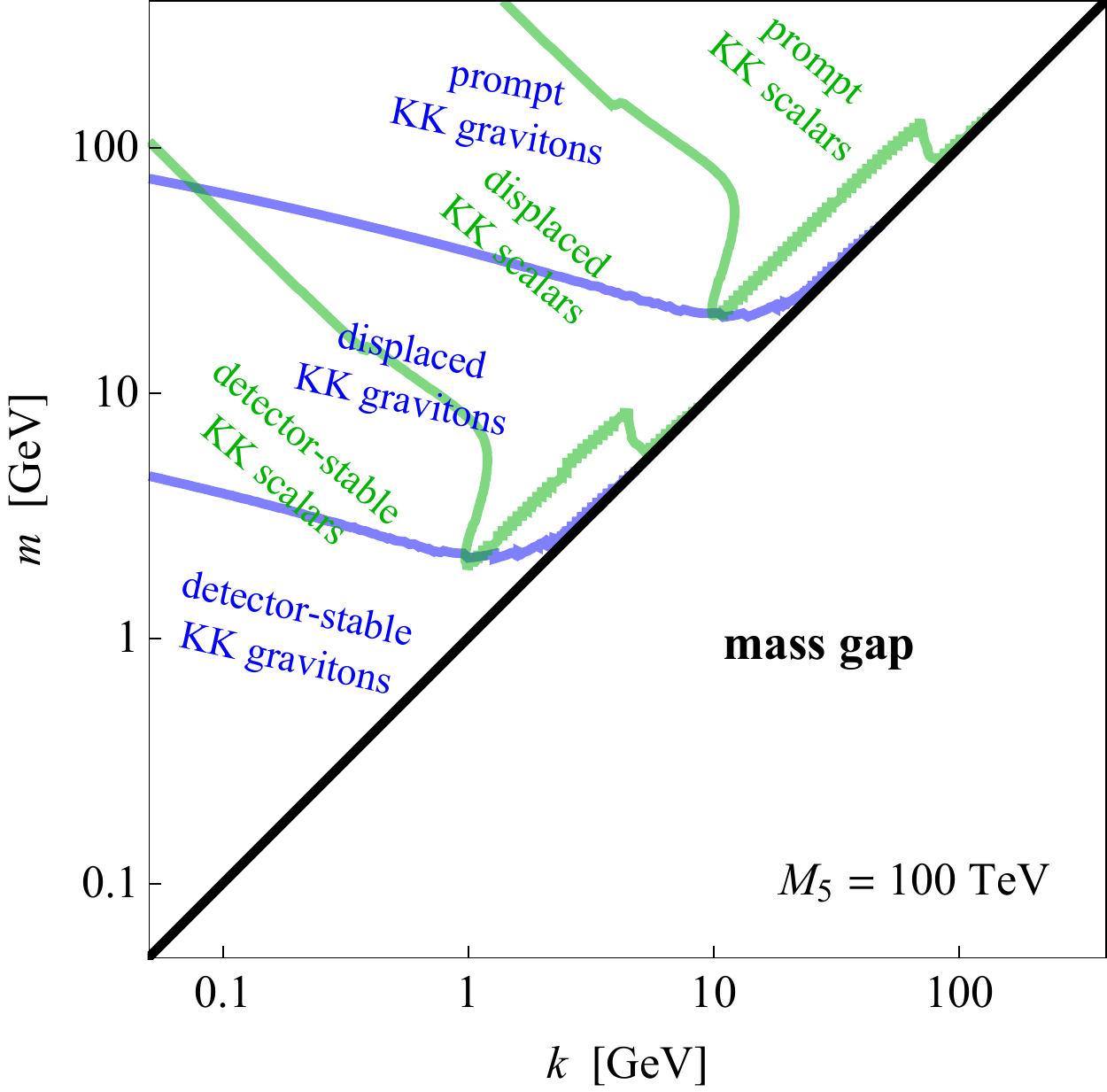}\\
\vspace{-5mm}
\end{center}
\caption{Mass ranges in which the KK gravitons (blue) or KK scalars (green) are detector-stable, displaced, or prompt, for $M_5 = 10$~TeV (left) and $100$~TeV (right), as functions of $k$. The boundaries between detector-stable/displaced/prompt were defined as $c\tau = 10$\,m and $1$\,mm. For the KK scalars, the results are presented for the case of rigid boundary conditions and vanishing curvature-Higgs interaction from eq.~\eqref{xi}.}
\label{lifetime-map}
\end{figure}

Since no symmetry prevents its appearance, it is natural to introduce on the SM brane a renormalizable  curvature-Higgs interaction 
\beq
{\cal L_\xi} = \xi R H^\dagger H \,,
\label{xi}
\eeq
where $\xi$ is a dimensionless coupling. After electroweak symmetry breaking,
this interaction induces a kinetic mixing, proportional to $\xi v/M_5$, between the Higgs and the radion component of the KK scalars. Through a field redefinition (see ref.~\cite{Cox:2012ee} for the explicit calculation) the kinetic mixing can be turned into a mass mixing. It is not difficult to check that, unless a KK scalar is accidentally nearly mass degenerate with the Higgs,\footnote{One may worry that the KK modes are so densely distributed that one of them is always sufficiently degenerate in mass with the Higgs. However, this is not the case. The condition of significant mixing requires that the mass square difference between the Higgs and the KK scalar ($|m_n^2-m_h^2|$) be not much larger than the off-diagonal mass entry, given by $\xi k^{1/2}vm_h^2/M_5^{3/2}$. It is easy to see that the mass square difference between KK states, $m_{n+1}^2 - m_n^2=(2n+1)/R^2$ (for $n$ such that $m_n \approx m_h$) is always larger than both the off-diagonal entry (as long as $\xi$ is not a very large number) and the Higgs width (as long as $k\gtrsim 50$~MeV). Thus, the condition of significant Higgs-dilaton mixing can only be the result of a fortuitous accident.} the mixing angle is very small, once we take into account the collider limits on $M_5$ that we will discuss later. This simplifies greatly the expressions of ref.~\cite{Cox:2012ee} since, in light of the present LHC bounds, today we can safely neglect the mixing angle between the Higgs and the KK scalars.

The term in eq.~(\ref{xi}) can also modify the Higgs couplings from their SM values. However, this occurs only at $\mathcal {O} (v^2/M_5^2)$ and the effects are too small to be measured in any present or planned experiment. 

The only practically relevant effect from the curvature-Higgs term is a modification of the KK dilaton couplings that survives even in the limit $v/M_5\to 0$. Its origin can be understood by considering the interactions of the Higgs in unitary gauge ($h$) and the KK dilaton mass eigenstates ($\phi_n$), which can be written in the following form
\beq
{\cal L} =\frac{h}{v}\, J_{\rm SM} +\sum_n \frac{\phi_n}{M_5}\, J_n \, ,
\label{correnti}
\eeq
where $J_{\rm SM}$ and $J_n$ are currents of canonical dimension four. The term in eq.~(\ref{xi}) implies that mass eigenstates are obtained after a field redefinition of $h$ and $\phi_n$ that involves mixing angles of order $\xi v/M_5$. Once this field redefinition is applied in eq.~(\ref{correnti}), we find that the Higgs current is modified only at $\mathcal {O} (v^2)$, while the dilaton currents receive additional contributions proportional to $J_{\rm SM}$ at $\mathcal {O} (v^0)$. These effects have been computed in ref.~\cite{Cox:2012ee}, and the results for the KK scalar decay rates are presented in appendix~\ref{app:scalar-SM-decays}, including the leading contribution from the Higgs-curvature mixing $\xi$. 

\subsection{KK mode production cross sections at the LHC}

KK gravitons can be produced from both $gg$ and $q\bar q$ via their coupling to $T^{\mu\nu}$. The cross section for the $n$-th KK graviton mode is given by
\beq
\sigma_n = \frac{\pi}{48\Lambda_G^{(n)2}}\left[3\,{\cal L}_{gg}(m_n^2)
+ 4\,\sum_q{\cal L}_{q\bar q}(m_n^2)\right] ,
\label{single-KKG-xsec}
\eeq
where
\beq
{\cal L}_{ij}(\hat s) = \frac{\hat s}{s}\,\int_{\hat s/s}^1 \frac{dx}{x}\, f_i(x)\, f_j\left(\frac{\hat s}{xs}\right) \,
\eeq
are the parton luminosities, for which we will be using the leading-order MSTW2008 parton distribution functions~\cite{Martin:2009iq}. Approximating the closely-spaced spectrum of KK modes by a continuum, one can write
\beq
\frac{d\sigma}{dm} \simeq \,\theta(m - k)\,\frac{1}{48M_5^3}\,\sqrt{1-\frac{k^2}{m^2}} \left[3\,{\cal L}_{gg}(m^2)
+ 4\,\sum_q{\cal L}_{q\bar q}(m^2)\right] .
\label{xsec-cont}
\eeq
Notably, while the parameter $k$ sets the turn-on of the spectrum, the cross section at $m \gg k$ is independent of $k$. Figure~\ref{xsecs} shows the cross section for several choices of $M_5$ and $k$.

\begin{figure}[t]
\begin{center}
\includegraphics[width=0.49\textwidth]{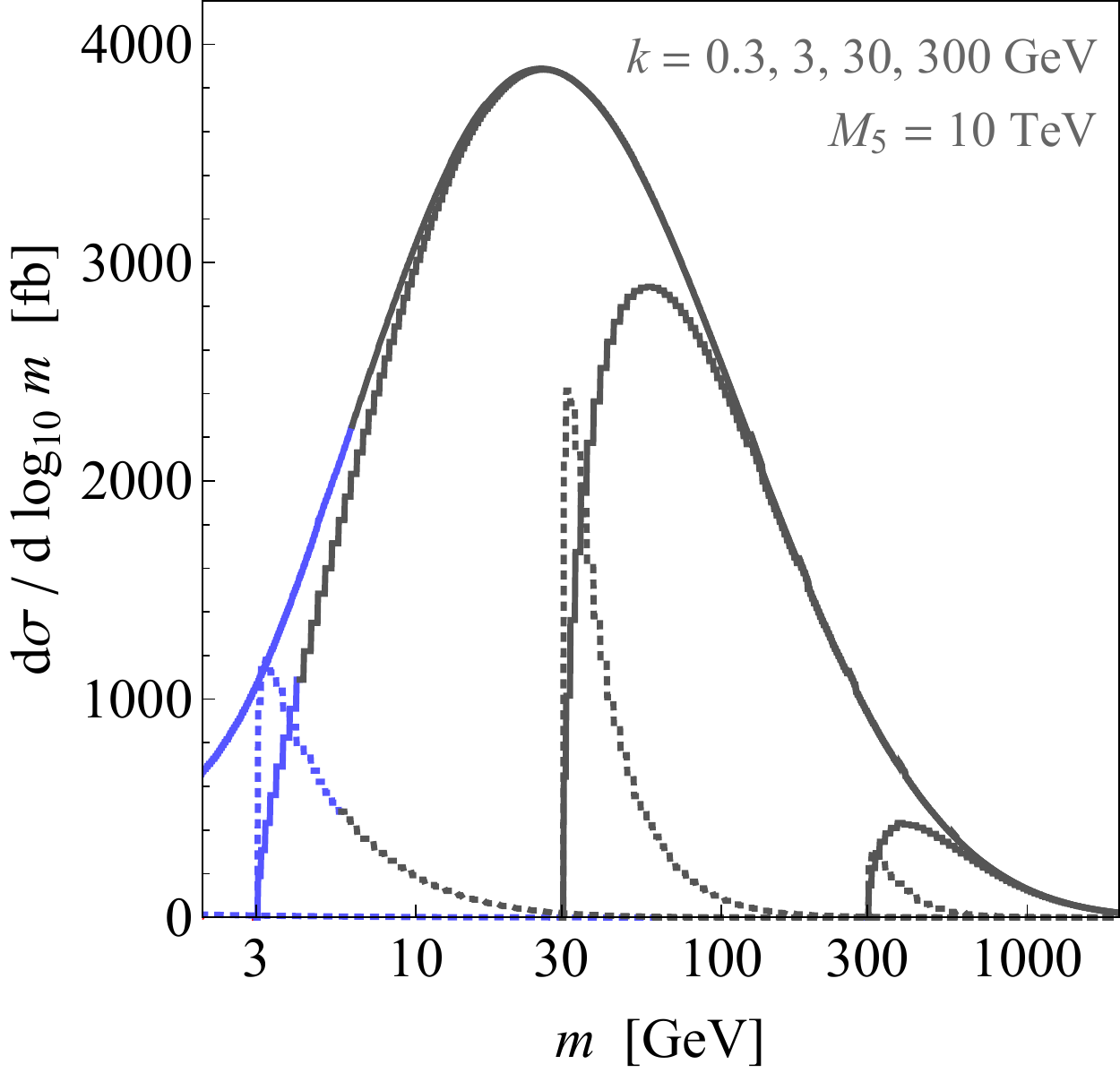}\quad
\includegraphics[width=0.48\textwidth]{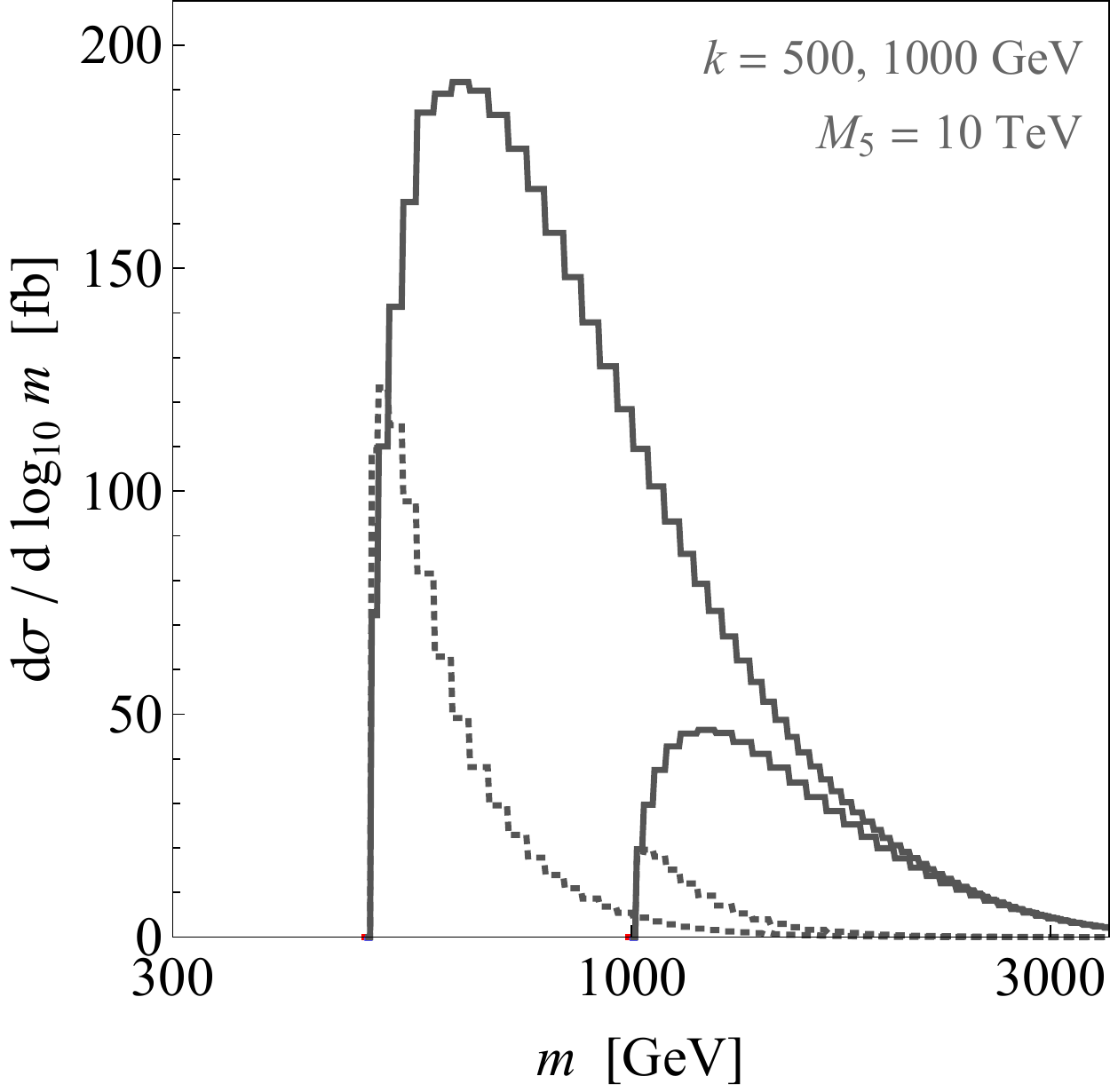}\\
\vskip 5mm
\hskip 5mm
\includegraphics[width=0.46\textwidth]{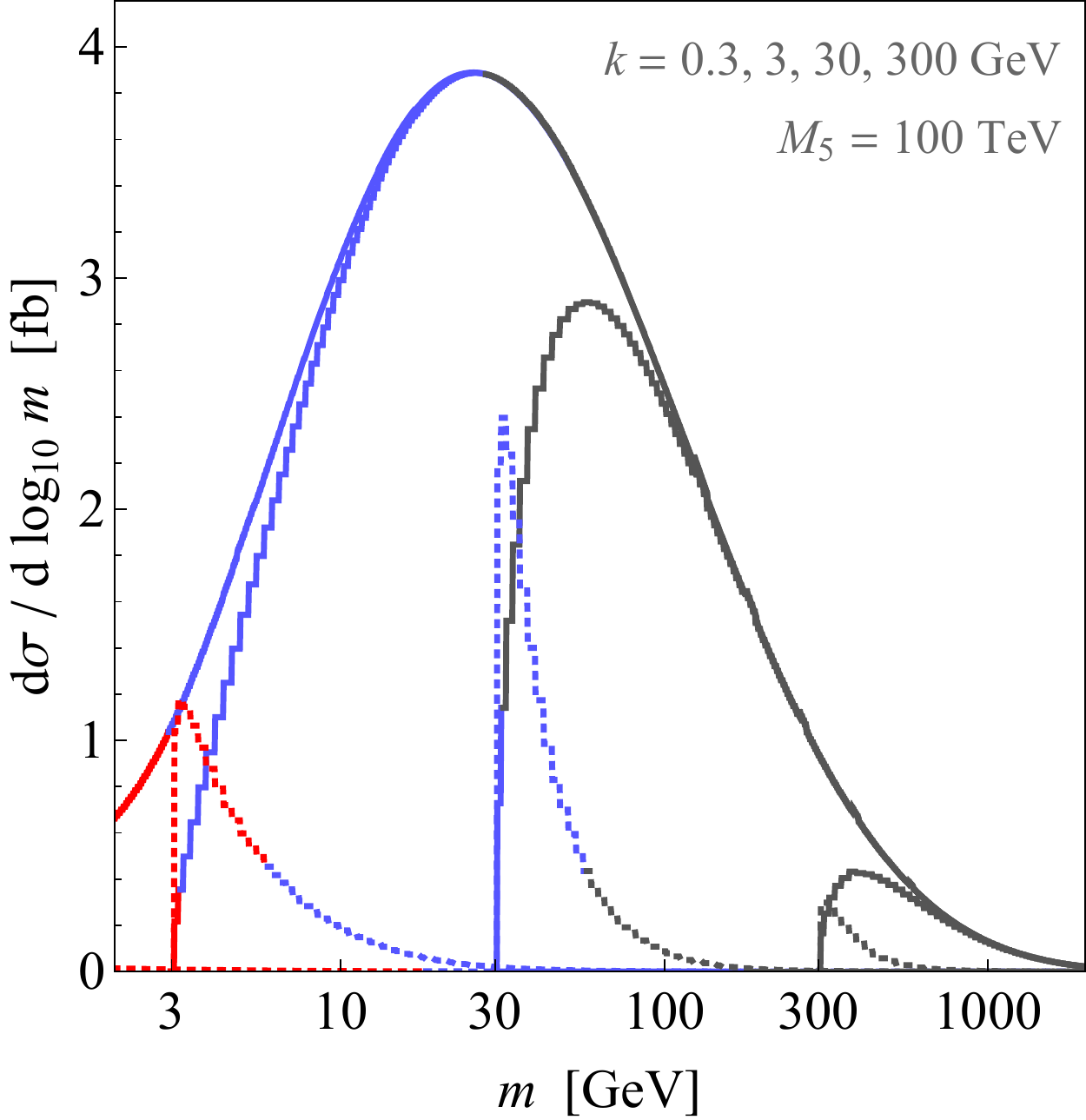}\quad
\includegraphics[width=0.48\textwidth]{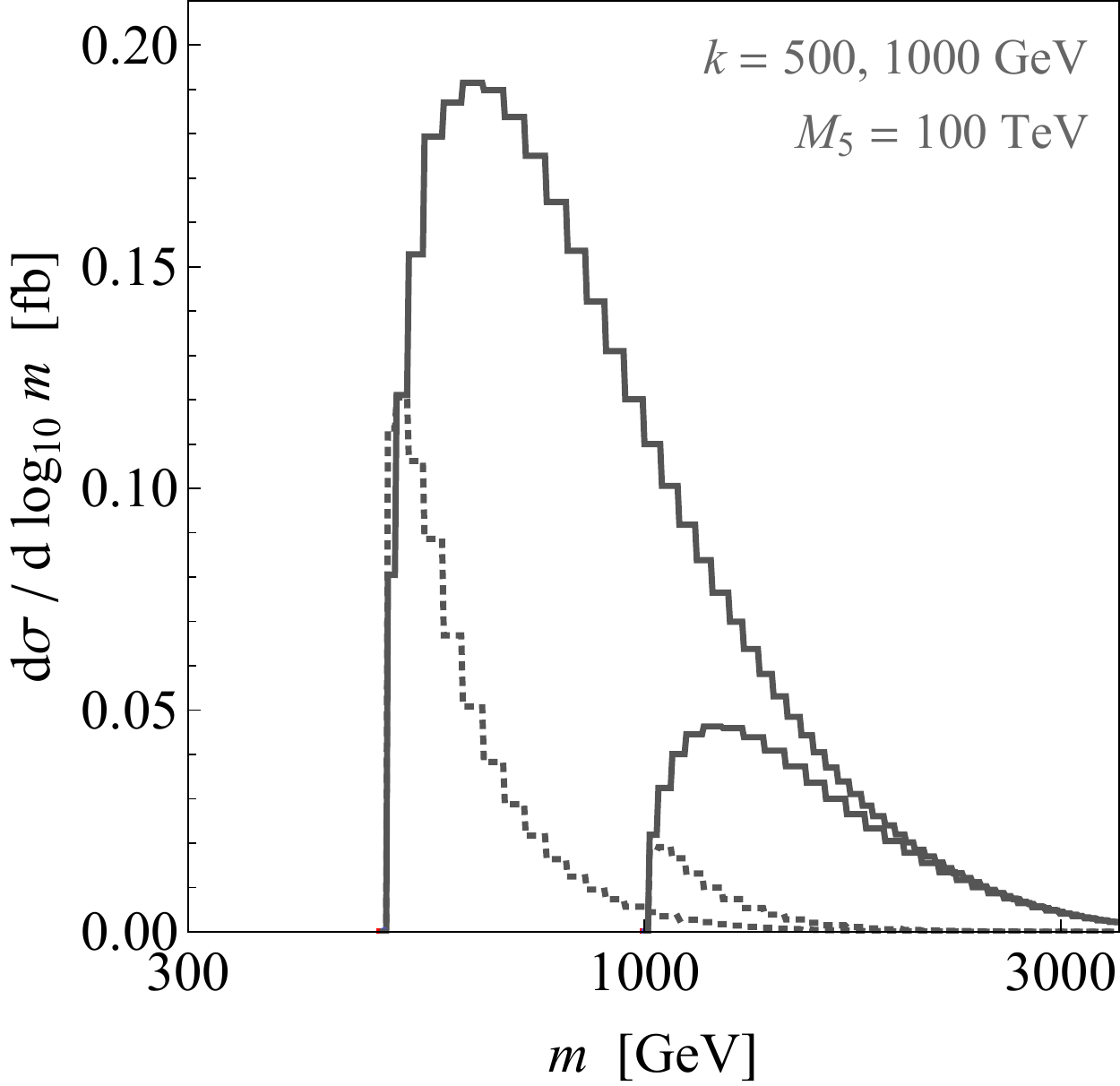}
\end{center}
\caption{The KK graviton (solid) and KK scalar (dashed, multiplied by a factor of $500$) LHC production cross sections at $\sqrt s = 13$~TeV for $M_5 = 10$~TeV (top) and $100$~TeV (bottom), $k = 0.3$, $3$, $30$ and $300$~GeV (left) and $500$ and $1000$~GeV (right). The KK mode lifetimes are color-coded: decays close to the interaction vertex ($c\tau < 1$\,mm) are in black, displaced decays ($c\tau < 10$\,m) are in blue, and the detector-stable regime is in red. For the KK scalars, the results are presented for the case of rigid boundary conditions and $\xi = 0$.}
\label{xsecs}
\end{figure}

The production cross section of KK scalars, similarly to their decay rates,  depends on the boundary conditions on the dilaton field. In the limit of rigid boundary conditions, the cross section of the $n$-th scalar mode in the $gg$ channel is given by
\beq
\sigma_n = \frac{b_{\rm QCD}^2\alpha_s^2}{256\pi\Lambda_\Phi^{(n)2}}\,{\cal L}_{gg}(m_n^2) \,.
\label{gg2KKdilaton}
\eeq
The $\alpha_s^2/4\pi^2$ suppression factor arises because the gluon contribution to $T_\mu^\mu$ vanishes at the classical level, so the coupling needs to rely on the $\beta$ function. The $\beta$ function factor is $b_{\rm QCD}=11-2n_Q/3$, where $n_Q$ is the number of quarks lighter than $m_n$; so $b_{\rm QCD}=7$ for $m_n \gg m_t$. Production from $q\bar q$ is even more suppressed due to the smallness of the light-quark masses. In the continuum approximation, the dilaton tower cross section becomes
\beq
\frac{d\sigma}{dm} \simeq \,\theta(m - k)\,\frac{b_{\rm QCD}^2\alpha_s^2}{1728\pi^2 M_5^3}\,\sqrt{1-\frac{k^2}{m^2}}\left(1 - \frac{8k^2}{9m^2}\right)^{-1}\frac{k^2}{m^2}\,{\cal L}_{gg}(m^2) \,.
\eeq
Note the $k^2/m^2$ suppression, which is absent in the KK graviton case and appears here due to the $n$ dependence of $\Lambda_\Phi^{(n)}$ in eq.~\eqref{KKdilaton-couplings-n}. The combination of the two suppression factors makes the cross section very small. The cross section is shown in figure~\ref{xsecs}, where it has been multiplied by a factor of $500$ to be visible.

For relaxed boundary conditions ($0<\mu<\infty$), the KK scalars couple directly to ${\cal L}_{\rm SM}$ in addition to their coupling to $T_\mu^\mu$. The production cross section due to this additional coupling alone, in the unstabilized limit, for $m \gg k$, is given by
\beq
\frac{d\sigma}{dm} \simeq \frac{1}{192 M_5^3}\,{\cal L}_{gg}(m^2) \,,
\eeq
which is still a factor of $12$ smaller than the KK graviton productions cross section from $gg$ in eq.~\eqref{xsec-cont}. It should also be noted that the KK dilatons diphoton branching fraction is smaller than or comparable to that of the KK gravitons~\cite{Cox:2012ee}.

Importantly, the above discussion does not imply that KK dilatons are irrelevant for phenomenology. Even though their direct production is insignificant, they can be produced in nonnegligible amounts in decays of heavy KK gravitons (see figure~\ref{BRs-KKG-to-KKS}). Such processes can lead to interesting signatures. For example, as evident from figure~\ref{lifetime-map} (and~\ref{xsecs}), one can have a situation in which the KK gravitons decay promptly throughout most of the mass range, while the KK scalars that are sometimes produced in KK graviton decays have displaced decays.

For simplicity, throughout the paper, we do not include QCD corrections to the productions cross sections or the partial decay widths of the KK modes. For the KK gravitons, the corresponding $K$ factors would be the same as in other models with the SM on a brane, and they have been computed at the next-to-leading order (see ref.~\cite{Das:2016pbk} and references therein). For example, the $K$ factor for production varies between $1.6$ for $m = 100$~GeV and $1.3$ for $m = 2$~TeV~\cite{Das:2016pbk}.

\section{LHC signatures}
\label{sec2}

As obvious from the previous section, CW/LD has a variety of signatures at the LHC. We will start our discussion with standard signatures, {\it i.e.}\ those that are being covered (whether optimally or not) by existing searches, and then move to signatures that are unique to CW/LD and motivate examining new search strategies. Our estimates for the sensitivity of part of the searches that we discuss below will be summarized in figure~\ref{sensitivity} which is contained in the conclusions. Limits (at low $k$) from $e^+e^- \to f\bar f$ scattering at LEP2 (not shown) are weak: $M_5 \gtrsim 1.8$~TeV, based on the results of refs.~\cite{Giudice:2004mg,Franceschini:2011wr}. Bounds on light KK gravitons from beam dump experiments, the neutrino burst of supernova 1987A, and BBN are, according to the estimates in ref.~\cite{Baryakhtar:2012wj}, irrelevant in the range of parameters included in figure~\ref{sensitivity}. 

\subsection{Standard signatures}
\label{sec21}

\subsubsection{Continuum $s$-channel effects at high invariant mass}

The quasi-continuous KK graviton spectrum that extends to high masses, at least all the way to the cutoff of the theory (at a scale of order $M_5$), gives contributions to spectra of various SM final states (see table~\ref{BRs}) that decrease with mass slower than the SM contributions. Similar signatures exist in the LED scenario~\cite{Giudice:1998ck,Han:1998sg} and, as in the LED case, the $\gamma\gamma$, $e^+e^-$ and $\mu^+\mu^+$ channels likely have the best sensitivity to this effect.

However, while searches for high-mass excesses exist in both the diphoton~\cite{Aaboud:2017yyg,Chatrchyan:2011fq} and the dilepton~\cite{Aaboud:2017buh,Khachatryan:2014fba} channels, there is an important difference between the LED benchmark models that are being used in these searches and CW/LD. In the LED case, one always considers scenarios with multiple extra dimensions because a single flat extra dimension is excluded by astrophysics, and then the dominant contribution to the signal is coming from cutoff-dominated off-shell processes. Conversely, in the case of a single extra dimension, which is viable in CW/LD, the dominant effect is an on-shell production of the KK modes, and there are no divergent contributions that need to be cut off. Therefore, unlike LED, the effect in CW/LD can be reliably computed within the effective theory.
The shapes of the high-mass excesses in CW/LD are very different from those in LED models with multiple extra dimensions,\footnote{%
At the high masses that are typically being used in searches for the diphoton signatures of LED, $m_{\gamma\gamma} \gtrsim 500$~GeV, and approximating the CW/LD spectrum by a continuum, the diphoton rate is rescaled relative to LED by the factor
\beq
\theta(m_{\gamma\gamma} - k)\,\frac{30\Lambda_T^8}{283\pi M_5^3}\,\sqrt{1-\frac{k^2}{m_{\gamma\gamma}^2}}\;\frac{1}{m_{\gamma\gamma}^5}
\left[1 + \frac{2975}{283 \times 2^8}\, \left(1 - \frac{k}{m_{\gamma\gamma}}\right)^9 \sqrt{\frac{m_{\gamma\gamma}}{k}}\right]^{-1} \,,
\nonumber
\eeq
where for the purpose of this formula the LED cutoff $\Lambda_T$~\cite{Giudice:1998ck} should be taken to be sufficiently low so that interference with the SM is negligible. The factor in the square brackets approximately parameterizes the decrease in the diphoton branching fraction that occurs for $k \ll m_{\gamma\gamma}$ due to KK graviton decays to lighter KK modes. The same formula applies in the dilepton channels.}
as shown in figure~\ref{high-mass-diphoton}. Another special property of CW/LD, also demonstrated in figure~\ref{high-mass-diphoton}, is the mass gap, controlled by the parameter $k$, which has an obvious impact on the search if the turn-on of the spectrum occurs within the mass range being examined.

\begin{figure}[t]
\begin{center}
\includegraphics[width=0.7\textwidth]{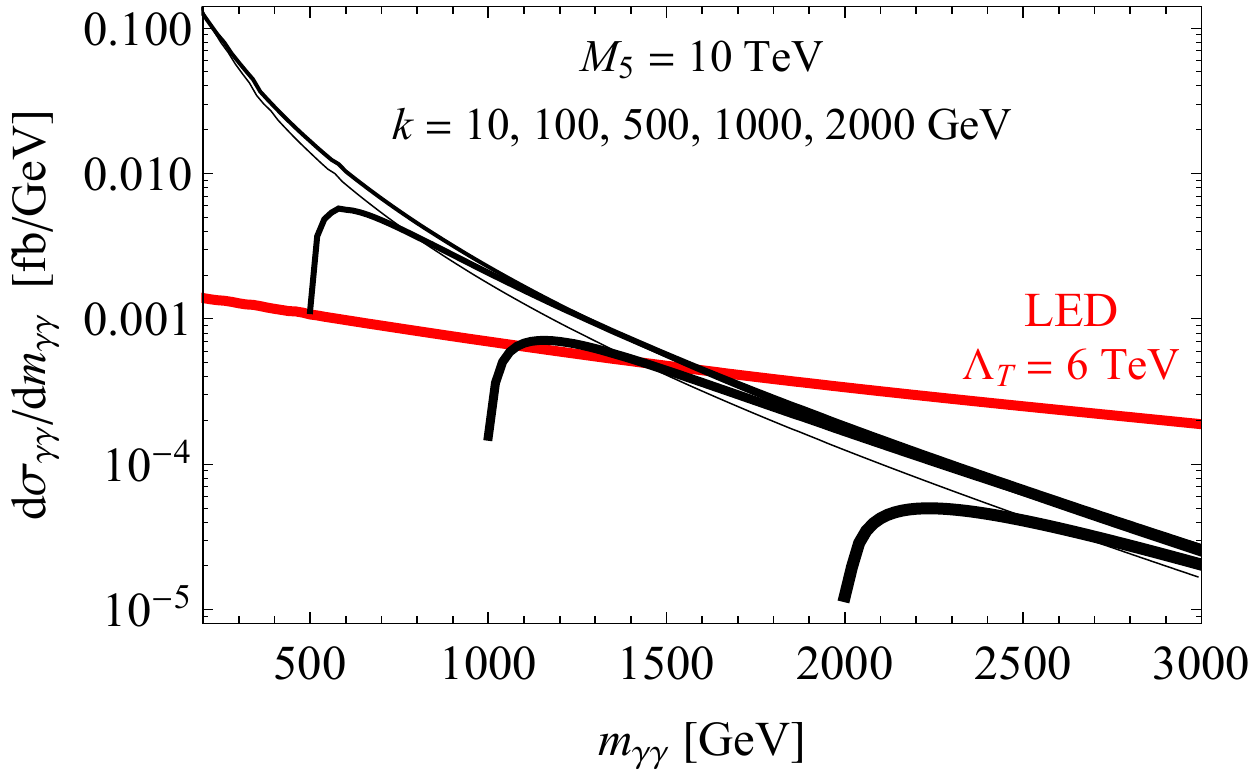}
\end{center}
\caption{Cross section of the diphoton signal at the $13$~TeV LHC for $k = 2000$, $1000$, $500$, $100$, $10$~GeV (from thickest to thinnest) and $M_5 = 10$~TeV. Also shown (in red) is the signal from LED with $> 2$ extra dimensions with $\Lambda_T = 6$~TeV. In the $k = 10$~GeV case, the signal is reduced because of the diphoton branching ratio suppression due to significant decay rates into lighter KK gravitons.}
\label{high-mass-diphoton}
\end{figure}

Our expected limit\footnote{Here and in the following, we present expected rather than observed limits because our goal is to compare the sensitivity of the various methods, without being affected by fluctuations in particular datasets.} on CW/LD from reinterpreting the ATLAS dilepton search~\cite{Aaboud:2017buh} is denoted $\ell\ell_{\rm cont}$ in figure~\ref{sensitivity}. To obtain this limit, we defined inclusive search regions with $m_{\ell\ell} > 400$, 500, 700, 900, 1200, 1800, 3000~GeV, summing the electron and muon events. For each point in the parameter space the search region with the best expected limit was used. Systematic uncertainties from the different bins of ref.~\cite{Aaboud:2017buh} that contribute to each of our search regions were conservatively assumed to be fully correlated.

In the diphoton channel, we reinterpret the ATLAS search~\cite{Aaboud:2017yyg}, which gives the limit denoted 
$\gamma\gamma_{\rm cont}$ in figure~\ref{sensitivity}. It is unfortunate that, in the context of our scenario, this search uses just a single search region ($m_{\gamma\gamma} > 2240$~GeV). Scenarios with $k > 2240$~GeV would obviously benefit from a harder cut on $m_{\gamma\gamma}$ since there is no signal for $m_{\gamma\gamma} < k$, while the currently used search region is not background-free. Scenarios with low $k$ would likely benefit from search regions with softer cuts on $m_{\gamma\gamma}$.

Even though these dilepton and diphoton searches rely only on the high-mass part of the spectrum, and for low $k$ the production cross section at high masses is independent of $k$ (see eq.~\eqref{xsec-cont} or figure~\ref{xsecs}), one can see in figure~\ref{sensitivity} that the limits on $M_5$ weaken significantly at low $k$. This happens because the branching fractions to pairs of SM particles are reduced due to the presence of decays to lighter KK modes, as we have shown in figure~\ref{BRs-w-KKG-KKG} for the diphoton channel.

\subsubsection{Continuum $t$-channel effects in dijets at high invariant mass}

In the dijet channel, while $s$-channel contributions exist, a much bigger effect arises from $t$-channel exchange of KK gravitons (including interference with QCD), since such processes can take advantage of the $uu$ initial state, which is very abundant in $pp$ collisions (as analyzed in ref.~\cite{Giudice:2004mg}). One can see the signal either via its effect on the invariant mass spectrum at high masses, or via its effect on the angular distribution in high-mass events. Existing searches~\cite{Aaboud:2017yvp,Sirunyan:2017ygf,CMS-PAS-EXO-16-046} do the latter.

\begin{figure}[t]
\begin{center}
\includegraphics[width=0.46\textwidth]{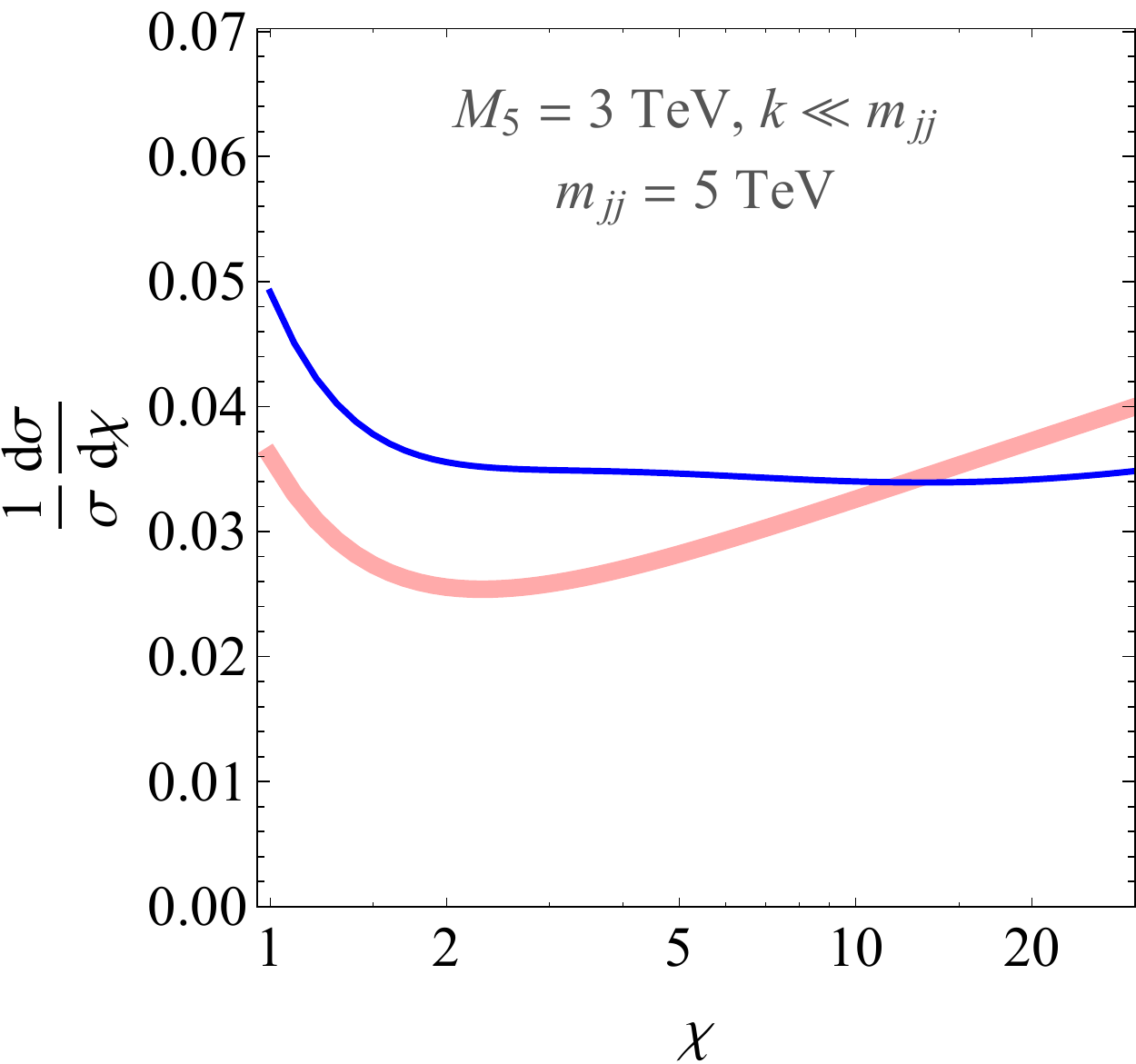}\qquad
\includegraphics[width=0.46\textwidth]{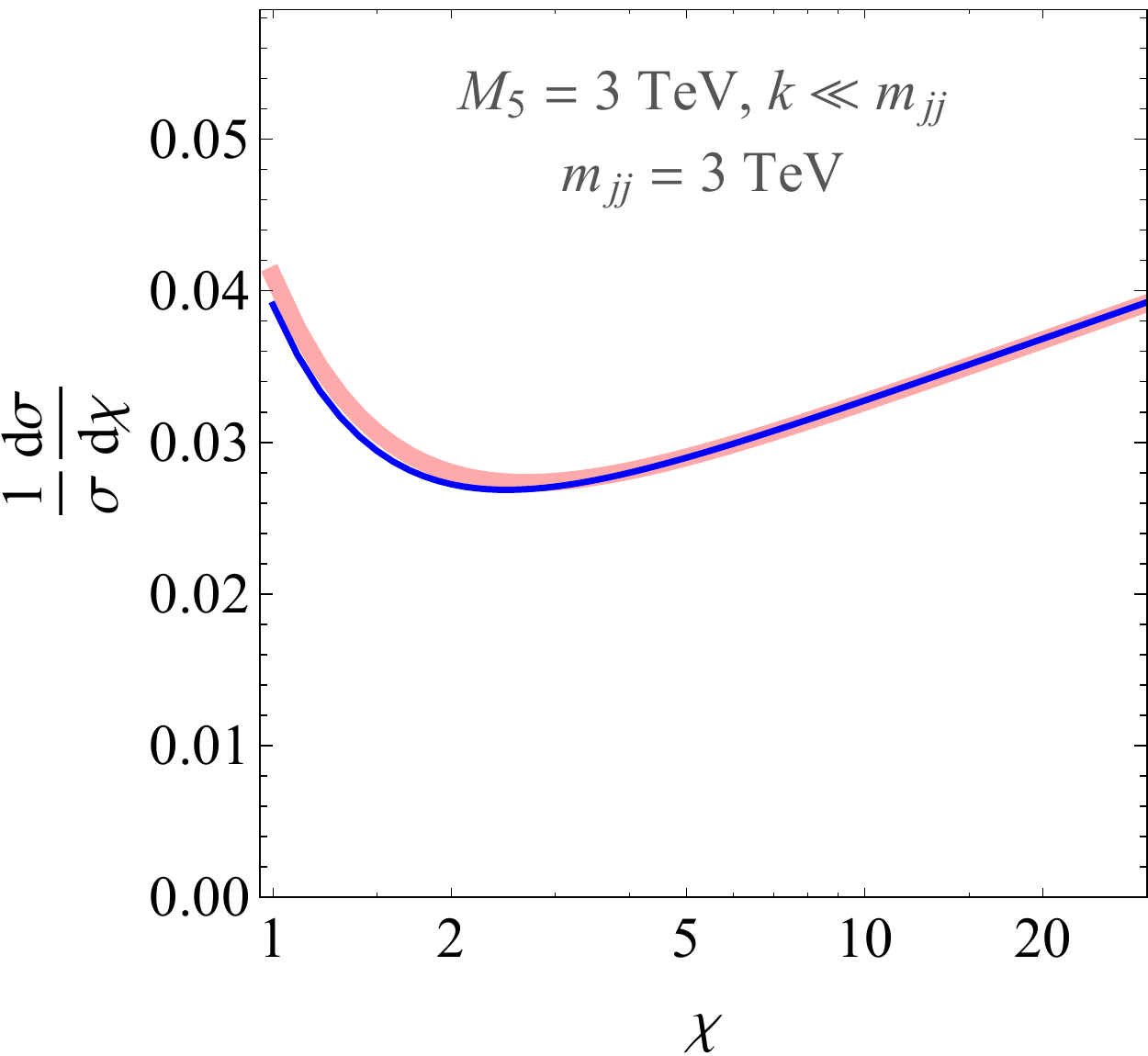}
\end{center}
\caption{Dijet angular distributions of the background (thick pink) and background+signal (thin blue), for $M_5 = 3$~TeV, in the limit of $k \ll m_{jj}$. The variable $\chi$, used in refs.~\cite{Aaboud:2017yvp,Sirunyan:2017ygf,CMS-PAS-EXO-16-046}, is defined in terms of the rapidities of the two jets as $\chi \equiv \exp(|y_1 - y_2|)$. As discussed in the text, while formally events with $m_{jj} = 5$~TeV (left plot) can be used to constrain the scenario, the interpretation is uncertain because $m_{jj} \gtrsim M_5$ is beyond the validity range of the theory. On the other hand, even events with $m_{jj} = 3$~TeV (right plot) already do not have much discriminating power.}
\label{jj-ang}
\end{figure}

The angular distributions (at the parton level) can be computed as described in appendix~\ref{app:jj-ang}. Examples of the angular distributions for CW/LD (including the SM contribution and interference) are shown in figure~\ref{jj-ang}. Figure~\ref{sensitivity} shows the expected limit we obtain ($jj_{\rm ang}$) based on the ATLAS search~\cite{Aaboud:2017yvp}. We use the first 1, 2, 3, 4 or 5 bins of $\chi$ for the mass ranges $m_{jj} > 5.4$~TeV and $4.9 < m_{jj} < 5.4$~TeV. While the resulting limit on $M_5$ is not as strong at high $k$ as the limits from the high-mass diphoton and dilepton continua, the search retains its power at low $k$ because the reduction of the SM branching fractions is irrelevant for $t$-channel processes, which, as expected, make the dominant contribution to the signal.

Note, however, that the $\sim 4$~TeV limit on $M_5$ that we obtained here by relying on events with $m_{jj} \sim 5$~TeV is not very reliable, considering that the theory is expected to break down at a scale of the order of $M_5$. This is not an issue in the diphoton and dilepton channels that we discussed above, because there, even stronger limits on $M_5$ are derived based on events at lower energies. This was possible because the SM diphoton and dilepton events end at lower energies than the dijet events, for a given integrated luminosity, due to cross section differences. One might consider setting a limit by relying only on search regions with $m_{jj} \lesssim M_5$, however such search regions are not sensitive to the signal, as shown in figure~\ref{jj-ang}.

\subsubsection{Distinct $\gamma\gamma$ and $e^+e^-$ resonances}
\label{sec:resonance-searches}

As we have seen in section~\ref{sec:KK-masses-couplings}, the splittings between the KK graviton modes near the bottom of the spectrum are typically sufficiently large for the separate modes to be seen as distinct peaks in the diphoton and dielectron invariant mass spectra. (In the dimuon channel, the peaks will merge into a broad excess due to the low resolution in $m_{\mu^+\mu^-}$, which is roughly between $5\%$ and $10\%$ in the relevant range of masses~\cite{Aaboud:2017buh,CMS-PAS-EXO-16-031}.)

\begin{figure}[t]
\begin{center}
\includegraphics[width=0.87\textwidth]{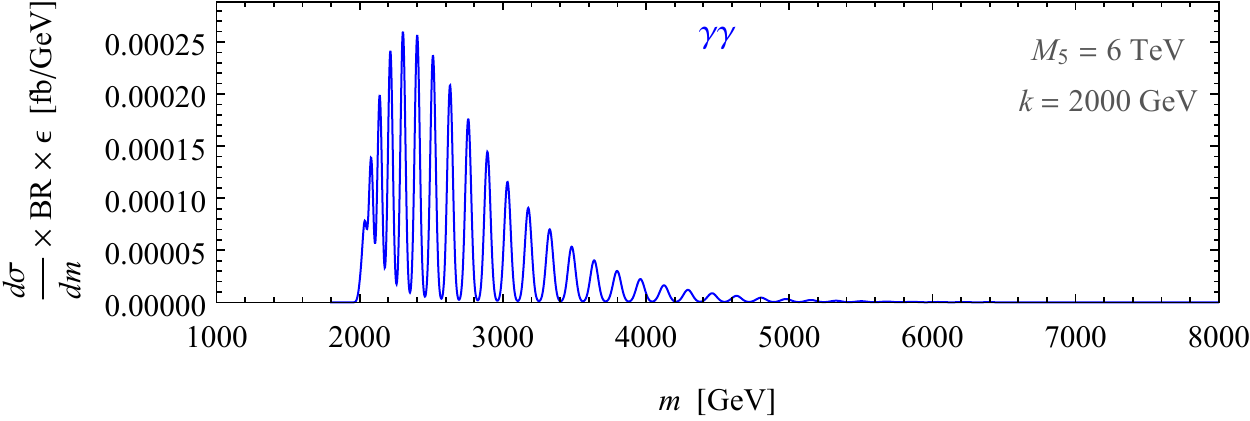}\\\vskip 5mm
\includegraphics[width=0.87\textwidth]{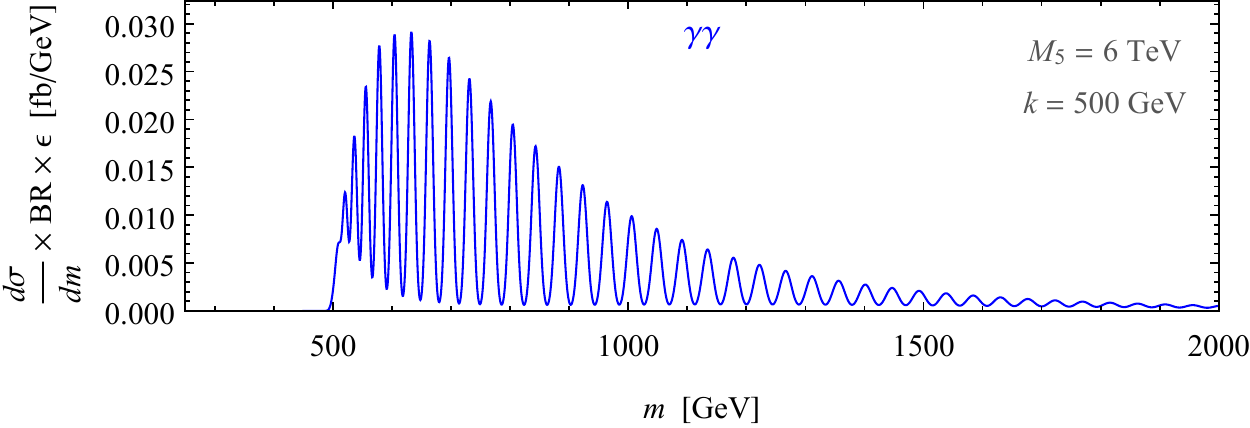}\\\vskip 5mm
\includegraphics[width=0.87\textwidth]{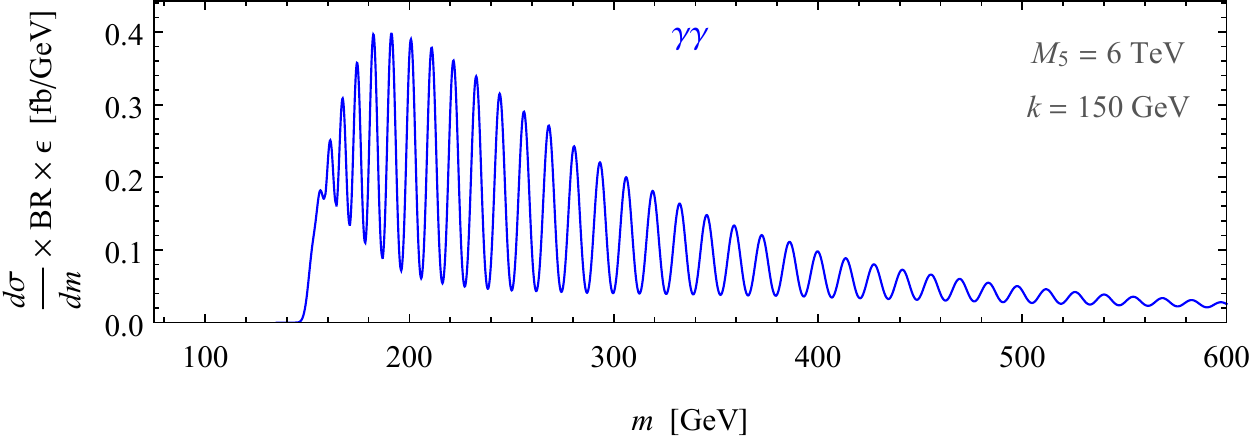}
\end{center}
\caption{Signal in the diphoton spectrum at the $13$~TeV LHC for $M_5 = 6$~TeV and $k = 2000$, $500$, $150$~GeV (from top to bottom) after accounting for the experimental resolution, eq.~\eqref{diphoton-resolution}.}
\label{full_mass}
\end{figure}

Several examples of the peaks in the diphoton channel are shown in figure~\ref{full_mass}. Here and below, we have assumed the experimental resolution in the diphoton invariant mass to be described by
\beq
\frac{\sigma(m_{\gamma\gamma})}{m_{\gamma\gamma}} = \sqrt{ \frac{a^2}{m_{\gamma\gamma}\,({\rm GeV})} + \frac{c^2}{ 2}}
\label{diphoton-resolution}
\eeq
with
\beq
a = 12\%\,,\quad c = 1\%\,,
\eeq
based on the partial information provided in refs.~\cite{ATLAS-CONF-2015-081,ATLAS-CONF-2016-059,Aaboud:2017yyg,Khachatryan:2016yec}.
The intrinsic widths of the resonances are also taken into account, although they are usually negligible relative to the resolution. For the product of acceptance and efficiency, the typical value of $\epsilon = 0.5$ has been assumed.

\begin{figure}[t]
\begin{center}
\includegraphics[width=0.47\textwidth]{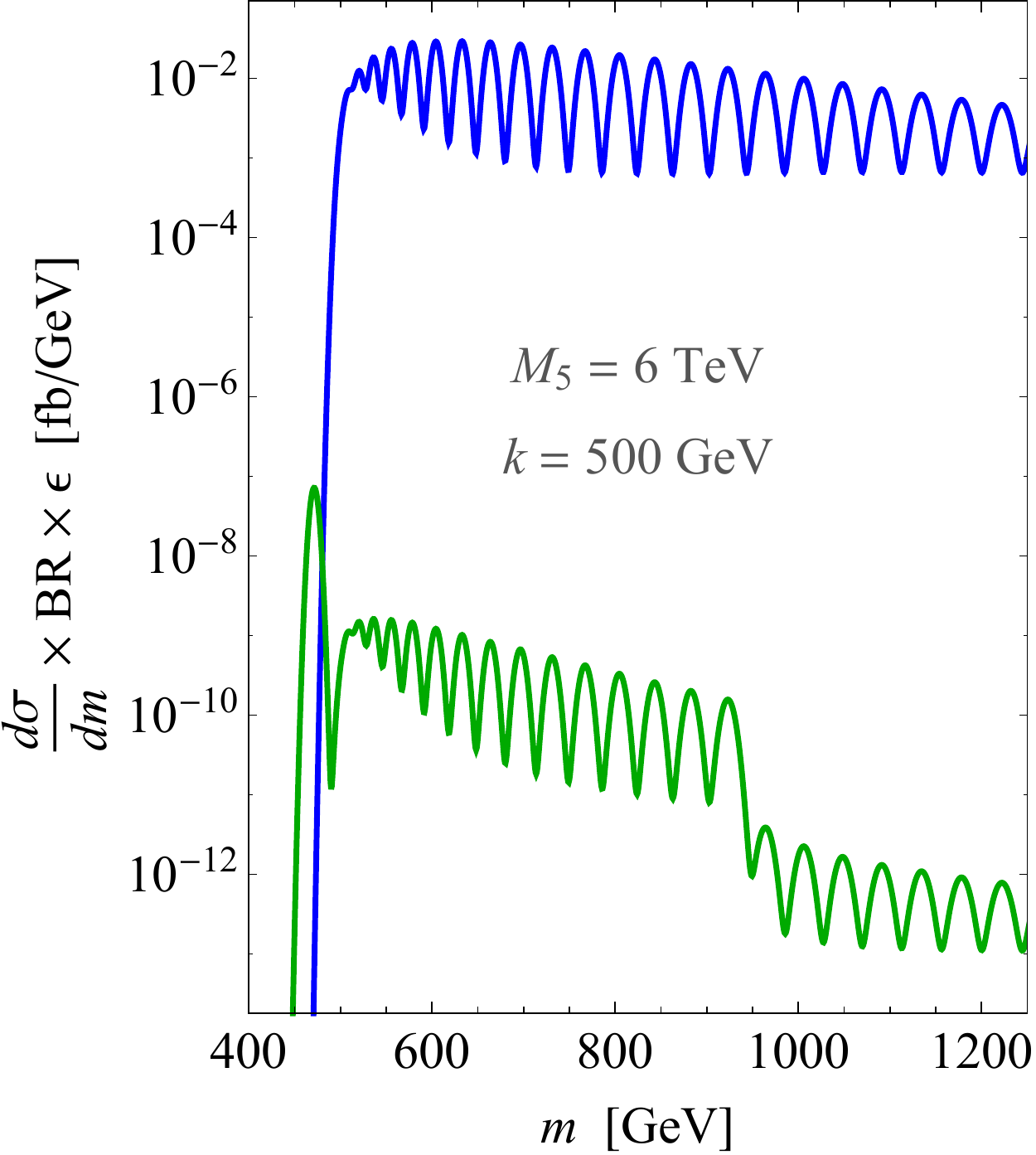}\qquad
\includegraphics[width=0.47\textwidth]{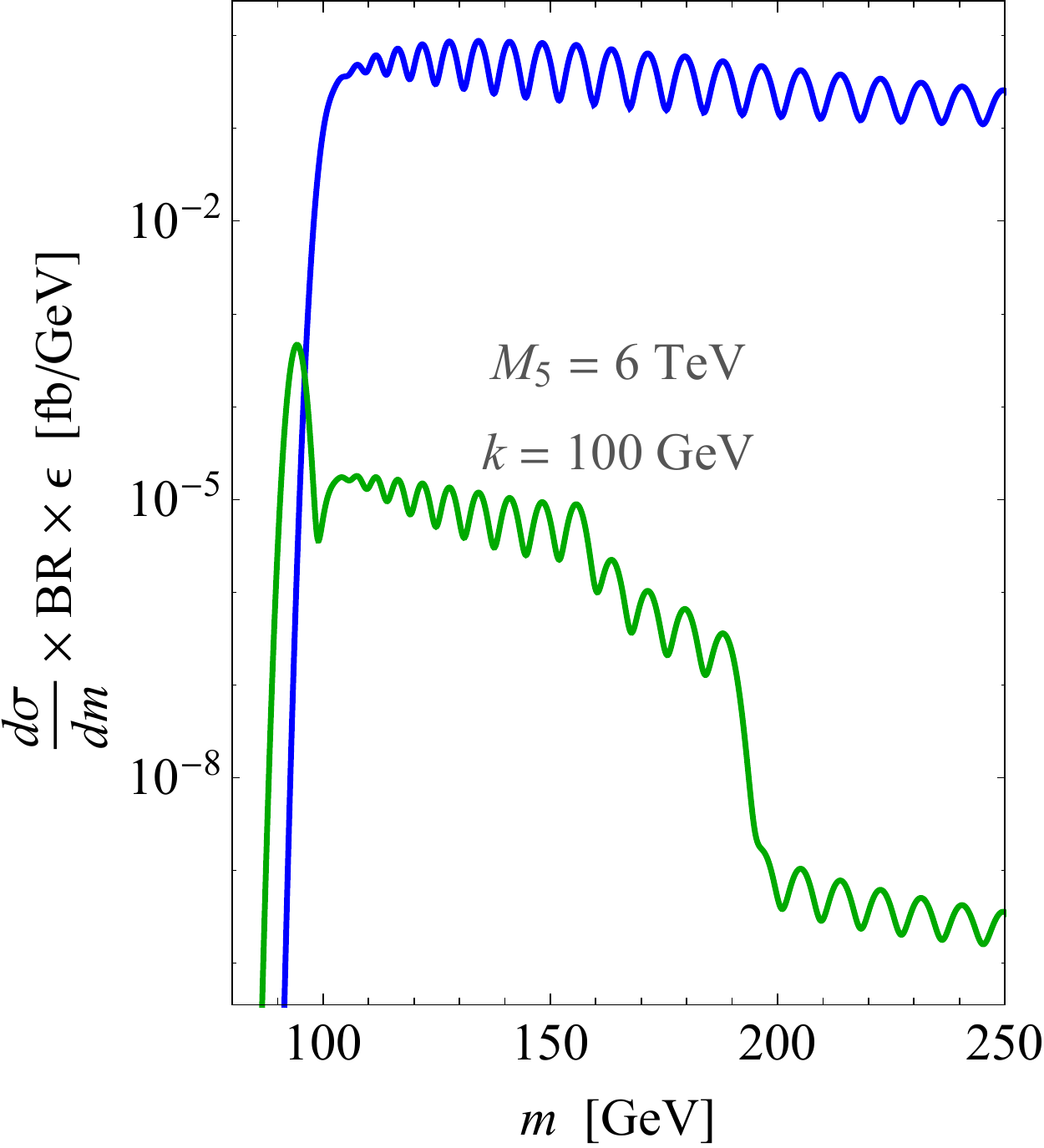}
\end{center}
\caption{The diphoton signal of the scalar KK modes (the lower curve, green) vs.\ that of the KK gravitons (the upper curve, blue) for $k = 500$~GeV (left) and $100$~GeV (right), in the case of rigid boundary conditions for the dilaton field.}
\label{scalars_diphoton}
\end{figure}

Standard resonance searches (diphoton~\cite{Aaboud:2017yyg,Khachatryan:2016yec}, dilepton~\cite{Aaboud:2017buh,CMS-PAS-EXO-16-031}) likely have sensitivity to such peaks, although there are several caveats. First, bump hunting in these searches might not work as expected due to the additional nearby peaks. Another caveat is that the intrinsic background due to the rest of the KK tower needs, in principle, to be taken into account. Finally, the $\mu^+\mu^-$ channel of the dilepton search will be very ineffective because nearby peaks will merge to a large extent due to the low resolution. (Note, however, that the worse resolution makes the $\mu^+\mu^-$ channel subdominant in sensitivity relative to $e^+e^-$ also in the usual situation of a single narrow resonance.) Disregarding these issues for a moment, we included in figure~\ref{sensitivity} approximate expected limits from the ATLAS diphoton search~\cite{Aaboud:2017yyg} supplemented by the CMS low-mass diphoton search~\cite{CMS-PAS-HIG-17-013} ($\gamma\gamma_{\rm res}$) and the ATLAS dilepton search~\cite{Aaboud:2017buh} ($\ell\ell_{\rm res}$), on a single peak from the spectrum, while assuming the signal efficiency to be similar to the benchmark models used in these searches. At each point in the parameter space, we used the peak that gives the best expected limit. The mode number corresponding to such a peak varied along the exclusion curves of both the diphoton and the dilepton channels from $n \approx 7$ at high $k$, up to $n \approx 30$ at $k \approx 100$~GeV, and further up at yet lower values of $k$. However, we have only considered modes up to $n = 40$ to ensure that neighboring peaks do not merge significantly (recall figure~\ref{dm}).  It should be noted that, due to the caveats above, these single-resonance limits are optimistic, since the presence of the other resonances is likely to weaken the sensitivity of the single-resonance searches.

Since the resonance searches~\cite{Aaboud:2017yyg,Aaboud:2017buh,CMS-PAS-HIG-17-013} do not address arbitrarily low masses, the above condition of $n \leq 40$ implies that below a certain value of $k$ no limit is set. Note, however, that as one goes lower in mass, the experimental resolution worsens, and at some point the distinct peaks will no longer be visible. For example, for $m_{\gamma\gamma} = 10$~GeV, the invariant mass resolution is about 4\% (much higher than indicated by the band in figure~\ref{dm}), so the separation between the peaks is largely lost also for $n \leq 40$. The situation is better in the $e^+e^-$ channel, where the degrading calorimeter resolution is somewhat compensated by an improving resolution on the track momentum, so that the overall $e^+e^-$ invariant mass resolution can be around 2\% even at low masses~\cite{Khachatryan:2015hwa}. Additionally, at very low masses, the resolution in the $\mu^+\mu^-$ channel becomes adequate, being approximately $1\%$ for $m_{\mu^+\mu^-} \sim 10$~GeV~\cite{Chatrchyan:2012xi,Aad:2016jkr}. In all three channels, however, standard triggers start losing efficiency at $m \lesssim 100$~GeV. Still, it may be useful to extend the searches down in mass, to the extent possible. The low-mass $\mu^+\mu^-$ spectrum may also be accessed by LHCb~\cite{Ilten:2016tkc,Aaij:2017rft}.

Of course, since the spectrum contains ${\cal O}(10)$ comparably strong peaks, an estimate for the sensitivity to the model as a whole would be actually higher than indicated by the limits we show. However, again, the caveats discussed above suggest it is unclear what the result of a full analysis would be. Moreover, one should consider searching for the periodic structure of peaks as a whole, rather than only attempting to detect individual peaks. We will analyze this possibility in detail in section~\ref{sec:FT}.

We note in passing that resonant signals due to the scalar (dilaton / radion) KK modes can be neglected, at least in the case of rigid boundary conditions on which we focus in this work, as shown in figure~\ref{scalars_diphoton} for the diphoton channel. This is true even at low masses, where the diboson decays of the scalar KK modes are kinematically forbidden so that the diphoton branching fraction is larger (right panel of the figure). In both cases, a further significant drop in the diphoton branching fraction of the KK scalars occurs at $m \approx 2k$ due to decays to pairs of scalar zero modes. The contributions in the dilepton channels are suppressed even more with respect to the KK graviton contributions.

\subsection{Novel signatures}
\label{sec22}

While some of the generic searches discussed in the previous section set limits on $M_5$ of around $5$--$10$~TeV, they do not get even close to the kinematic limit of potential exclusion. The total KK graviton production cross section is sizable even for $M_5 = 100$~TeV. For small $k$, it is given by
\beq
\sigma_{\rm total} \approx 5~{\rm fb} \times \left(\frac{100~{\rm TeV}}{M_5}\right)^3 \,.
\eeq
Even if we restrict ourselves to invariant masses above $200$~GeV, making the final-state objects triggerable with standard techniques, the cross section is still as large as
\beq
\left.\sigma_{\rm total}\right|_{m\,>\,200\,{\rm GeV}} \approx 0.5~{\rm fb} \times \left(\frac{100~{\rm TeV}}{M_5}\right)^3 \,,
\eeq
{\it i.e.}, $\sim 20$ events in the currently available datasets even for $M_5$ as high as 100~TeV. This motivates examining possibilities for more dedicated searches for any special signatures that the scenario has to offer.

\subsubsection{Periodicity in the diphoton and dilepton spectra}
\label{sec:FT}

The approximately periodic structure of the diphoton signal, as shown in figure~\ref{full_mass}, and similarly in the dielectron channel, motivates searching for these signals in the Fourier transform of the invariant mass spectrum.

More precisely, considering the mass formula in eq.~\eqref{KKG-masses}, it would be useful to search for a periodicity in the variable $\sqrt{m^2-k^2}$, for a signal hypothesis with a given $k$. We therefore define the power spectrum as
\beq
P(T) \equiv \left|\frac{1}{\sqrt{2\pi}}\int_{m_{\rm min}}^{m_{\rm max}} dm\;\frac{d\sigma}{dm}\exp\left({i\frac{2 \pi \sqrt{m^2-k^2}}{T}}\right)\right|^2 ,
\label{ps-defn}
\eeq
and expect the signal to produce a peak centered near $T = 1/R$. We take the lower limit of integration to be $m_{\rm min} = k$, since starting the integration at a lower mass would contaminate the sample with extra background events. The optimal upper limit of integration, $m_{\rm max}$, depends on a number of factors, including the signal-to-background ratio, the statistical fluctuations of the signal, the resolution / mass splitting ratio, and other factors that vary with $m$, and would in general be different for each point in the parameter space.

\begin{figure}[t]
\begin{center}
\includegraphics[width=0.43\textwidth]{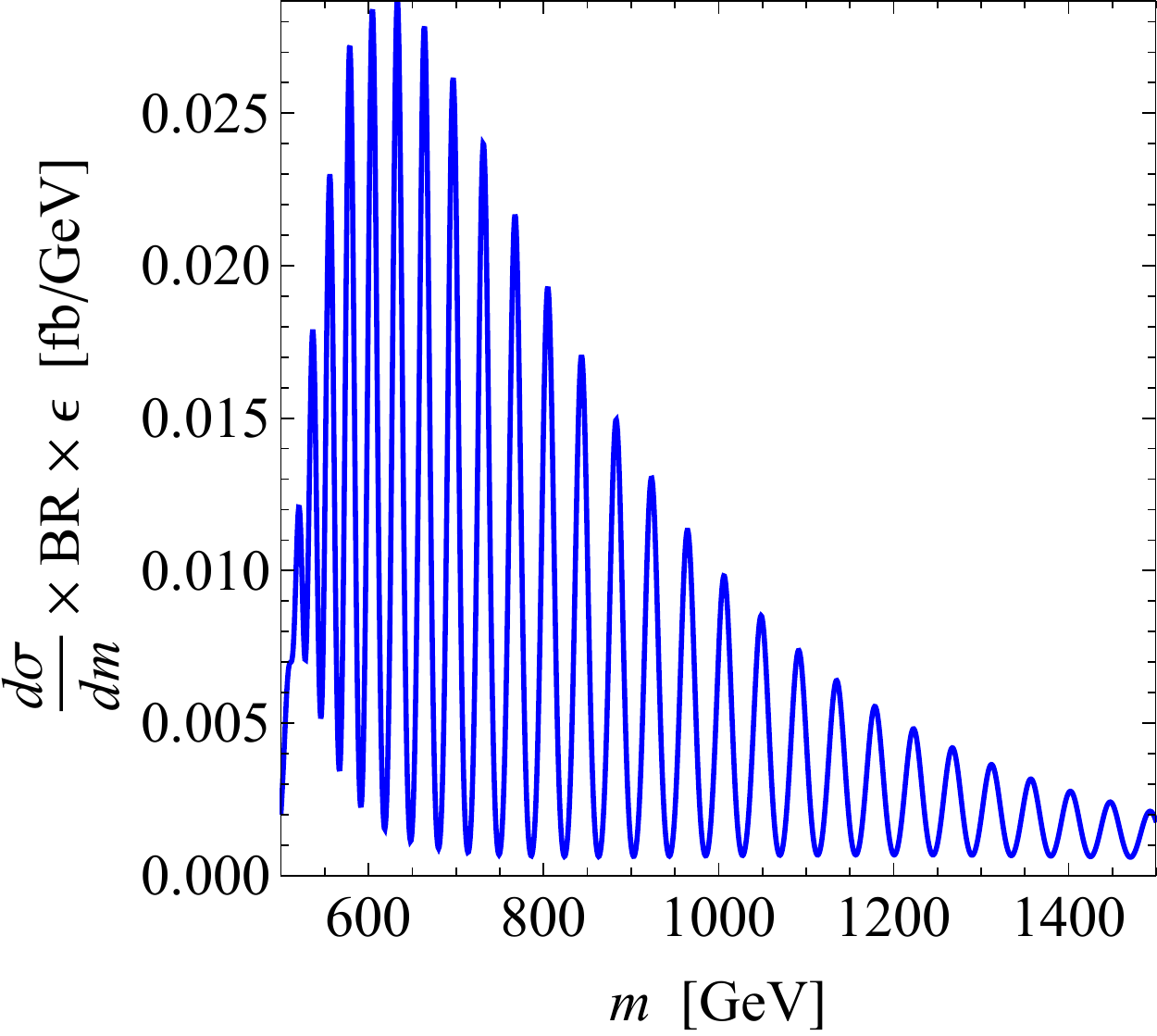}\qquad
\includegraphics[width=0.38\textwidth]{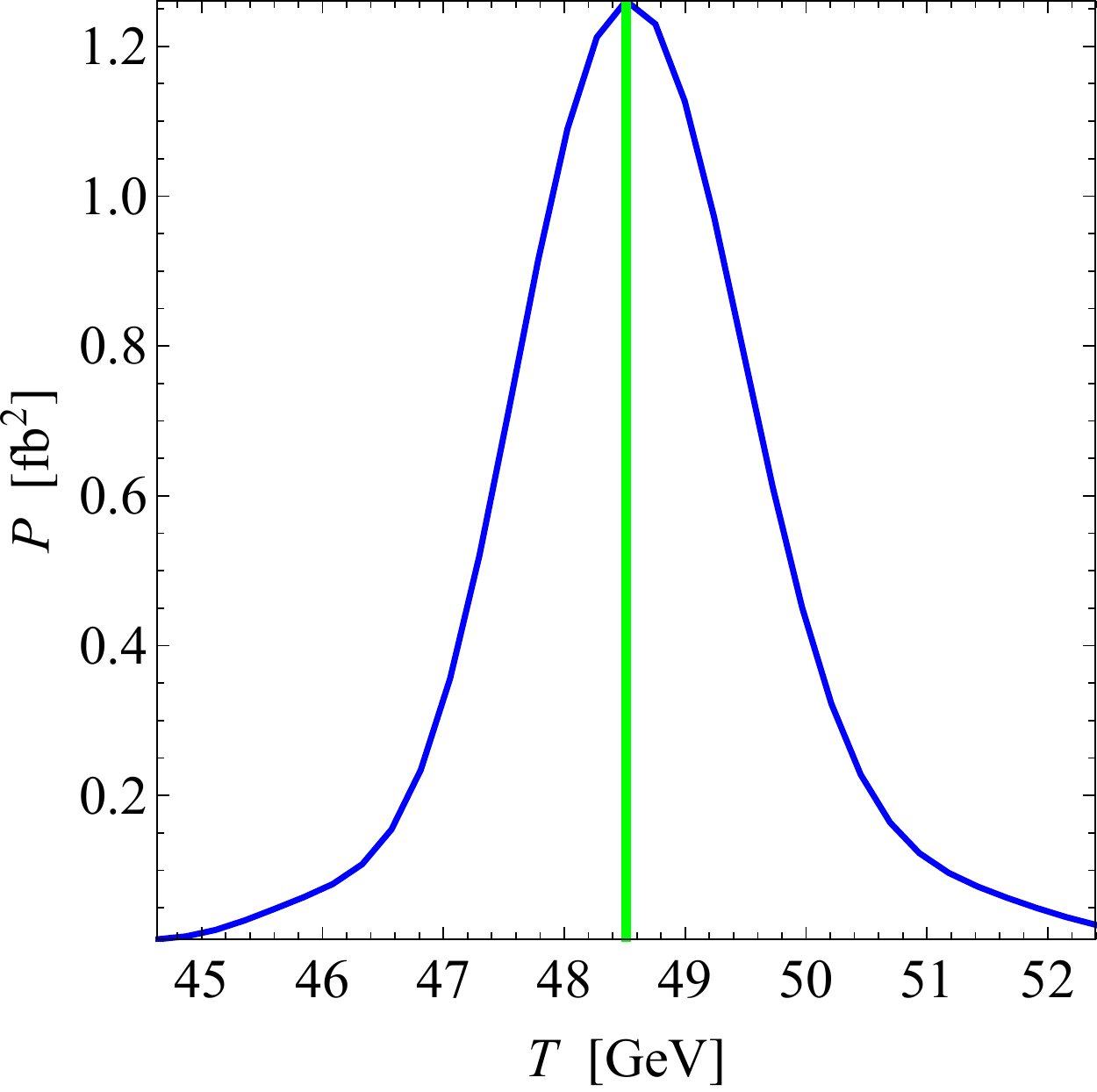}
\end{center}
\caption{The diphoton mass spectrum for $M_5 = 6$~TeV, $k = 500$~GeV in the range $k < m < 3k$ (left) and its power spectrum obtained from eq.~\eqref{ps-defn} (right). The expected periodicity, $1/R$, is indicated by the green vertical line.}
\label{input_mass_FT}
\vskip 10mm
\begin{center}
\includegraphics[width=0.43\textwidth]{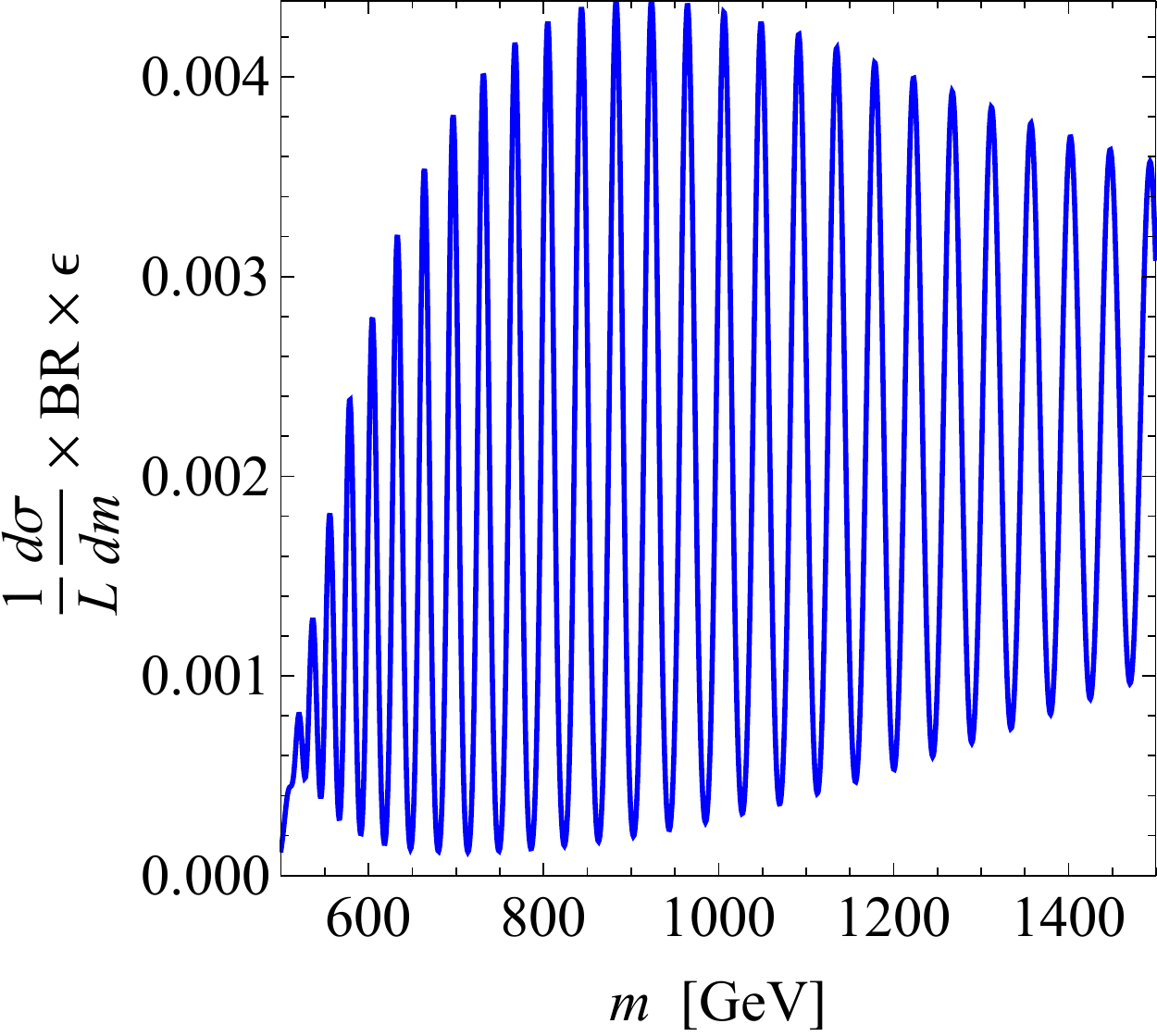}\qquad
\includegraphics[width=0.39\textwidth]{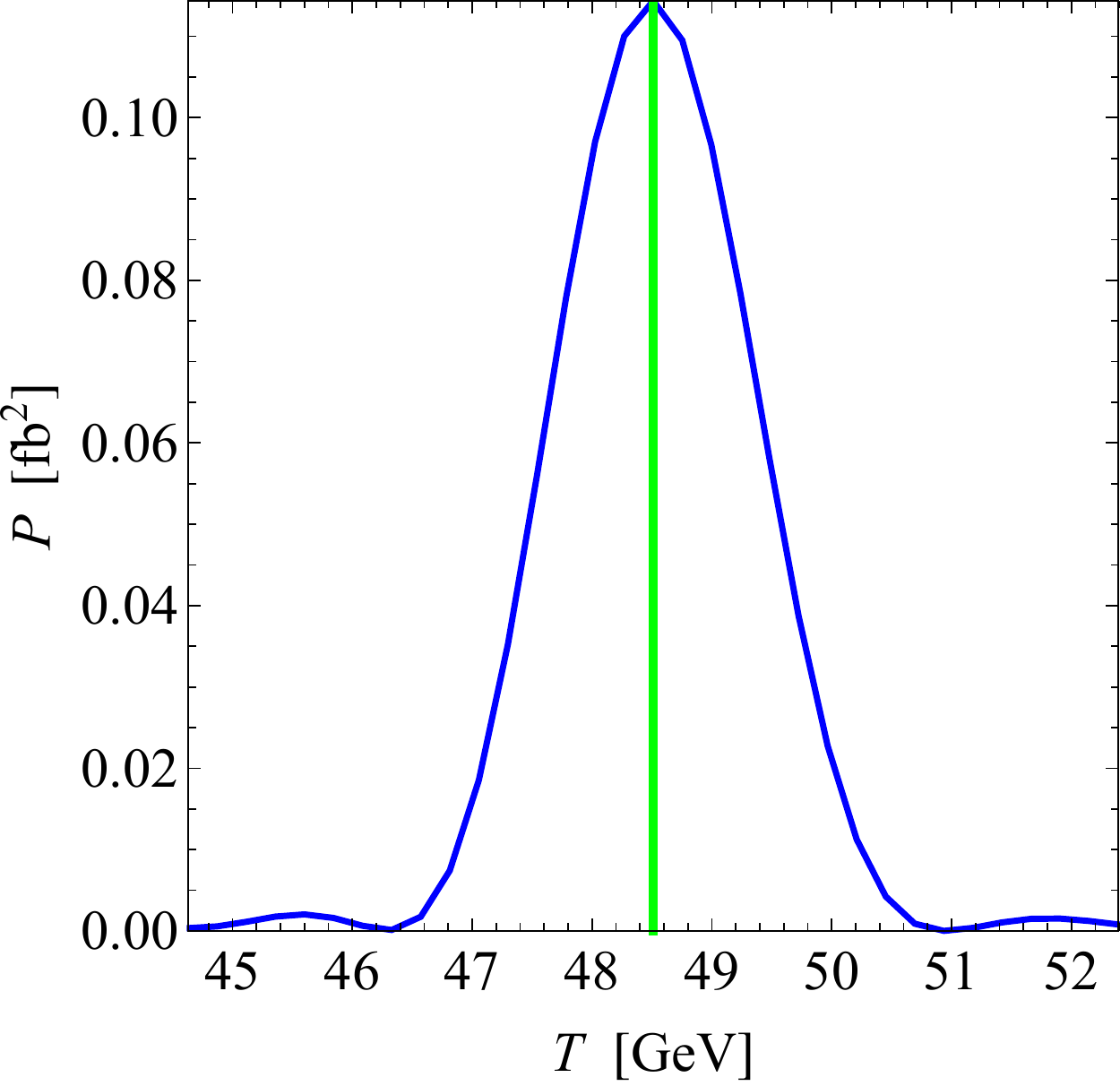}
\end{center}
\caption{Same as figure~\ref{input_mass_FT}, but after dividing out the parton luminosity.}
\label{input_norm_mass_FT}
\end{figure}

\begin{figure}[t]
\begin{center}
\includegraphics[width=0.41\textwidth]{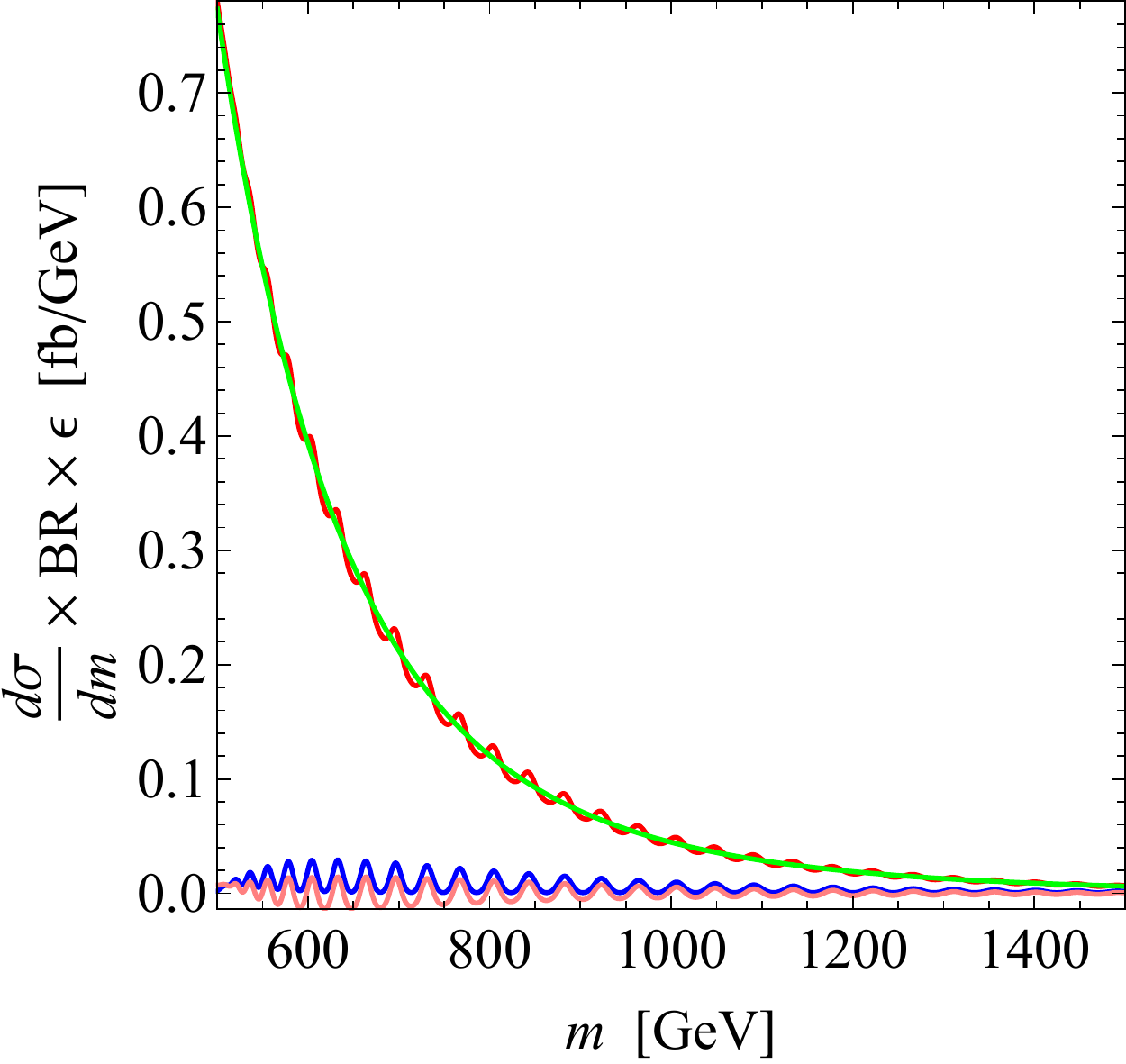}\qquad
\includegraphics[width=0.39\textwidth]{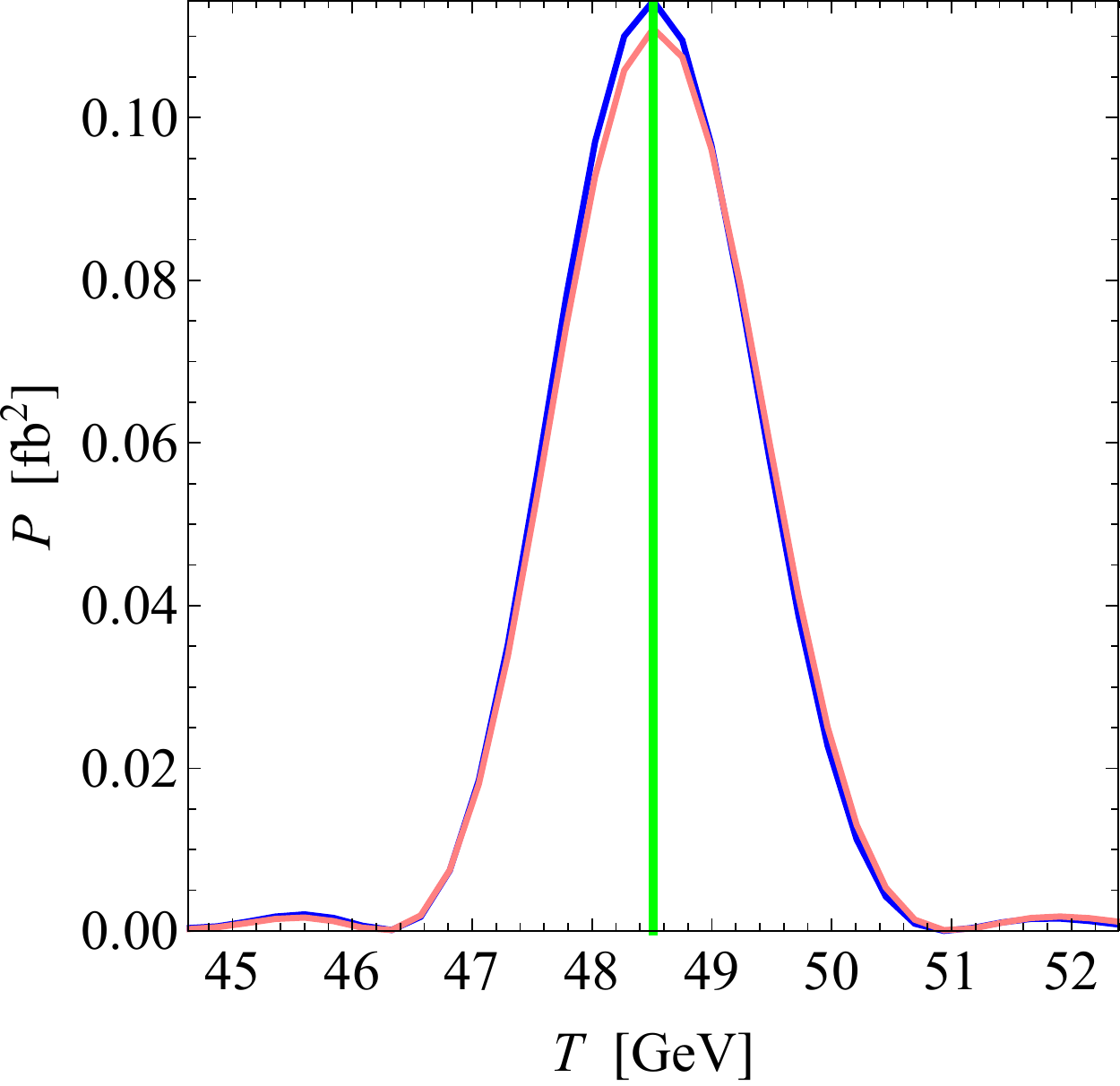}
\end{center}
\caption{Left: The signal+background curve (red) is fit to a non-oscillating function (green), which is then subtracted to leave just the oscillations (pink). The original oscillations are shown in blue. Right: the power spectrum of the extracted oscillations (pink) vs.\ the original one (blue).}
\label{add_and_subtract_bg}
\vskip 10mm
\begin{center}
\includegraphics[width=0.39\textwidth]{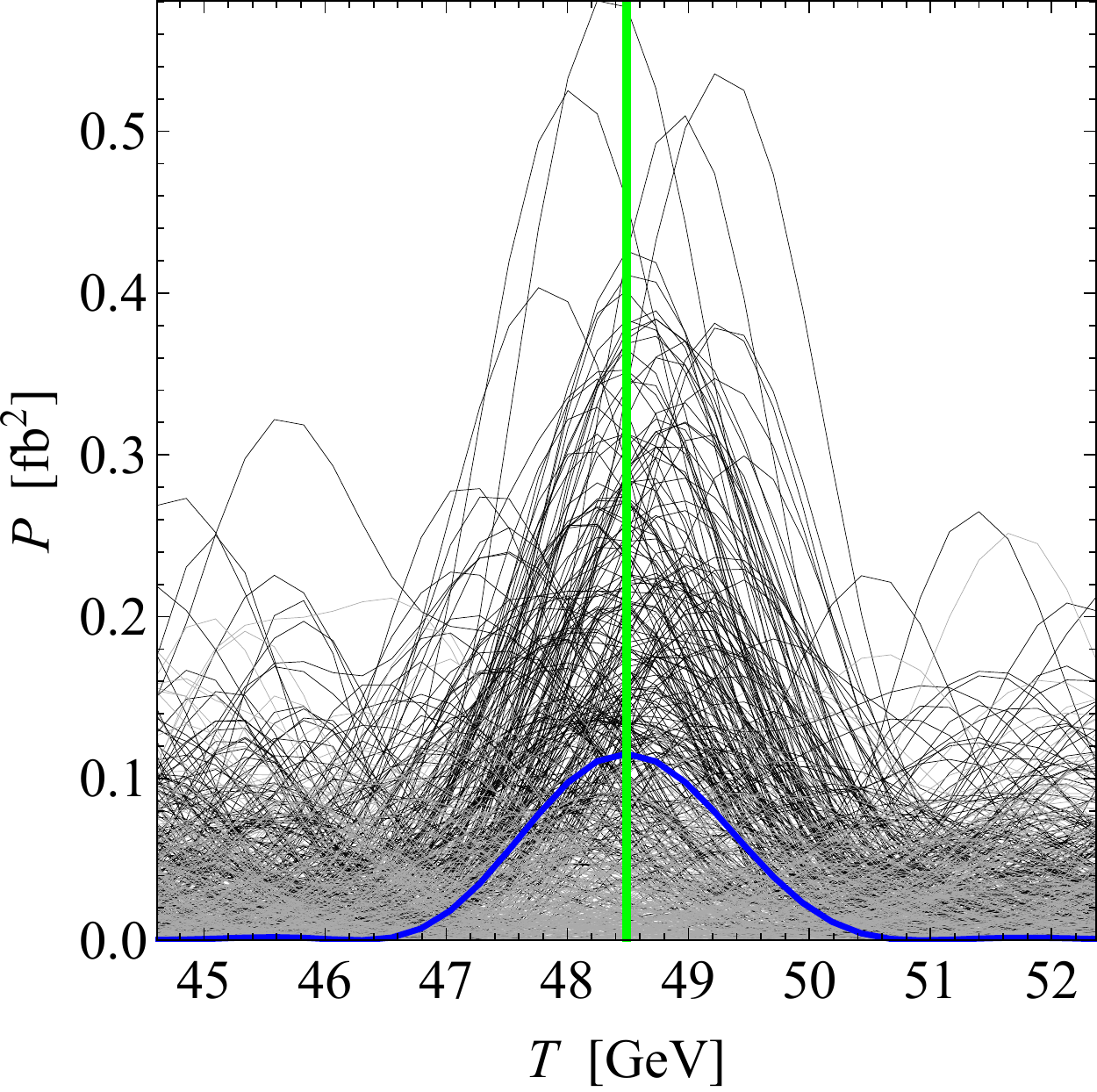}
\end{center}
\caption{Power spectra for multiple realizations of the signal+background (black lines) compared with the input signal (blue line), and multiple realizations of the background in the absence of any signal (gray lines).}
\label{fluctuated_FT}
\end{figure}

In figure~\ref{input_mass_FT}, we return to one of the examples from figure~\ref{full_mass} and show the power spectrum based on the mass range $k < m < 3k$. It peaks at $1/R$, as expected. An even narrower peak is obtained if the parton luminosity, which is responsible for the fast decrease of the signal with mass, is divided out,\footnote{More specifically, we divide by $L(m^2) \equiv {\cal L}_{gg}(m^2)
+ \frac43\,\sum_q{\cal L}_{q\bar q}(m^2)$, in accordance with eq.~\eqref{single-KKG-xsec}. More optimally, one would weigh the parton luminosity of each initial state by its QCD $K$ factor.} as shown in figure~\ref{input_norm_mass_FT}.

\begin{figure}[t]
\begin{center}
\includegraphics[width=0.87\textwidth]{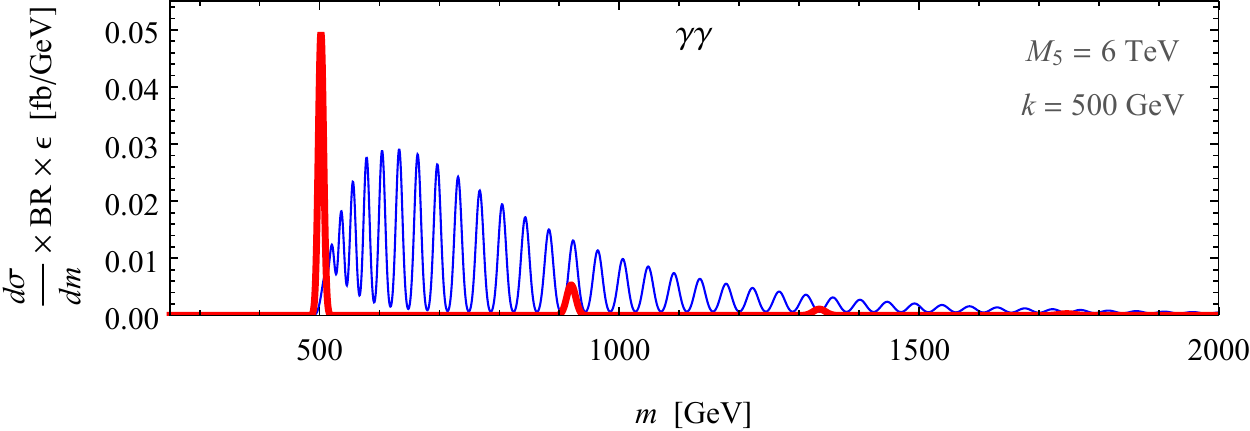}
\end{center}
\caption{Comparison between the diphoton spectra of the KK gravitons in CW/LD (thin blue, as in figure~\ref{full_mass}) and the RS model (thick red), after the RS curve has been scaled down by a factor of $30$ for presentation purposes. The parameters of the RS model were chosen such that the values of $M_5$ and the mass of the first KK mode, $m_1$, are the same in both models. The latter is obtained by setting $k_{\rm RS} \approx 0.26k$.}
\label{CW_vs_RS}
\end{figure}

In figure~\ref{add_and_subtract_bg}, background is added, based on the ``Spin-2 selection'' of ref.~\cite{Aaboud:2017yyg},\footnote{We have used this particular selection because the background is conveniently available from ATLAS. There may be room for optimization for the purposes of the particular analysis we describe here.} and then a fit to a non-oscillating function is subtracted. Nicely, the power spectrum of the extracted oscillations is almost identical to the original one.

To account for statistical fluctuations, we generate multiple realizations of fluctuated spectra for 36.7~fb$^{-1}$ of data (which is the typical amount of data used in the other searches we mention, and in particular in the ATLAS diphoton search~\cite{Aaboud:2017yyg} from which we take the background). This is done by finely binning the expected mass spectrum (that describes background + signal, or background only) and fluctuating the expected number of events in each bin according to Poisson statistics. An example of the resulting power spectra is shown in figure~\ref{fluctuated_FT}. To quantify the signal significance, we define the signal as the integrated power spectrum within the range of one (expected) width around the (expected) center of the peak, for a given $k$ and $M_5$ hypothesis. The significance is computed by dividing the average signal-induced excess on top of the average background by the background uncertainty. We also require the expected number of signal events in the mass range analyzed to be at least $3$.

For each point in the $M_5$--$k$ plane, we optimized the choice of $m_{\rm max}$ in eq.~\eqref{ps-defn} to obtain the best expected limit. The optimal values are usually in the range $2k \lesssim m_{\rm max} \lesssim 3k$. To address scenarios with low values of $k$, considering that the diphoton background curve in ref.~\cite{Aaboud:2017yyg} is shown only down to 200~GeV (although measurements at lower masses are definitely possible), and an independent prediction of the background shape is beyond the scope of this work, we have set $m_{\rm min} = \mbox{max}(k,\, 150~{\rm GeV})$. The background in the mass range $150$--$200$~GeV was obtained by extrapolation. The resulting expected limit is shown as $\gamma\gamma_{\rm FT}$ in figure~\ref{sensitivity}. The Fourier transform method seems to be competitive with the other methods.

We would like to note that even though multiple KK modes can be accessible at the LHC also in the RS scenario if the parameter $k$ of the RS scenario is taken to be small (as has been considered in refs.~\cite{Giudice:2004mg,Kisselev:2008xv,Franceschini:2011wr}), their spectrum will be quite different. It will not feature a series of comparable-size peaks that would motivate a Fourier analysis. In the RS scenario, the KK mode masses are given by $m_n \simeq x_n k$, where $x_n$ is the $n$-th zero of the Bessel function $J_1(x)$, and the interaction scales are $\Lambda_G^{(n)2} \simeq M_5^3/k$. As one can infer from figure~\ref{CW_vs_RS}, the resonant experimental signature of the RS scenario is dominated by the lightest mode, like it is usually assumed.\footnote{Much smaller splittings, \emph{e.g.}\ comparable to those of CW/LD, would be obtained for an RS model with a lower value of $k_{\rm RS}$. However, the beginning of the spectrum in such case will be correspondingly lower, possibly leading to limits from past experiments. We have not included such a case in figure~\ref{CW_vs_RS} also because the dilution of the rate due to the cascade decays (which we have not computed for the RS model) might be non-negligible if the spectrum begins at a low mass.}

\subsubsection{Turn-on of the spectrum at low invariant mass}

For $k \lesssim 100$~GeV, it is becoming increasingly difficult to see distinct resonances. While the relative mass splittings remain approximately the same as for high $k$, the relative experimental resolution on the invariant mass worsens as one goes to lower masses, according to eq.~\eqref{diphoton-resolution} in the diphoton channel, for instance. As a result, the resonances broaden and merge with neighboring ones, as demonstrated in figure~\ref{full_mass_20} for the case of $k = 20$~GeV.

While the separate KK mode peaks are difficult to see, it is interesting to know whether one can design a search that would be sensitive to the turn-on of the spectrum as a whole. One could contemplate doing such a search in either the $\gamma\gamma$, the $e^+e^-$ or the $\mu^+\mu^-$ channel. In all the three channels, triggering will become an issue as one goes significantly below $m \sim 100$~GeV. However, one could consider relying on initial-state radiation, as in refs.~\cite{ATLAS-CONF-2016-070,Sirunyan:2017dnz,Sirunyan:2017nvi}, or doing a trigger-level analysis / data scouting, as in refs.~\cite{ATLAS-CONF-2016-030,CMS-PAS-EXO-16-056}. Additionally, if such approaches make $\mu^+\mu^-$ events triggerable at low masses, it may even be possible to see the individual peaks, since the dimuon invariant mass resolution improves at low masses~\cite{Chatrchyan:2012xi,Aad:2016jkr}, as we have mentioned in section~\ref{sec:resonance-searches}. The resolution in the $e^+e^-$ channel~\cite{Khachatryan:2015hwa} might also allow individual peaks to be seen. Additionally, as we have already mentioned, the low-mass $\mu^+\mu^-$ spectrum may also be accessed by LHCb~\cite{Ilten:2016tkc,Aaij:2017rft}.

\begin{figure}[t]
\begin{center}
\includegraphics[width=0.87\textwidth]{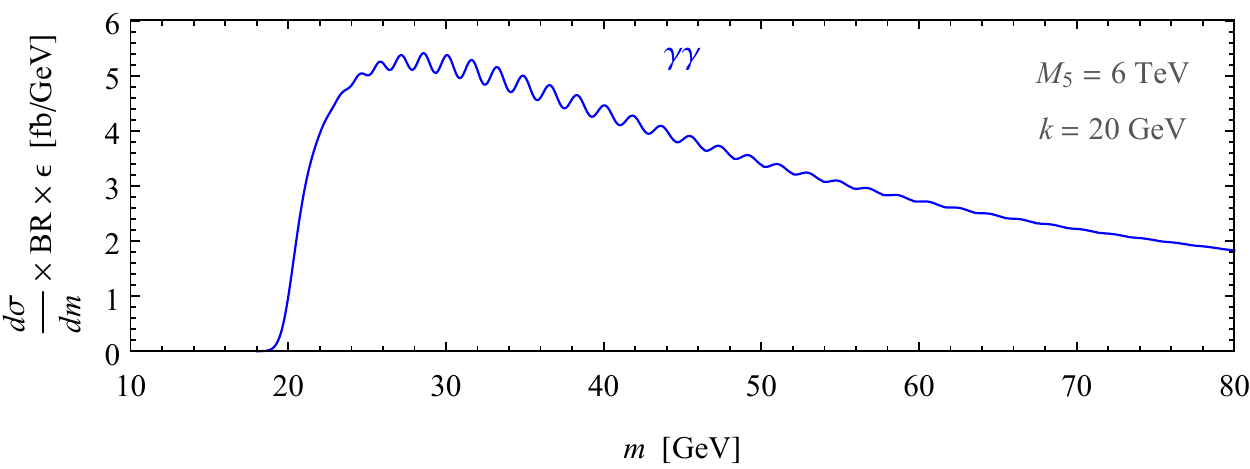}
\end{center}
\caption{Signal in the diphoton spectrum at the $13$~TeV LHC for $M_5 = 6$~TeV and $k = 20$~GeV, after accounting for the experimental resolution according to eq.~\eqref{diphoton-resolution}.}
\label{full_mass_20}
\end{figure}

\subsubsection{Cascades within the KK graviton and KK dilaton towers}
\label{pheno-cascades}

As we have seen in figure~\ref{BRs-w-KKG-KKG} (right), KK gravitons have sizable branching fractions to pairs of lighter KK gravitons if $k$ is much smaller than the KK graviton mass. We also find that the matrix elements for these decays are such that mode $n$ decays preferentially to modes $\ell$ and $m$ that approximately satisfy the conservation of the extra-dimensional momentum, $n = \ell + m$, despite phase space suppression, as demonstrated in figure~\ref{KKG2KKG_PS}. This implies that multi-step cascades can be common. Furthermore, because of this property of the matrix elements, such cascades will lead mostly to genuine multi-object final states, without boost-induced merging of the lighter KK graviton decay products into single objects. CW/LD therefore provides a new interesting benchmark model for high-multiplicity final states. It would be useful to examine to what extent it is being addressed by existing high-multiplicity searches, such as those originally motivated by black holes~\cite{Sirunyan:2017anm}, RPV supersymmetry~\cite{ATLAS-CONF-2016-057,Aaboud:2017faq,Khachatryan:2016xim}, or models with colorons or axigluons~\cite{Khachatryan:2016xim}. Additionally, a small fraction of the multi-object events will contain high multiplicity of special objects with low backgrounds. In particular, searches for multiple leptons ({\it e.g.}~refs.~\cite{Aaboud:2017dmy,ATLAS-CONF-2016-075,Sirunyan:2017lae,Sirunyan:2017hvp}), photons ({\it e.g.}~ref.~\cite{Aad:2015bua}) or $b$ jets ({\it e.g.}~ref.~\cite{CMS-PAS-SUS-16-013}) could have some sensitivity to this scenario.

\begin{figure}[t]
\begin{center}
\includegraphics[width=0.47\textwidth]{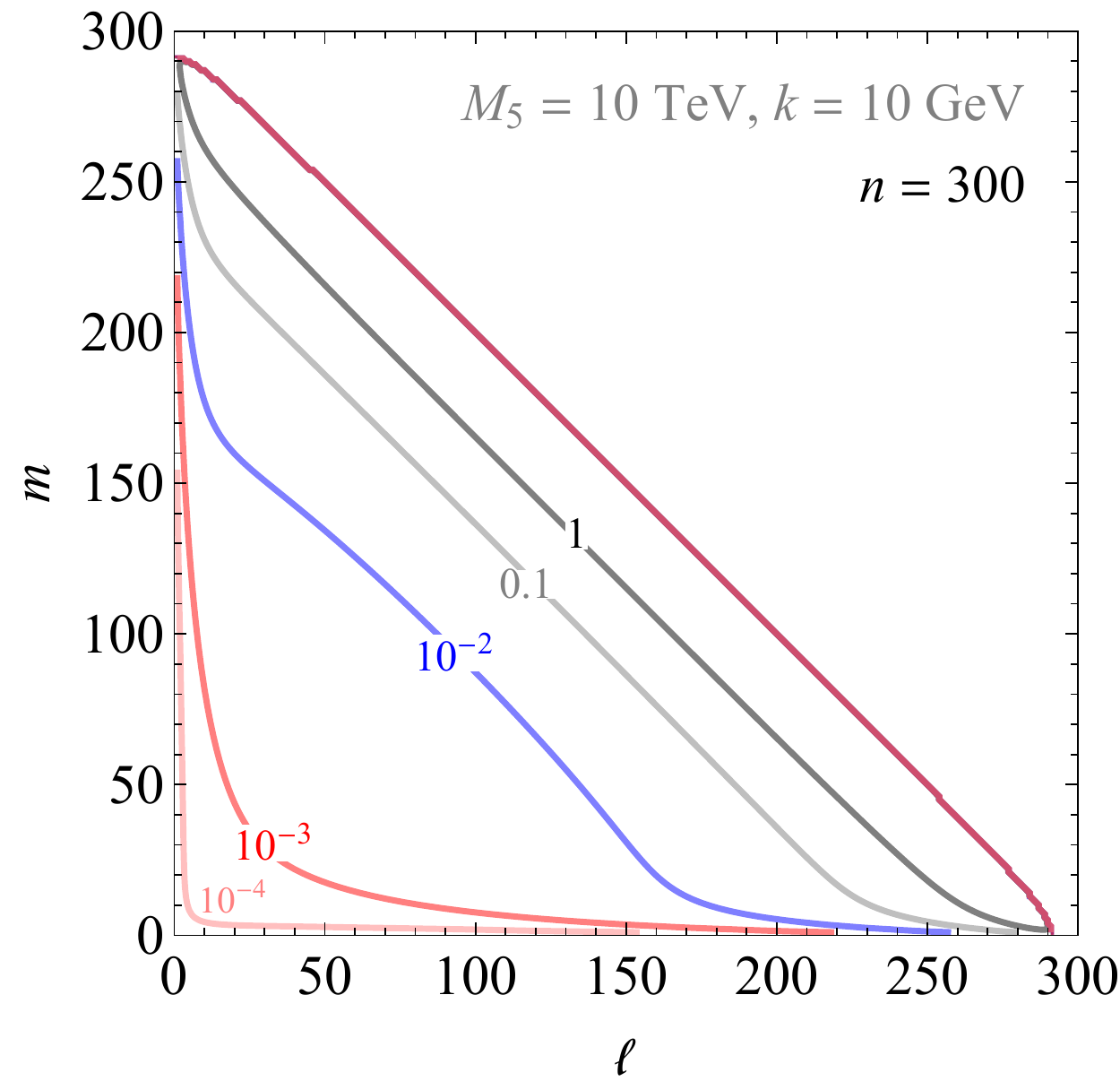}\quad
\includegraphics[width=0.49\textwidth]{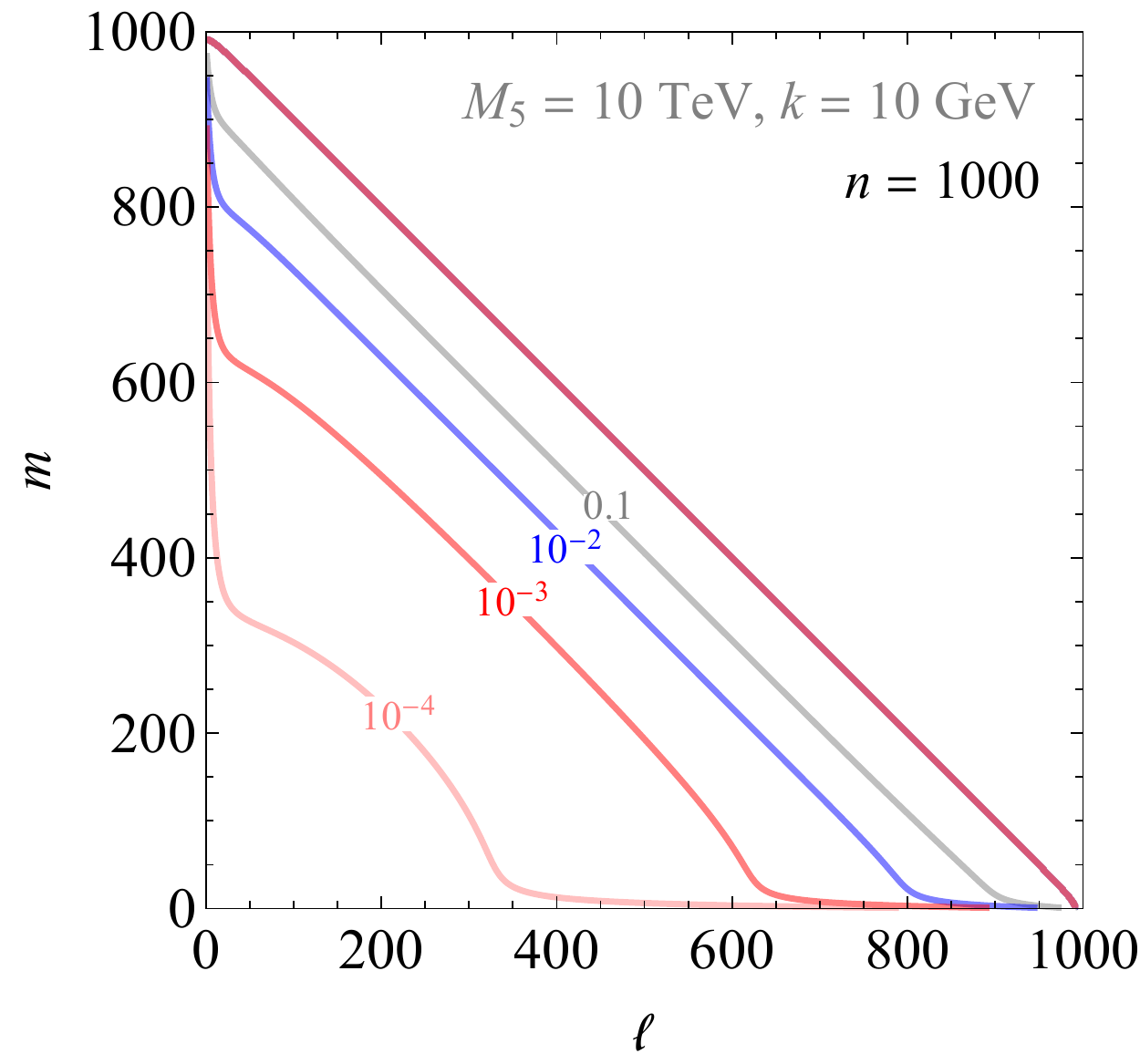}
\end{center}
\caption{Contours of the partial width for a KK graviton mode $n$ to decay to a pair of modes $\ell$ and $m$, $\Gamma_{G_n \to G_\ell G_m}$, normalized by $\Gamma_{G_n \to \sum G_\ell G_m} / n^2$. The example shown is $M_5 = 10$~TeV, $k = 10$~GeV, for mode $n = 300$ (left) and $n = 1000$ (right). The masses of these modes are $m_{300} \approx 318$~GeV and $m_{1000} \approx 1059$~GeV.}
\label{KKG2KKG_PS}
\end{figure}

\begin{figure}[t]
\begin{center}
\includegraphics[width=0.48\textwidth]{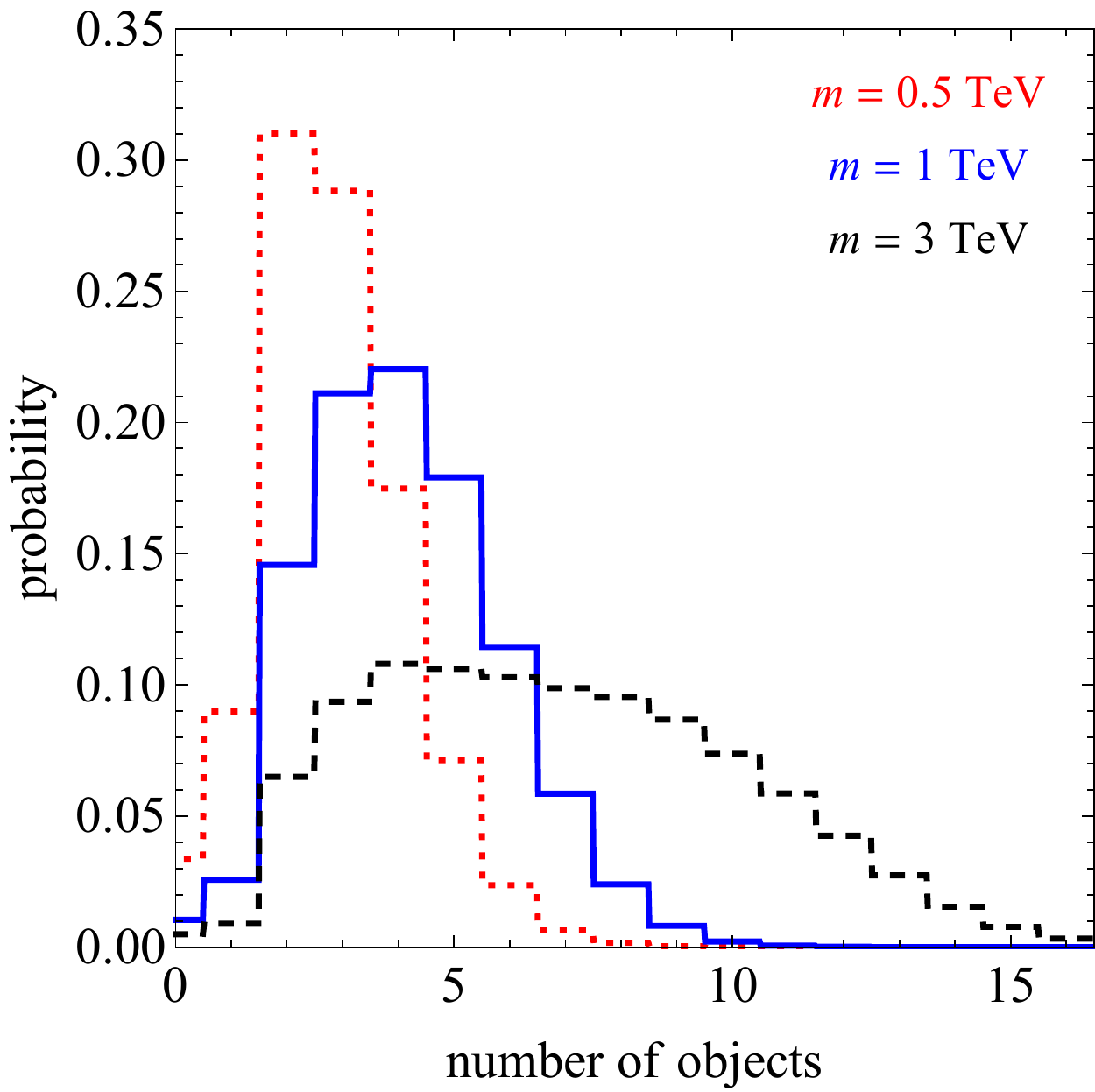}\quad
\includegraphics[width=0.49\textwidth]{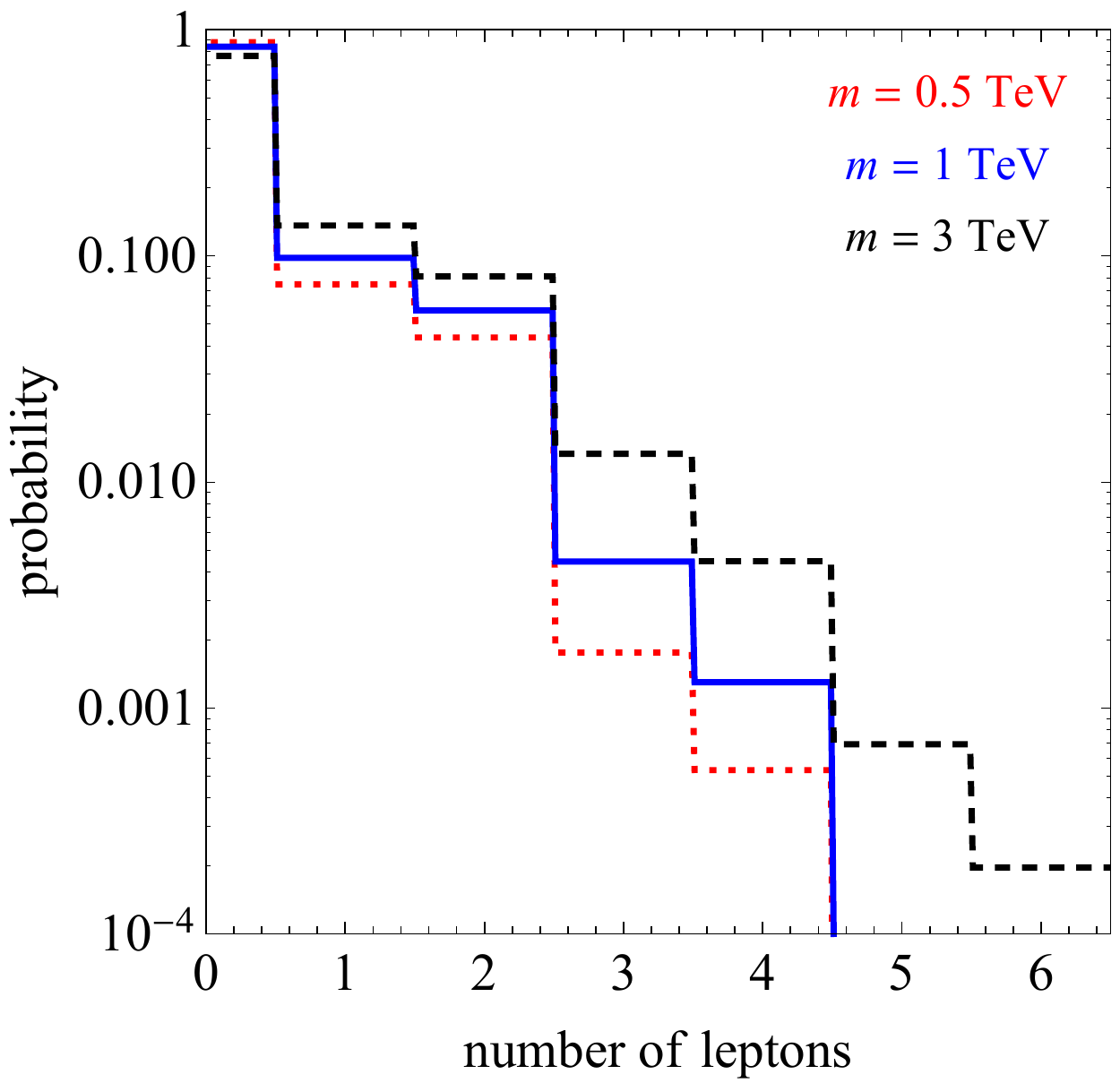}\\
\vskip 5mm
\includegraphics[width=0.48\textwidth]{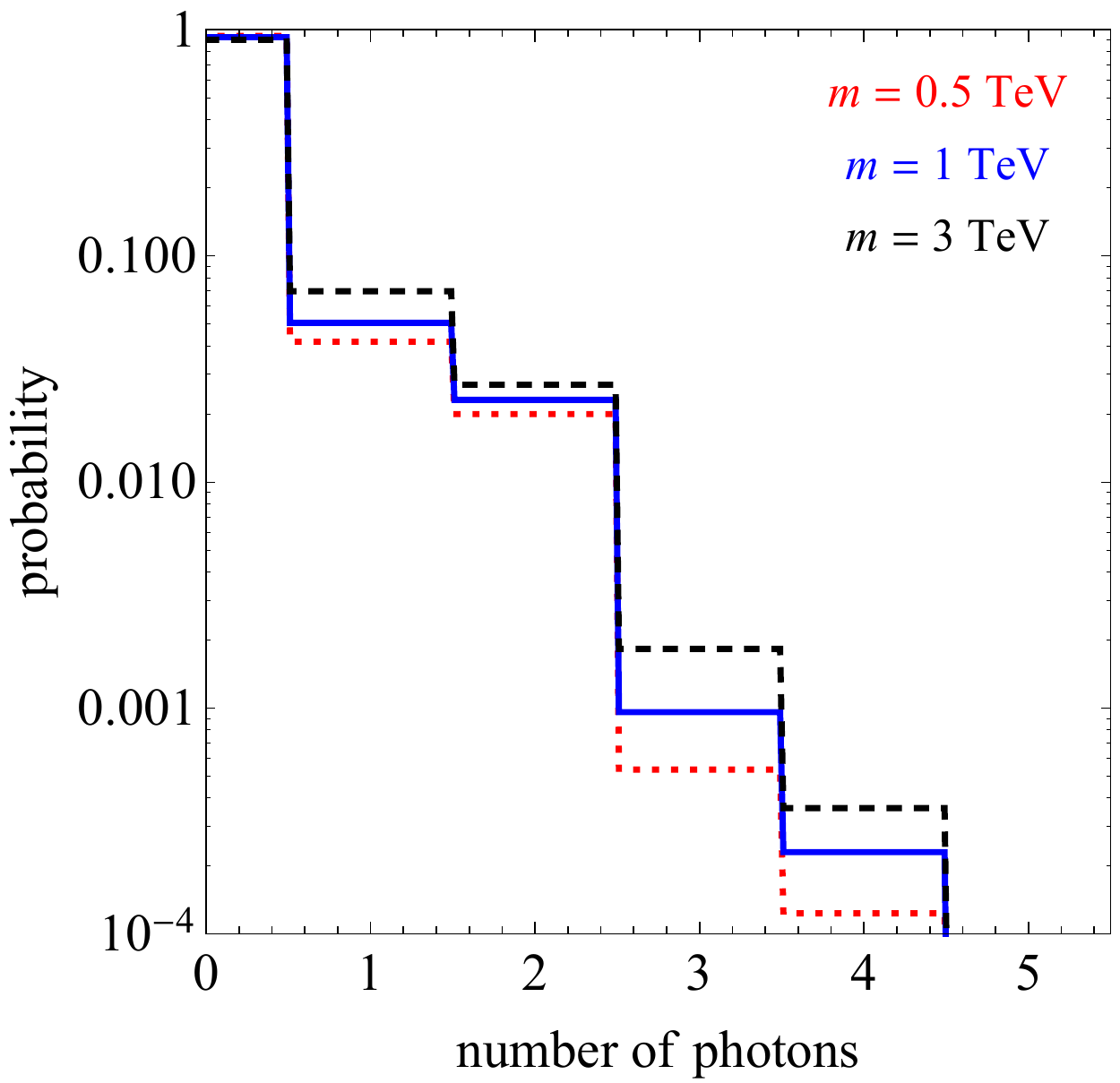}\quad
\includegraphics[width=0.48\textwidth]{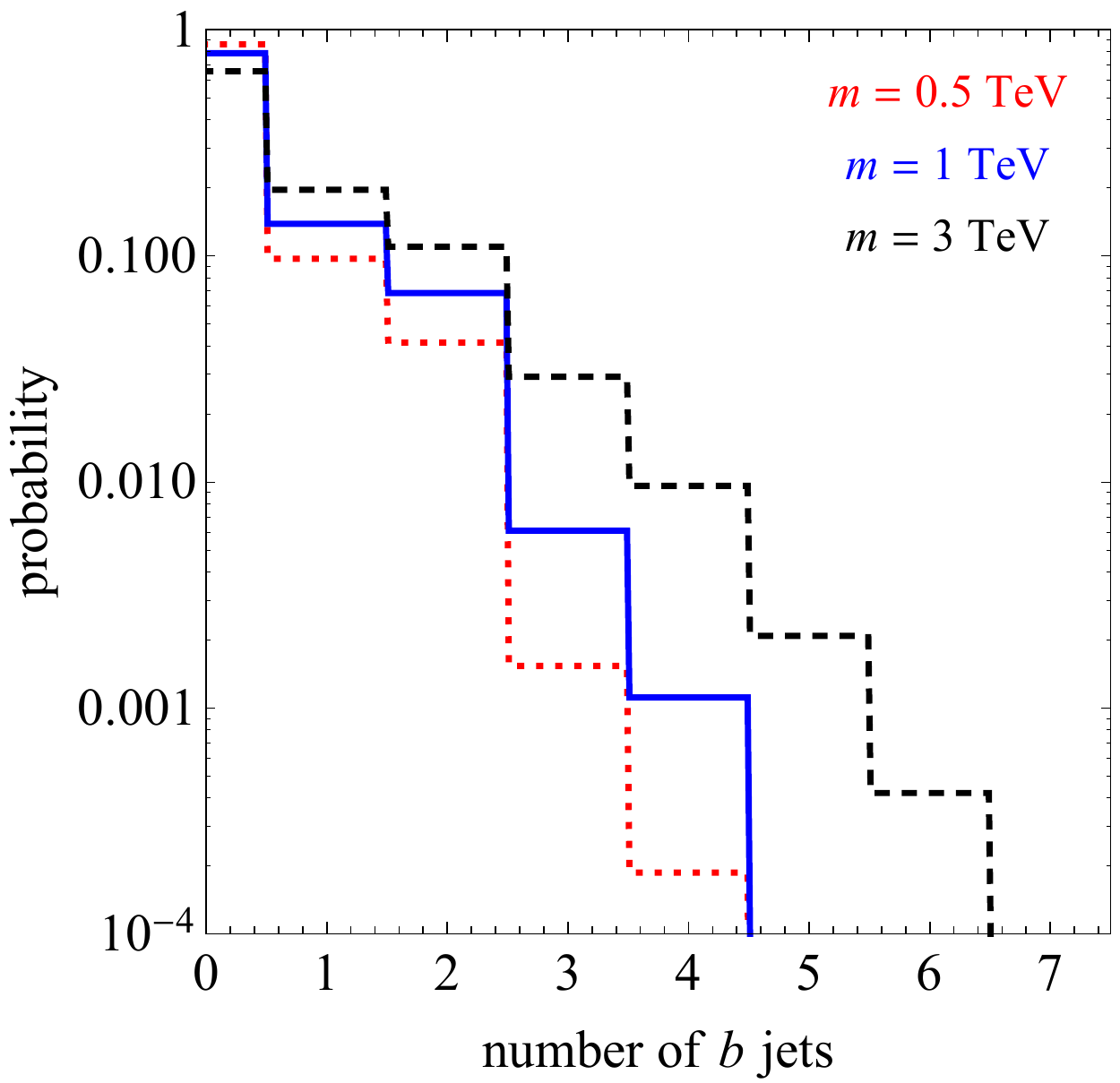}
\end{center}
\caption{For $M_5 = 10$~TeV, $k = 1$~GeV, and KK gravitons with mass 0.5~TeV (dotted red), 1~TeV (solid blue) and 3~TeV (dashed black), distributions of the number of objects (including jets, leptons and photons) with $p_T > 50$~GeV (top-left), leptons with $p_T > 20$~GeV (top-right), photons with $p_T > 20$~GeV (bottom-left) and $b$ jets with $p_T > 50$~GeV (bottom-right). In all cases, only objects with $|\eta| < 2.5$ are counted. Typical isolation and identification efficiencies have been taken into account. Leptons include electrons and muons.}
\label{cascades-distribs}
\end{figure}

To learn more about the properties of the final states produced in scenarios with KK graviton cascades, we have set up a Monte Carlo simulation. The KK gravitons were defined as new particles in {\sc Pythia}~8.219~\cite{Sjostrand:2014zea} with the appropriate masses and partial decay widths to pairs of lighter KK gravitons and to SM particles. After production, decays, parton showering and hadrozination, the Monte Carlo output is passed through a detector simulation software that was developed, validated and used (with small variations) in refs.~\cite{Kats:2011qh,Evans:2012bf,Evans:2013jna}. Jets are formed with the anti-$k_T$ algorithm~\cite{Cacciari:2008gp} with cone size $R = 0.45$ using {\sc FastJet}~\cite{Cacciari:2005hq}.

Figure~\ref{cascades-distribs} shows the multiplicity distributions of various types of objects in the case with $k = 1$~GeV, which is representative of scenarios in which cascades lead to final states with a high multiplicity of hard SM objects. We have taken $M_5 = 10$~TeV, although the dependence of the results on the value of $M_5$ is weak. We see that only relatively heavy KK gravitons feature high overall object multiplicities ($\sim 10$). The production cross section of such heavy KK gravitons is relatively small, {\it e.g.}\ for $m > 3$~TeV,
\beq
\left.\sigma_{\rm total}\right|_{m\,>\,3\,{\rm TeV}} \approx 0.4~{\rm fb} \times \left(\frac{10~{\rm TeV}}{M_5}\right)^3 \,,
\eeq
{\it i.e.}, only $\sim 16$ events in the currently available datasets for $M_5 = 10$~TeV. We also see that the fractions of events with multiple ($3$ or preferably more) leptons, photons or $b$ jets are always small, especially at the lower masses where the production cross section is sizable.

Taking other values of $k$ does not make the signatures significantly more spectacular. In particular, much lower values of $k$ make the cascades proceed down to much lower masses (in accordance with figure~\ref{BRs-w-KKG-KKG}, right), making the SM particles produced in them too soft. We thus see that the high-multiplicity signatures due to the cascades do not obviously set strong limits. Still, such scenarios suggest an interesting new target for the multilepton searches, and motivate further development of the relatively few existing multi-object~\cite{Sirunyan:2017anm,ATLAS-CONF-2016-057,Aaboud:2017faq,Khachatryan:2016xim}, multiphoton~\cite{Aad:2015bua} and multi-$b$~\cite{CMS-PAS-SUS-16-013} searches.

Another interesting possibility is for part of the cascade to be displaced. Indeed, as can be seen in figures~\ref{lifetime-map} and~\ref{xsecs}, one can have a situation in which a heavy KK graviton produces a cascade that contains a KK graviton or KK dilaton whose decay is displaced. It is worth examining whether the existing search strategies for displaced decays~\cite{Aad:2015uaa,Aad:2015rba,ATLAS-CONF-2016-103,Aaboud:2017iio,CMS-PAS-EXO-16-003,CMS:2014hka,CMS-PAS-EXO-14-012,Aaij:2017mic}, which are motivated by completely different scenarios, provide sufficiently optimal coverage of these signatures of CW/LD.

\begin{figure}[t]
\begin{center}
\includegraphics[width=0.65\textwidth]{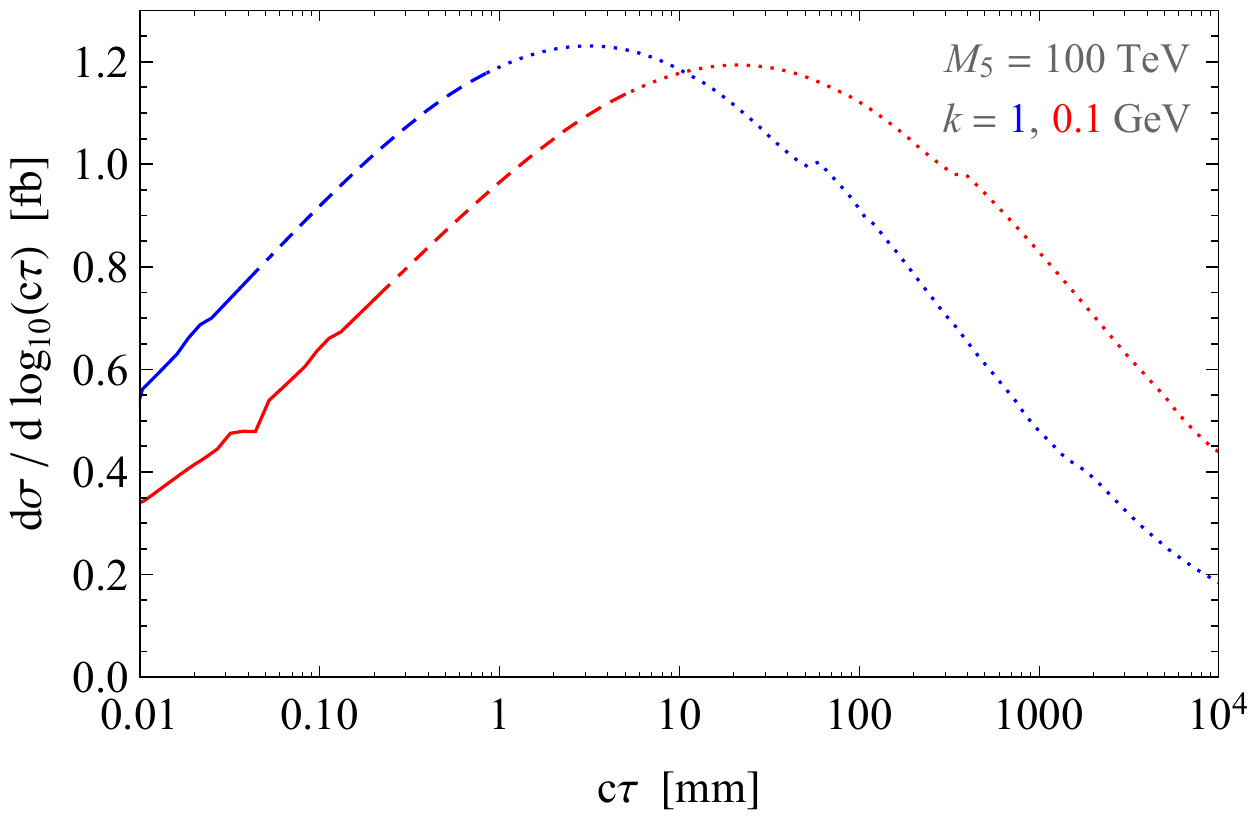}
\end{center}
\caption{KK gravitons resonant production cross section at the 13~TeV LHC, differential in the mode lifetimes, for $M_5 = 100$~TeV and $k = 1$~GeV (blue) and $0.1$~GeV (red). Dotted segments indicate very light modes  ($m < 40$~GeV) which will likely be difficult to see, dashed segments describe intermediate masses ($40 < m < 100$~GeV), while solid segments describe heavy KK modes ($m > 100$~GeV). For higher values of $M_5$, all the displacements would be larger (behaving approximately as $c\tau \propto M_5^3$), but the production cross sections would be smaller (behaving approximately as $\sigma \propto 1/M_5^3$).}
\label{xsec-vs-lifetime}
\end{figure}

\subsubsection{Resonant production of particles with displaced decays}

It is usually believed that a particle produced resonantly with a currently observable rate at the LHC must decay promptly, due to the possibility for it to decay back to the production mode. The KK gravitons of CW/LD provide an interesting exception, with the help of their dense spectrum. Even though the production cross section of any single KK mode whose decay is displaced is small, when summing over all modes with mass around a typical value, the rate can become sizable, as can be seen from figure~\ref{xsecs} (bottom-left plot) or figure~\ref{xsec-vs-lifetime}. While at leading order the KK graviton is produced with zero $p_T$, initial-state radiation can easily give it a sufficient boost for its prolonged lifetime to manifest itself in a displaced decay. While it is far from obvious to us that this signature will be competitive in sensitivity with the other signatures we have discussed, it would be interesting to explore whether in some parameter range this is the case.

\section{Conclusions}
Since being thrust into the spotlight over two decades ago, weak scale extra dimensional models have continued to provide new insights into the theoretical questions raised by the structure of the Standard Model, as well as being a rich source of novel collider signatures.  The majority of both theoretical and phenomenological studies have focussed on either LED or RS scenarios, or small deformations thereof.  However the theoretical and phenomenological vistas may be very different in scenarios with alternative geometries in the extra dimension.  One example that has been essentially dormant since it was first proposed is the CW/LD setup.  We have revisited this scenario anew and studied both theoretical and phenomenological aspects.

In this work we have reconsidered the UV motivation for the CW/LD action and studied its structure from a ground-up perspective.  In particular, we have studied the stability of the geometry to deformations of the action by bulk and brane cosmological constant terms, as such terms are difficult to avoid from an effective field theory perspective.  We find that a bulk CC must be exponentially small to avoid significantly altering the geometry.  As a consequence bulk supersymmetry is called for, in accordance with the top-down purview of superstring theory.  On the other hand, the setup does not require a significant tuning of the CC term on the TeV brane unless $k\ll M_5$, which is convenient as the TeV brane which accommodates the SM cannot be supersymmetric.

\begin{figure}[t]
\begin{center}
\vspace{6mm}
\includegraphics[width=0.55\textwidth]{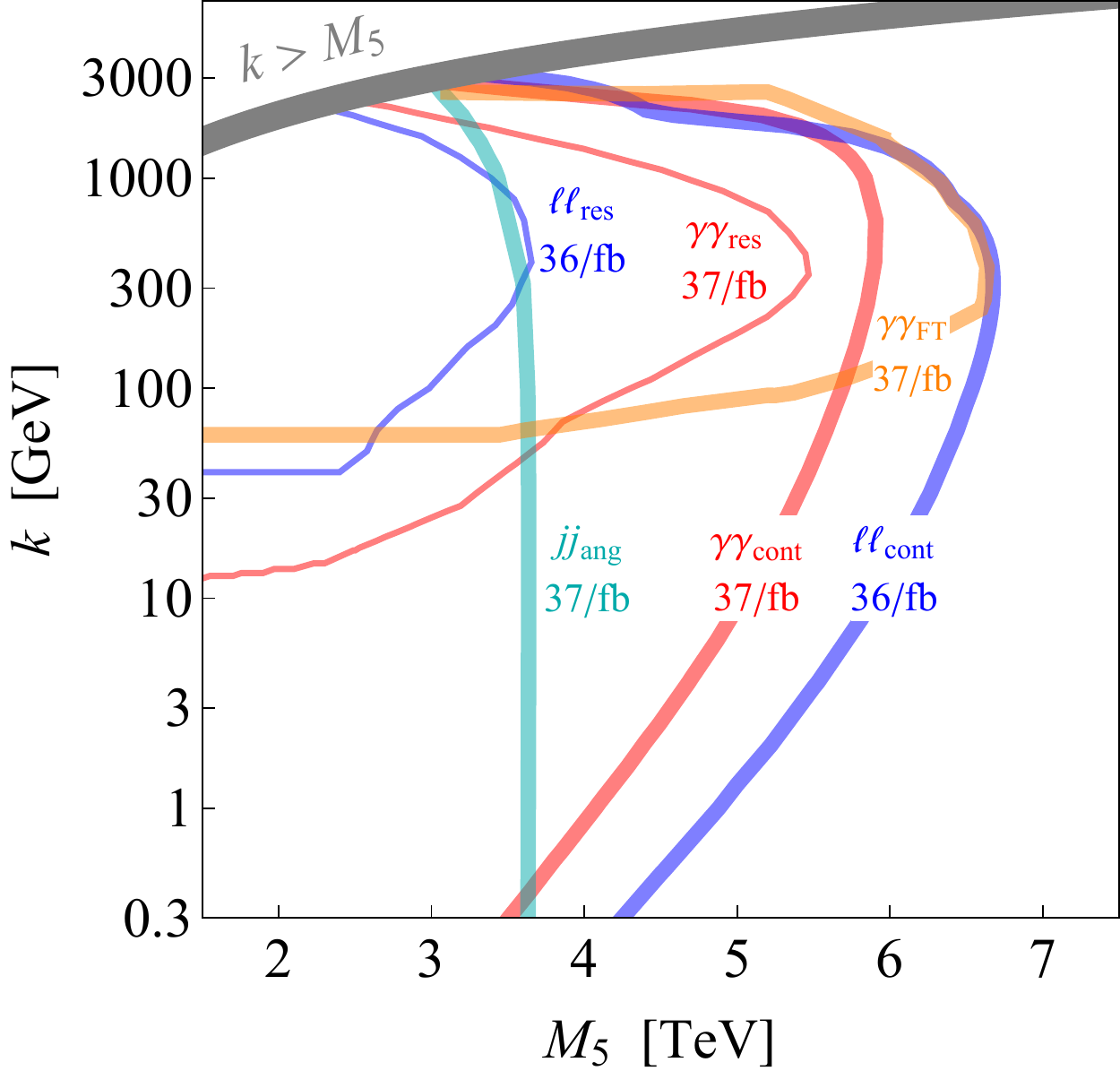}\\
\vspace{-4mm}
\end{center}
\caption{Estimated sensitivity of some of the search channels, with the caveats discussed in the text. The $jj_{\rm ang}$ curve refers to the search in the dijet angular distributions~\cite{Aaboud:2017yvp}, $\ell\ell_{\rm res}$ to the dilepton resonance search~\cite{Aaboud:2017buh}, $\gamma\gamma_{\rm res}$ to the diphoton resonance searches~\cite{Aaboud:2017yyg,CMS-PAS-HIG-17-013}, $\ell\ell_{\rm cont}$ to the search in the high-mass dilepton continuum~\cite{Aaboud:2017buh}, $\gamma\gamma_{\rm cont}$ to the search in the high-mass diphoton continuum~\cite{Aaboud:2017yyg}, and $\gamma\gamma_{\rm FT}$ to our proposed Fourier transform analysis.  Note that, as discussed in \Sec{sec1}, the region of lower $k$ is likely to be more fine-tuned in a UV-completion of the CW/LD.}
\label{sensitivity}
\end{figure}

From a phenomenological perspective, in figure~\ref{sensitivity} we show existing collider constraints on the model, finding that the 5D Planck scale is typically required to be $M_5 \gtrsim 6$ TeV, as in the case of other extra dimensional scenarios.  Without the full matching from the UV to IR it is impossible to quantitatively estimate the fine-tuning implications for the Higgs sector; however, it is unlikely that the prospects are any more rosy than alternative extra-dimensional models.  More optimistically we have identified phenomenological signatures which are of particular interest and warrant highlighting.  These signatures are not just relevant to the CW/LD setup but are also worth considering in their own right as they may also arise in other new-physics scenarios.  We list these signatures here, beginning with a particularly novel possibility.
\begin{itemize}
\item  \textbf{Frequency-space searches}.  The closely spaced resonances of the CW/LD setup, which are a model-independent prediction, lead to oscillating signatures in diphoton and dilepton final states that can efficiently be searched for by inspecting for bumps in `Fourier space'.  The latter being defined through the Fourier transform, or some close cousin, of the invariant mass spectrum.  Densely packed resonances are not restricted to the CW/LD setup, thus a broader range of frequency space searches may capture evidence of other weakly coupled hidden sectors.  This motivates a general set of `Fourier space' analyses and suggests that, while we usually search for new states through tell-tale bumps in the invariant-mass domain, it may be that the bumps are lurking in the Fourier domain.
\item  \textbf{$s$-channel long lived particles (LLPs)}:  The CW/LD setup contains extremely large numbers of weakly coupled states.  Since they are very weakly coupled they may be long-lived and give displaced signatures in colliders.  Due to the weak coupling the production cross section of any given particle is extremely small, which is why the $s$-channel production of LLPs is not often considered at colliders.  However, this suppression for a single state is counteracted by the large number of states and the inclusive cross section for producing such particles can become relevant at the LHC.  Such possibilities are not restricted to extra dimensions and, as the LLP frontier at the LHC is coming under increasing scrutiny, it would be worthwhile to consider general searches for $s$-channel produced LLPs, alongside more conventional production mechanisms.
\item  \textbf{Graviton cascades}:  In the CW/LD scenario the cascade decays of heavy gravitons to a series of lighter gravitons and scalars eventually terminate in very high multiplicity SM final states.  Some of these final decays may even be displaced.  Thus the CW/LD is a calculable scenario in which the searches for \textit{e.g.}\ black holes have applicability.
\end{itemize}
There are additional defining features of the CW/LD scenario of phenomenological relevance, particularly in comparison to LED or RS.  While high-mass excesses in the diphoton, dilepton and other channels are present, their shapes are very different from the LED benchmark models that are being used for searches. This happens because in LED scenarios, which necessarily involve multiple large extra dimensions, the signal is coming from cutoff-dominated off-shell production, rather than the on-shell production that dominates in the case of a single extra dimension.  Another interesting feature of the model is that the KK tower may start at a low mass if $k$ is small. This motivates low-mass searches for the turn-on of the spectrum in various channels, as well as a variety of searches for cascade signatures due to heavy KK modes decaying into lighter ones.

In summary, the CW/LD setup motivates new classes of LHC searches that will extend the experimental coverage of physics beyond the SM into new territory.  More broadly, while the CW/LD scenario is in some sense a second cousin to LED or RS, this work demonstrates that, since the phenomenological signatures and theoretical perspectives depend sensitively on the extra dimensional geometry, the study of further deformations of these models may reveal additional unconsidered collider signatures.

\acknowledgments

We thank Ofer Aharony, Luis \'{A}lvarez-Gaum\'{e}, John Paul Chou, Amit Giveon, Massimo Porrati, Riccardo Rattazzi, Javi Serra, Matt Strassler and Luca Vecchi for useful discussions.
The research of YK was supported in part by the Israel Science Foundation (grant no.~780/17).

\appendix

\section{Background solutions}\label{app:background}
We start from the full general action \eqref{eq:Einstein}. We focus on solutions compatible with 4D Poincar\'e invariance, which, in conformally flat coordinates, can be parametrized by the line element
\beq\label{lineelement2}
ds^{2}=e^{2\sigma(y)}\left(\eta_{\mu\nu}dx^{\mu}dx^{\nu}+dy^{2}\right)\,.
\eeq
Without loss of generality we also assume $\sigma(y=y_{\rm T}=0)=0$. The stress energy tensor corresponding to the action \eqref{eq:Einstein} is
\beq
T_{MN}=T^{S}_{MN}+ T^{\text{SM}}_{MN}\,,
\eeq 
where we defined the stress energy tensors of the bulk scalar $S$ and of the SM, respectively $T^{S}_{MN}$ and $T^{\text{SM}}_{MN}$. We assume that the SM stress energy tensor does not contribute to the background solutions. This corresponds to the assumption that the tuning of the 4D cosmological constant cancels the SM contribution to the vacuum energy proportional to $m_{h}^{2} v^{2}$, with $m_{h}$ the Higgs mass and $v$ the Higgs vev. The stress energy tensor for $S$ is given by
\beq\label{tmunu1}
\displaystyle T^{S}_{MN}=-\frac{2}{\sqrt{-g}}\frac{\delta\left(\sqrt{-g}\mathcal{L}_{S}\right)}{\delta g^{MN}}\,,
\eeq
with
\beq\label{Slagr}
\mathcal{L}_{S}=-\frac{M_{5}^{3}}{6}g^{MN}\partial_{M}S\partial_{N}S-\frac{M_{5}^{3}}{2}V(S)-\sum_{i={\rm T,P}}\delta(y-y_{i})\frac{1}{\sqrt{g_{55}}}\lambda_{i}(S)\,.
\eeq
Plugging this equation into eq.~\eqref{tmunu1} we get
\beq
T^{S}_{MN}= \displaystyle g_{MN}\mathcal{L}_{S}+\frac{M_{5}^{2}}{3}\partial_{M}S\partial_{N}S+\sum_{i={\rm T,P}}\delta(y-y_{i})\lambda_{i}(S)\sqrt{g_{55}}\delta_{M}^{5}\delta_{N}^{5}\,.
\eeq
The linearized Einstein equations are given by
\beq
\mathcal{R}_{MN}=\frac{1}{M_{5}^{3}}\widetilde{T}_{MN}\,,\qquad \widetilde{T}_{MN}=T_{MN}-\frac{1}{3}g_{MN}g^{PQ}T_{PQ}\,,
\eeq
while the equation of motion of the scalar field $S$ is given by
\beq
\frac{M_{5}^{3}}{3}\partial_{M}\left(\sqrt{-g}g^{MN}\partial_{N}S\right)=-\sqrt{-g}\frac{\partial \mathcal{L}_{S}}{\partial S}\,,
\eeq
where the $M_{5}^{3}/3$ factor is due to the non-canonical normalization of $S$ in the action \eqref{eq:Einstein}. Setting the scalar field to its background value $S=S_{0}(y)$ and plugging the metric ansatz \eqref{lineelement2} into the action \eqref{eq:Einstein} we get the linearlized field equations (respectively the $\mu\nu$ and $55$ Einstein equations and the scalar equation of motion)
\begin{subequations}\label{eq:BG}
\begin{align}
&  \displaystyle\sigma^{\prime\prime}+3(\sigma^{\prime})^2 =- \frac{e^{2\sigma}}{3}V(S_0)  - \sum_{i={\rm T}, {\rm P}}\frac{e^{\sigma}}{3M_5^3}\lambda_i(S_0)\delta(y - y_i)\,,\label{eq:BG1}\\
& \displaystyle\sigma^{\prime\prime} + \frac{1}{12}(S_0^{\prime})^2= -\frac{e^{2\sigma}}{12}V(S_0) 
  - \sum_{i={\rm T}, {\rm P}}\frac{e^{\sigma}}{3M_5^3}\lambda_i(S_0)\delta(y - y_i)\,,\label{eq:BG2}\\
& \displaystyle S_0^{\prime\prime}+3S_0^{\prime}\sigma^{\prime}  =  \frac{3}{2}e^{2\sigma}V^{\prime}(S_0)+\sum_{i={\rm T}, {\rm P}}\frac{3e^{\sigma}\lambda^{\prime}(S_0)}{M_5^{3}}\delta(y - y_i)\,.\label{eq:BG3}
\end{align}
\end{subequations}
In these equations the primes on the background fields indicate derivatives with respect to $y$, while the primes on the potentials indicate derivatives with respect to $S$. The argument of the potentials and their derivatives indicates that they are computed on the background $S_{0}$. Equations \eqref{eq:BG} can be written as a set of bulk differential equations
\begin{subequations}\label{eq:BGbulk}
\begin{align}
& \displaystyle \left[\sigma^{\prime\prime}+3(\sigma^{\prime})^2\right]\eta_{\mu\nu} =- \frac{e^{2\sigma}}{3}V(S_0)\eta_{\mu\nu}  \,,\label{eq:BG1bulk}\\
& \displaystyle \sigma^{\prime\prime} + \frac{1}{12}(S_0^{\prime})^2= -\frac{e^{2\sigma}}{12}V(S_0) \,,\label{eq:BG2bulk}\\
& \displaystyle S_0^{\prime\prime}+3S_0^{\prime}\sigma^{\prime} = \frac{3}{2}e^{2\sigma}V^{\prime}(S_0)\,,\label{eq:BG3bulk}
\end{align}
\end{subequations}
supplemented by initial conditions obtained by matching the jumps in the derivative of the functions with the coefficients of the $\delta$-functions (junction conditions)
\begin{subequations}\label{eq:BGjump}
\begin{align}
& \displaystyle \left[\sigma^{\prime} \right]_{i} = -\frac{e^{\sigma}\lambda_{i}(S_0(y_i))}{3M_5^3}~, \label{eq:BG1jump}\\
& \displaystyle \left[S_0^{\prime} \right]_{i} = \frac{3e^{\sigma}\lambda_{i}^{\prime}(S_0(y_i))}{M_5^3}\,,
~~~~i={\rm T}, {\rm P}\, .
\label{eq:BG2jump}
\end{align}
\end{subequations}
The jump in a function at the point $y_{i}$ is defined as $[f]_{i}=f(y_{i}+\epsilon)-f(y_{i}-\epsilon)$.
The solution to these equations can be found using the ``superpotential'' method~\cite{DeWolfe:1999cp}. Defining a functional $\mathcal{W}(S)$ such that
\be\label{superpotential}
V(S)=\frac{3}{4}\left(\mathcal{W}'(S)\right)^{2}-\frac{1}{3}\mathcal{W}^{2}(S)\,,
\ee
the solution to eqs.~\eqref{eq:BGbulk} is given by the solution to the system of coupled first order differential equations
\begin{subequations}\label{solutions}
\begin{align}
&\displaystyle \sigma'=-\frac{e^{\sigma}}{6}\,\mathcal{W}(S_{0})\,,\label{solutions1}\\
&\displaystyle S_{0}'=\frac{3\,e^{\sigma}}{2}\, \mathcal{W}'(S_{0})\,,\label{solutions2}
\end{align}
\end{subequations}
where the prime on $\mathcal{W}$ indicates a derivative with respect to $S$, complemented by the junction conditions
\begin{subequations}\label{solutionsjump}
\begin{align}
& \displaystyle \left[\mathcal{W}(S_{0}) \right]_{i} = \frac{2\lambda_{i}(S_0(y_i))}{M_5^3}\,,\label{solutionsjump1}\\
& \displaystyle \left[\mathcal{W}'(S_{0}) \right]_{i} = \frac{2\lambda_{i}^{\prime}(S_0(y_i))}{M_5^3}\,,
~~~~i={\rm T}, {\rm P}
\label{solutionsjump2}
\end{align}
\end{subequations}
which directly follow from eqs.~\eqref{eq:BGjump}.
For a discussion about the generality of this ``superpotential method'' in generating all the solutions to the initial set of differential equations the reader is referred to ref.~\cite{DeWolfe:1999cp}. Notice that the background solutions only depend on the value of the boundary potentials and their derivatives evaluated at the boundaries. These do not fix uniquely the interactions of the fluctuations of the dilaton around the background solution. The second derivative of the boundary potentials evaluated at the boundaries also becomes relevant for the stabilization of the size of the extra dimension as discussed in Appendix~\ref{app:stabilization}. Higher derivatives are only relevant for higher order interactions. As an example of the method, we report the values of the relevant quantities for RS and LD (focusing illustratively on the unstabilized case) with the choice $\sigma(y=y_{\rm T}=0)=0$ and $S_{0}(y=y_{\rm T}=0)=0$:
\begin{itemize}
\item {\bf Randall-Sundrum}
\be
\begin{array}{lll}
\displaystyle \mathcal{W}_{\text{RS}}(S)=-6k_{\text{RS}}\,,\qquad &\displaystyle V_{\text{RS}}(S)=-12k_{\text{RS}}^{2} \,,\qquad &\displaystyle \lambda_{i}(S)=\pm 6k_{\text{RS}} M_{5}^{3}\,,\vspace{2mm}\\
\displaystyle \sigma_{\text{RS}}(y)=-\log\left(\frac{1}{1-k_{\text{RS}}y}\right)\,,\qquad &\displaystyle S_{0\,\text{RS}}(y)=0\,,\vspace{2mm}\\
\displaystyle y_{T}=0\,,\qquad & \displaystyle y_{P}=\frac{1-e^{-\pi k_{\text{RS}} R_{\text{RS}}}}{k_{\text{RS}}}\,.
\end{array}
\ee
\item {\bf Linear Dilaton}
\be
\begin{array}{ll}
\displaystyle \mathcal{W}_{\text{LD}}(S)=-4ke^{-S/3}\,,\qquad &\hspace{-2cm}\displaystyle V_{\text{LD}}(S)=-4k^{2}e^{-2S/3}\,,\vspace{2mm}\\
\displaystyle  \lambda_{i}(S)= e^{-\frac{S}{3}}\, M_5^3\left[\pm 4k + \frac{\mu_{i}}{2}(S - S_{i})^2\right]\,,&\vspace{2mm}\\
\displaystyle \sigma(y)=\frac{2ky}{3}\,,\qquad &\hspace{-2cm}\displaystyle S_{0}(y)=3 \sigma(y)=2ky\,,
\label{superpotential-LD}
\end{array}
\ee
where the plus and minus signs in the boundary potentials correspond respectively to $y=y_{\rm P}=\pi R$ and $y=y_{\rm T}=0$.
\end{itemize}

\section{Radius stabilization and comparison with RS}\label{app:stabilization}

Having a mechanism able to guarantee the stability of the size of the extra dimension is a point of fundamental importance.
An extra dimension with a size $R$ that is not stabilized, { \it i.e.}\ not dynamically fixed to some 
specific value, is simply unacceptable because in the quantum theory 
there will be a scalar field -- the radion, describing 
fluctuations of the radius of the extra dimensions -- that  remains massless.
 A massless radion would contribute to the Newton potential in the form of a fifth force, and this is phenomenologically excluded.

The CW/LD scenario is defined by the presence of a scalar field with a non-trivial
profile along the bulk of the extra dimension. 
For this reason, 
it can reasonably be expected that the CW/LD setup is well-suited for a stabilization mechanism along the lines of 
 the Goldberger-Wise mechanism~\cite{Goldberger:1999uk} in RS.
 Let us sketch the generic picture. A non-trivial scalar profile along the extra dimension leads to
radion stabilization with a preferred value for $R$. This is because a non-trivial scalar profile 
 generates a kinetic energy in the fifth direction that would prefer (since it scales like $1/R$) an extra dimension as large as possible in order to be minimized. This is in contrast with a bulk potential energy term which (since it scales like $R$) would prefer the size of the extra dimension to be as small as possible. The balance between these two opposite effects selects a specific value of $R$. 
 Bearing in mind this general motivation, it is important to discuss the specific implementation of this idea in the CW/LD model in order to highlight the difference with respect to RS.

To this end, let us start considering the brane potential in eqs.~\eqref{eq:Smart1}--\eqref{eq:Smart2} without the inclusion of the $\mu_i$-dependent terms. 
At this level of the analysis the size of the extra dimension is not stabilized even if the dilaton features a non-trivial profile along the extra dimension.
It is true that one can write a relation of the type 
$S(\pi R) = 2\pi kR$ (using the freedom to choose $\sigma_0 = S_0 = 0$) but there is no specific dynamics selecting a preferred value of $R$ 
since $S(y)$ is not fixed at the branes.  

We now include the brane contributions $\delta\lambda_{i = {\rm T,P}}(S) \equiv e^{-S/3}M_5^3 \mu_i (S - S_i)^2/2$.
These terms do not affect Einstein's equations in the bulk, and only enter at the level of the junction conditions. However, before discussing this point, it is instructive to write down the 
brane potential $\delta\lambda_{i = {\rm T,P}}(S)$
 on the LD background. 
 It is straightforward to see that the following potential for $R$ is generated
 \begin{equation}
 V_{\rm eff}(R) = \frac{M_5^3}{2}e^{4\sigma_0 - S_0/3}\left[
 e^{2\pi kR e^{\sigma_0 - S_0/3}}
 \left(
 2\pi kRe^{\sigma_0 -S_0/3} + S_0 - S_{\rm P}
 \right)^2\mu_{\rm P} + (S_0 - S_{\rm T})^2\mu_{\rm T}
 \right]~.
 \end{equation}
 This potential has a minimum at 
 \begin{equation}\label{eq:Rmin}
 R = \frac{e^{\sigma_0 - S_0/3}}{2\pi k}(S_{\rm P} - S_0)~.
 \end{equation}
irrespective of the specific values of $\mu_{i = {\rm T,P}}$. 
This means that the addition of the brane contributions $\delta\lambda_{i = {\rm T,P}}(S)$ 
generates a potential that dynamically selects a specific value of $R$.

We can now come back to the role of the brane contributions $\delta\lambda_{i = {\rm T,P}}(S)$ 
 within Einstein's equations. The potential terms $\delta\lambda_{i = {\rm T,P}}(S)$ enter in the junction conditions and eq.~(\ref{eq:BG1jump}) -- evaluated both at the Planck and TeV brane -- implies
 \begin{equation}\label{eq:BoundaryP}
  S_0 = S_{\rm T}~,~~~~~~~~~ S(\pi R) = S_{\rm P}~.
 \end{equation}
 This means that the addition of the brane potential $\delta\lambda_{i = {\rm T,P}}(S)$ provides a boundary condition that fixes the value of $S(\pi R)$ to a given number, naturally expected to be of order one, as the specific form of $\delta\lambda_{i = {\rm T,P}}(S)$ is expected to be determined by physics at the cut-off scale. 
 
 All in all, by choosing $\sigma_0 = 0$, we find from eq.~(\ref{eq:Rmin}) the fundamental relation
 \begin{equation}\label{eq:StabTuning}
 2\pi k R = (S_{\rm P} - S_{\rm T}) e^{S_{\rm T}/3} \approx 63~,
 \end{equation} 
where the numerical value is determined by eq.~(\ref{kR}).
The stabilization of the size of the extra dimension 
is achieved in the sense that now $R$ is fixed in terms of 
the fundamental parameters of the model $S_{i = {\rm T,P}}$.  The latter are expected to be 
order-one numbers generated on the TeV and Planck branes by the unknown UV physics, and 
must satisfy the condition in eq.~(\ref{eq:StabTuning}) to correctly address the
 electroweak-Planck hierarchy in the CW/LD model.
 
Some important comments need to be made.

First of all, one may wonder about the specific functional form of the brane potential 
 in eqs.~(\ref{eq:Smart1})--(\ref{eq:Smart2}).
Let us consider for simplicity the TeV brane. 
What are the consequences for the stabilization mechanism, had we chosen a generic polynomial
potential
 $\lambda^{\rm pol}_{\rm T}(S) = V_0 + V_1(S - S_{\rm T}) + V_2(S - S_{\rm T})^2$?
To answer this question, we go back to the junction conditions
 in eqs.~(\ref{eq:BG1jump})--(\ref{eq:BG2jump}). We find
 \begin{eqnarray}
 -4kM_5^3e^{-S_0/3} &=& V_0 + V_1(S_0 - S_{\rm T}) + V_2(S_0 - S_{\rm T})^2~,\label{J1Pol}\\
 \frac{4kM_5^3}{3}e^{-S_0/3} &=& V_1 + 2V_2(S_0 - S_{\rm T})~.\label{J2Pol}
 \end{eqnarray}
 From eq.~(\ref{J2Pol}) we see that $V_1$ is not an independent parameter but is fixed to the value
$V_1 = 4kM_5^3 e^{-S_0/3}/3 - 2V_2(S_0 - S_{\rm T})$. 
Using this condition we find that eq.~(\ref{J1Pol}) is satisfied by 
\begin{equation}
V_0 = -4kM_5^3e^{-S_0/3}~,~~~~~~~~~S_0 = S_{\rm T}~.
\end{equation}
The first equality is the usual tuning of the brane tension needed to cancel the 4D cosmological constant. 
The second condition fixes the value of the background scalar field at the TeV brane. 
We see that the polynomial potential $\lambda^{\rm pol}_{\rm T}(S)$ does the same job as the brane term  $\lambda_{\rm T}(S)$ considered in eq.~(\ref{eq:Smart1}). 
The equivalence between the two expressions -- at least up to quadratic order, which is enough as far as stability is concerned -- can be formally obtained by considering the 
potential $\lambda_{\rm T}(S)$ on the TeV brane (that is with $S = S_0$), and expanding 
around $S=S_{\rm T}$. We find
\begin{equation}
\lambda_{\rm T}(S_0) = kM_5^3 e^{-S_{\rm T}/3} \left[ -4+
\frac{4}{3}(S_0 - S_{\rm T})  + \left(  \frac{\mu_{\rm T}}{2k}- \frac29 \right)
(S_0 - S_{\rm T})^2\right] ~,
\end{equation}
which coincides with the polynomial form $\lambda^{\rm pol}_{\rm T}(S)$  after including the 
constraint on $V_1$ and a proper redefinition of $V_2$.

Second, a stabilized extra dimension corresponds to a massive radion. 
This is indeed what happens in the CW/LD model in the presence of the stabilizing potential in 
eqs.~(\ref{eq:Smart1})--(\ref{eq:Smart2}).
In this respect, it is worth emphasizing 
that the role of $\delta\lambda_{i = {\rm T,P}}(S)$ is twofold.
Besides stabilizing the extra dimension and giving a mass to the radion, 
  $\delta\lambda_{i = {\rm T,P}}(S)$ breaks the shift symmetry 
  of  $S$ under which 
 the action is rescaled by an overall factor, and, as a consequence, the dilaton becomes massive.
In order to compute the mass of the radion and the full spectrum of the scalar sector it is necessary to include quantum fluctuations. We postpone this computation to appendix~\ref{app:ScalarSpectrum}.

Finally, let us comment about the difference with respect to the stabilization in RS.
 The most important difference is of course structural. In RS, the metric is generated by the bulk cosmological 
 constant and the Goldberger-Wise scalar $\Phi_{\rm GW}$ is added to this background.  
On the contrary, in CW/LD the same bulk scalar field that sources the background geometry also stabilizes the extra dimension.
In RS, working in the coordinate system where $g_{55}=1$, the
stabilization condition reduces to
\begin{equation}\label{eq:StabTuningRS}
k_{\rm RS}R_{\rm RS} = \frac{k_{\rm RS}}{u}\ln\frac{\Phi_{\rm GW,P}}{\Phi_{\rm GW,T}}\approx 30~,
\end{equation}
where $k_{\rm RS}$ controls the RS warp factor, $u$ is a bulk mass term for $\Phi_{\rm GW}$, and
$\Phi_{\rm GW,T/P}$ are the values of the Goldberger-Wise scalar fixed at the TeV and Planck branes. 
 Assuming a logarithm of order one, the correct hierarchy can be achieved with $u/k_{\rm RS} \approx 1/30$.

Neither eq.~(\ref{eq:StabTuning}) nor eq.~(\ref{eq:StabTuningRS}) require large hierarchies in the 
input parameters. However, we stress that these two formulas are crucially different. 
In the CW/LD model there is no analogue for the ratio of the two independent parameters $k_{\rm RS}/u$ since 
the bulk potential of the dilaton field sources the background geometry, and the 
 correct hierarchy is entirely obtained by means of the relation between the fixed values of the scalar 
  field $S$ at the two branes.  
  This is in net contrast with eq.~(\ref{eq:StabTuningRS}) where the log-dependence on the ratio 
   $\Phi_{\rm GW,P}/\Phi_{\rm GW,T}$ makes the stabilization condition quite insensitive 
   to the specific values of the Goldberger-Wise field at the two branes.
 
\section{Graviton decays to SM particle pairs}
\label{app:grav-SM-decays}

The partial decay widths of a KK graviton of mass $m_n$ are (see {\it e.g.}~ref.~\cite{Das:2016pbk})
\begin{align}
&\Gamma_{\gamma\gamma} = \Gamma_0 \,,\qquad
\Gamma_{gg} = 8\,\Gamma_0 \,,\\
&\Gamma_{ZZ} = \frac12\left(\frac{13}{6} + \frac{7}{3}x_Z + \frac{1}{2}x_Z^2\right)\left(1-x_Z\right)^{1/2}\Gamma_0 \,,\\
&\Gamma_{W^+W^-} = \left(\frac{13}{6} + \frac{7}{3}x_W + \frac{1}{2} x_W^2\right)\left(1-x_W\right)^{1/2}\Gamma_0 \,,\\
&\Gamma_{hh} = \frac{1}{12}\left(1-x_h\right)^{5/2}\Gamma_0 \,,\\
&\Gamma_{f\bar f} = \frac{N_f}{2}\left(1+\frac23 x_f\right)\left(1-x_f\right)^{3/2}\Gamma_0 \,,
\end{align}
where
\beq
\Gamma_0 \equiv \frac{m_n^3}{80\pi \Lambda_G^{(n)2}} \,,\qquad
x_i \equiv \frac{4m_i^2}{m_n^2} \,,
\eeq
and $N_f = 3$ for each quark, $1$ for each charged lepton and $1/2$ for each (Majorana) neutrino. In our model, $\Lambda_G^{(n)}$ is given by eq.~\eqref{KKgraviton-couplings}.

\section{Graviton decays to gravitons}
\label{app:grav-grav-decays}
For a metric of the form 
\beq
ds^2 = 
e^{2{ \sigma}(y)}\left[ \tilde{g}_{\mu\nu} (x) \, dx^\mu dx^\nu + dy^2\right]
\label{eq:metric}
\eeq
the resulting 5D gravitational action is
\begin{eqnarray}
&&{\mathcal S} = \int d^4x \, dy \, e^{3{ \sigma}(y)} \sqrt{-\tilde{g}} \, \frac{M_5^3}{2} \left[ \mathcal{R}_4 (\tilde{g}) +\frac{1}{4} \left((\tilde{g}^{\mu\nu} \partial_y \tilde{g}_{\mu\nu})^2 +(\partial_y \tilde{g}^{\mu\nu}) (\partial_y \tilde{g}_{\mu\nu}) \right ) 
 \right] \, .
 \label{eq:vertices}
\nonumber
\end{eqnarray}
Terms without $y$-derivatives or with single $y$-derivatives in the bulk action are precisely cancelled by the boundary potential.  To determine the graviton trilinear interactions we make the usual expansion
\beq
\tilde{g}_{\mu\nu} =  \eta_{\mu\nu}+\frac{2}{M_5^{3/2}} \, h_{\mu\nu}  ~~~~,~~~~ \tilde{g}^{\mu\nu} =  \eta^{\mu\nu}-\frac{2}{M_5^{3/2}}\,  h^{\mu\nu}+\frac{4}{M_5^{3}} \, h^{\mu\lambda} h_{\lambda}^{\nu} +{\mathcal O}(h^3) ~~.
\eeq
The trilinear vertex and corresponding Feynman rule arising from the 4D part $\mathcal{R}_4 (\tilde{g})$ was calculated long ago~\cite{DeWitt:1967yk,DeWitt:1967ub,DeWitt:1975ys}.  To understand the $y$-derivative piece let us make the usual expansion in KK modes
\beq
h_{\mu \nu} (x,y) =\sum_{n=0}^{\infty} \frac{{\tilde h}^{(n)}_{\mu \nu}(x)\, \psi_n(y)}{\sqrt{ \pi R}} ~ ,
\eeq
and consider the decays $n\to m+l$.  The wavefunctions $\psi_n(y)$ are given in \cite{Giudice:2016yja}.  We now see that the $y$-derivative vertex will simply be of the form ${\tilde h}^{(n)\mu \nu} {\tilde h}^{(m)}_{\nu \kappa} {\tilde h}^{(l)\kappa}_{\mu}$, with a coefficient $\psi_n(y) \partial_y \psi_m(y) \partial_y \psi_l(y)$, permutations implied.

In the scattering amplitude we will encounter the integral over the wavefunctions
\begin{eqnarray}
I_{nml} & = & 2 \int^{\pi R}_0 dy \, e^{2 k y} \psi_n(y) \psi_m(y) \psi_l(y) \nonumber\\
& = & \frac{32}{R^3} \frac{k (n m_n) (m m_m) (l m_l) }{m^2_{l+m-n} m^2_{l-m+n} m^2_{m+n-l} m^2_{l+m+n}} ~~,
\label{eq:integral}
\end{eqnarray}
where $m_j = \sqrt{k^2 + j^2/R^2}$ and we have discarded an additional factor of $\left( 1-(-1)^{n+m+l} e^{-k \pi R} \right)$ which is essentially equal to one for our purposes.  This integral is symmetric under permutations of the labels.  When averaged over permutations, the second integral of interest is trivially related to the first
\begin{eqnarray}
I''_{nml} & = &\frac{1}{6} \sum_{\text{perms}} 2 \int^{\pi R}_0 dy\, e^{2 k y} \psi_n(y) \psi'_m(y) \psi'_l(y) \nonumber\\
& = & \frac{1}{6} (m_n^2 + m_m^2 + m_l^2) I_{nml} ~~.
\end{eqnarray}
A massless 5D (massive 4D) graviton has 5 physical polarizations, thus in principle there are 125 physically distinct helicity amplitudes.  However, these amplitudes simplify greatly and there are in fact only 4 distinctive analytic forms for the amplitudes.  When squared and integrated over the azimuthal angle\footnote{The sum of all squared amplitudes is isotropic, however individual amplitudes are not.} these amplitudes are expressed analytically through the four following functions:
\begin{align}
A^2_1 (x,y) &= \frac{2 [x (1+y^2)-2]^2 }{3 (x^2-y^2)}  \,,\\
A^2_2 (x,y) &= \frac{2 [y^2 -2 y (1+x)+3]^2 }{x-y} \,, \\
A^2_3 (x,y) &=  \frac{8 [2 x^2 + 2 x y + (1+y)^2]^2 }{3 (x+y)^2} \,, \\
A^2_4 (x,y) &= \left[ \frac{6 - 4 x + x^2 + 2 x^3 +y^4 +y^2 x (5 x-6)-5y^2 }{9 (x^2-y^2)} \right]^2 .
\end{align}
The last amplitude corresponds to the purely longitudinal polarizations $LL \to LL + LL$.  The final squared amplitude may then be expressed as
\beq
A^2_T (n,m,l) =  \frac{I^2_{nml}}{5(M_5 \pi R)^3} \sum_{\text{cyc}} m_n^4 \sum_j A^2_j \left(\frac{m_l^2+m_m^2}{m_n^2},\frac{m_l^2-m_m^2}{m_n^2}  \right) ,
\eeq
where the factor $1/5$ is due to the average over initial polarizations and the sum is over the three cyclic permutations.  The decay width is
\beq
\Gamma_{n\to m+l} = \frac{|\bold{p}_m|}{16 \pi m_n^2} \, A^2_T (n,m,l)  ~~.
\eeq
This expression has some interesting limits.  In particular, when $m^2_{n-l-m} \sim k^2$, the width is enhanced, as can be seen in figure~\ref{KKG2KKG_PS}.  In that region of phase space, namely for $\kappa \equiv n-m-l \lesssim kR$, the width is given by
\beq
\widetilde{\Gamma}_{n\to m+l} \simeq \frac{7 k^2}{3 \sqrt{2} M_5^3 \pi^4 R^2 } \frac{(n^2- m n +m^2) \sqrt{m n \kappa (n-m)}}{(k^2 R^2 + \kappa^2)^2} ~~.
\label{eq:approx}
\eeq
In this limit, the width scales as $k^{-2}$, thus the greatest partial decay widths are to the nearest modes as they have a parametric enhancement of $m_n^2/k^2$.  This can be understood already from the wavefunctions, and the integral in eq.~\eqref{eq:integral}.  In the limit $k\to 0$ we see that this integral becomes proportional to $\delta(n-m-l)$, hence in the limit of small $k$ this integral still greatly prefers the modes to be as nearby as possible.  This preference is enough to overcome the suppression due to phase space.

Armed with the knowledge that the nearby modes dominate the partial widths, we may now estimate the total decay width into lighter gravitons.  The width in eq.~\eqref{eq:approx} is dominated by $\kappa < kR$, and vanishes rapidly as $\kappa > k R$, thus we may approximate the summation over all $\kappa$ as an integral over $0<\kappa<\infty$.\footnote{Formally we should limit $\kappa < n$, however for large $n$, when $\kappa\sim n$ the width is already well approximated by $\kappa \to \infty$.  Thus this integral is a good approximation.}  We may also approximate the summation over all $m$ as an integral over all allowed modes $0 < m < n$.  Thus, avoiding double counting, we have
\begin{equation}\label{eq:TTTsum}
\Gamma_{G_n \to \sum G_lG_m}  \simeq  \frac{1}{2} \int^{\infty}_0 d \kappa \int^n_0 dm\, \widetilde{\Gamma}_{n\to m+l} 
 =  \frac{595}{3 \times 2^{14}} \frac{m_n^3}{M_5^3 \pi^2 R} \sqrt{\frac{m_n}{k}} ~~,
\end{equation}
which approximates the summation over modes of the full expression.

\section{Dilaton/radion spectrum and couplings with SM fields}
\label{app:ScalarSpectrum}

To isolate the scalar fluctuations we expand the scalar field as $S(x,y) = S_0(y) + \varphi(x,y)$ and the metric as $g_{MN}(x,y) = e^{2 \sigma(y)} \left[\eta_{MN} + h_{MN} (x,y) \right]$, where $S_0 =2k|y|$ and $\sigma =2k|y|/3$ are the background solutions in eq.~(\ref{eq:ScalarBackground}).
For the sake of clarity in this appendix we add the subscript $_0$ to the scalar field background solution.
The metric fluctuations are expanded in terms of 4D Lorentz representations as a tensor, vector, and scalar
\beq
h_{\mu\nu} (x,y)  ~~~,~~~  h_{5\nu} (x,y) = \epsilon_\nu (x,y) ~~~,~~~ h_{55} (x,y) = 2 G(x,y) ~~~.
\eeq
Thus we see that we have two separate 5D scalar fields given by $\varphi(x,y)$ and $G(x,y)$.  This means that in 4D there will be \emph{two} infinite towers of scalar perturbations.  However, one of these towers of KK scalars is eaten to become the longitudinal components of the infinite tower of massive spin-2 fields, along with the vectors $\epsilon_\nu (x,y)$, giving five degrees of freedom for each massive graviton. It thus remains to identify the uneaten scalar degrees of freedom.  We may achieve this by inspecting the Ricci scalar.

There is kinetic mixing between the tensors and scalars, with the 4D component given by
\beq
\mathcal{R}(g) \supset 2 G \left( \partial^\mu \partial^\nu h_{\mu\nu} - \partial^2 h_{\mu}^{\mu}  \right) ~~,
\eeq
which will in general lead to a tensor-scalar kinetic mixing in the 4D action.  To eliminate this mixing we may perform a Weyl rescaling of the 4D part of the metric, corresponding to
\beq
h_{\mu\nu} \to h_{\mu\nu} - G \eta_{\mu\nu}  ~~.
\eeq
The metric fluctuations now become
\beq\label{eq:linmix}
h_{\mu\nu} (x,y) -G(x,y) \eta_{\mu\nu}~,~~~  h_{5\nu} (x,y) = \epsilon_\nu (x,y)~,~~~ h_{55} (x,y) = 2 G(x,y)~,
\eeq
and we can see that after eliminating the tensor-scalar mixing the scalar fluctuations of the full 5D metric are traceless.

More important for the identification of the physical scalar fluctuations are the kinetic mixing terms between the vector and the scalar perturbations.  Expanding about the background these terms are
\begin{eqnarray}
\sqrt{-g} \left( \mathcal{R}(g) -\frac{1}{3} g_{MN} \partial^M S \partial^N S \right)& \to & 2 e^{3 \sigma(y)}  \bigg[ \epsilon_\mu \partial^\mu \left( -\frac{1}{2} G' +\frac{1}{3} S_0' \varphi' \right) + \partial^\mu G \epsilon_\mu' \bigg] \nonumber \\
& \to & -2 e^{3 \sigma(y)} \epsilon_\mu \partial^\mu \bigg[ \frac{3}{2} G' +3 \sigma' G -\frac{1}{3} S_0' \varphi'  \bigg] ~~, 
\end{eqnarray}
where in the last term we integrated by parts.  Thus, the scalar fluctuations satisfying the constraint equation
\beq
\frac{3}{2} ( G' + 2 \sigma' G ) = \frac{1}{3} S_0' \varphi'
\label{eq:const}
\eeq
do not mix with the vector $\epsilon_\mu$ and are therefore the physical scalar fluctuations that are \emph{not} eaten by the massive graviton modes.   One may be concerned that we may have omitted an important term in performing integration by parts, however the vector field carries odd parity under the orbifold symmetry, thus the boundary term vanishes.

If the previous constraint is employed we may continue without consideration of the massive graviton degrees of freedom as we have isolated the scalar fluctuations that are not kinetically mixed with them.  To proceed we will consider the following general metric
\begin{eqnarray}\label{eq:MainFluctuations}
ds^2 &=& e^{2\sigma(y)}
\left[
\left(
1+ 2F(x,y)\right)
\eta_{\mu\nu}dx^{\mu}dx^{\nu} + \left(
1 + 2G(x,y)
\right)dy^2
\right]~,\\
S(x,y) &=& S_0(y) + \varphi(x,y)~,
\end{eqnarray}
with $\sigma(y) = 2ky/3$.  We consider the linearized (\emph{i.e.}\ at the first order in the scalar fluctuations) Einstein's equations, which we recast in the form  
$\mathcal{R}_{MN} = \tilde{T}_{MN}/M_5^3 + B_{MN}$, where
$\tilde{T}_{MN} \equiv T_{MN} - \frac{1}{3}g_{MN}T_{PQ}g^{PQ}$ and $B_{MN}$ accounts for boundary contributions at the two branes.
As customary,
some equations are dynamical, some equations provide constraints. 
In particular, the off-diagonal $\mu\nu$ components enforce the condition
\beq
-2F(x,y) = G(x,y) \,,
\eeq
as found in eq.~(\ref{eq:linmix}), while 
the $\mu5$ component can be straightforwardly integrated, and we find eq.~(\ref{eq:const}).
As for the rest of the Einstein's equation, we find
\begin{eqnarray}
\delta \mathcal{R}_{\mu\nu} &=& -\eta_{\mu\nu}\left[
\Box F + F^{\prime\prime} +9\sigma^{\prime}F^{\prime} + 6F\left(
3(\sigma^{\prime})^2 + \sigma^{\prime\prime}
\right)
\right]~,\\
\delta \mathcal{R}_{55} &=& 2\left(
\Box F - 2F^{\prime\prime}  - 6\sigma^{\prime}F^{\prime}
\right)~,
\end{eqnarray}
\begin{eqnarray}
\delta\tilde{T}_{\mu\nu}/M_5^3 &=& \frac{e^{2\sigma}}{3}\left[
2FV(S_0) + \varphi V^{\prime}(S_0)
\right]\eta_{\mu\nu}~,\\
\delta\tilde{T}_{55}/M_5^3 &=& \frac{2}{3}S_0^{\prime}\varphi^{\prime} + \frac{e^{2\sigma}}{3}
\left[
2GV(S_0) + \varphi V^{\prime}(S_0)
\right]~,
\end{eqnarray}
\begin{eqnarray}
\delta B_{\mu\nu}^{i={\rm T}, {\rm P}} &=&   \frac{e^{\sigma}}{9M_5^3}\left[
\left(
6F - 3G
\right)\lambda_{i={\rm T}, {\rm P}}(S_0) + 3\varphi \lambda^{\prime}_{i={\rm T}, {\rm P}}(S_0)
\right]\delta(y - y_{i={\rm T}, {\rm P}})~,\\
\delta B_{55}^{i={\rm T}, {\rm P}} &=& \frac{4e^{\sigma}}{3M_5^3}\left[
G\lambda_{i={\rm T}, {\rm P}}(S_0) + \varphi \lambda^{\prime}_{i={\rm T}, {\rm P}}(S_0)
\right]\delta(y - y_{i={\rm T}, {\rm P}})~.
\end{eqnarray}
In addition, we need the equation of motion for the fluctuations of the dilaton field. We find
\begin{eqnarray}\label{eq:EOMvarphi}
&&\left[
\Box\varphi + \varphi^{\prime\prime}  + 3\sigma^{\prime}\varphi^{\prime} + 6S_0^{\prime}F^{\prime} + 4F\left(
3\sigma^{\prime}S_0^{\prime} + S_0^{\prime\prime}\right)  - \frac{3}{2}V^{\prime\prime}(S_0)\varphi e^{2\sigma}
\right]e^{-2\sigma} = \nonumber \\
&& \sum_{i={\rm T}, {\rm P}} \frac{3}{e^{\sigma}M_5^3}\left[
2F\lambda_{i}^{\prime}(S_0) + \varphi \lambda^{\prime\prime}_i(S_0)
\right]\delta(y - y_i)~.
\end{eqnarray}
Finally, we need to impose junction conditions at the boundaries. There is only one relevant junction condition, corresponding to the boundary term of eq.~(\ref{eq:EOMvarphi}).
We find 
\begin{equation}\label{eq:ScalarJump}
\left[\varphi^{\prime} \right]_{i={\rm T}, {\rm P}}  + 2F\left[S_0^{\prime} \right]_{i={\rm T}, {\rm P}} = \frac{3\varphi\lambda_{i={\rm T}, {\rm P}}^{\prime\prime}(S_0(y_i))
e^{\sigma}}{M_5^3}~,
\end{equation}
where, for a generic first derivative $A^{\prime}(y)$, the jump $\left[A^{\prime} \right]$ at the generic point $y_* $ is defined as $\left[A^{\prime} \right]_{*} \equiv A^{\prime}(y_* + \epsilon) - A^{\prime}(y_* - \epsilon)$. Using this definition, we can rewrite eq.~(\ref{eq:ScalarJump}) in the form
\begin{eqnarray}
\varphi^{\prime} + 2FS_0^{\prime} &=& 
\frac{3\varphi \lambda_{\rm T}^{\prime\prime}(S_0(y_{\rm T}))e^{\sigma}}{2M_5^3}~,~~~~~~~~{\rm at}~y = y_{\rm T}~,\label{eq:BC1}\\
\varphi^{\prime} + 2FS_0^{\prime} &=& -
\frac{3\varphi \lambda_{\rm P}^{\prime\prime}(S_0(y_{\rm P}))e^{\sigma}}{2M_5^3}~,~~~~~~{\rm at}~y = y_{\rm P}~.\label{eq:BC2}
\end{eqnarray}
We can now extract a dynamical equation for $F(x,y)$. Following ref.~\cite{Csaki:2000zn}, 
we consider {\it in the bulk} the combination 
\begin{equation}
\delta\mathcal{R}_{55} - \frac{\eta^{\mu\nu}}{4}\delta\mathcal{R}_{\mu\nu} = \frac{1}{M_5^3}
\left(
\delta\tilde{T}_{55} -  \frac{\eta^{\mu\nu}}{4}\delta\tilde{T}_{\mu\nu}
\right)~,
\end{equation}
and we find
\begin{equation}
3\Box F -3F^{\prime\prime} - 3\sigma^{\prime}F^{\prime} + 6F\left[
3(\sigma^{\prime})^2 + \sigma^{\prime\prime}
\right] =  \frac{2}{3}S_0^{\prime}\varphi^{\prime} - 2F\left\{
-6\left[
(\sigma^{\prime})^2 + \sigma^{\prime\prime}
\right] - \frac{1}{3}(S_0^{\prime})^2
\right\}~.
\end{equation}
Using the constraint in eq.~(\ref{eq:const}), and the explicit expressions for the background solution, we find
\begin{equation}
\Box F + F^{\prime\prime} + 2kF^{\prime} = 0~.
\end{equation}
This equation can be conveniently rewritten in the form
\begin{equation}\label{eq:EOM}
\left(
\Box + \frac{d^2}{dy^2} - k^2
\right)e^{ky}F(x,y) = 0~.
\end{equation}
Because of the relation $-2F(x,y) = G(x,y)$, the 5D field $G(x,y)$ satisfies the same equation of motion.
Similarly, starting from the eq.~(\ref{eq:EOMvarphi}) in the bulk, and using 
the constraint in eq.~(\ref{eq:const}), it is possible to show that also the scalar field fluctuation $\varphi(x,y)$ respects the same 
equation $\Box \varphi + \varphi^{\prime\prime} + 2k\varphi^{\prime} = 0$.
However, only a certain combination of these fields has a canonical kinetic term in the bulk. 
By direct diagonalization of the quadratic action, as 
illustrated in ref.~\cite{Kofman:2004tk}, one finds it to be
$v(x,y) \equiv  \sqrt{6}e^{ky}M_5^{3/2}[
F(x,y) - \varphi(x,y)/3]$,
and it also satisfies the 5D equation of motion $\Box v + v^{\prime\prime} + 2kv^{\prime} = 0$.
This is the equation we have to solve together with the appropriate boundary conditions.
To this end, we need an explicit expression for the brane potential.
The latter can be obtained by means of the solution-generating method championed 
in ref.~\cite{DeWolfe:1999cp}, and we find the explicit expressions
reported in eqs.~\eqref{eq:Smart1}--\eqref{eq:Smart2}.
 The mass parameters $\mu_{\rm T/P}$ do not affect the background solution (since they do not contribute to the background junction conditions) but they enter in the junction conditions 
 for the perturbations in eq.~(\ref{eq:ScalarJump}).
Notice the presence of a massless dilaton in the limit $\mu_{\rm T/P}\to 0$ since 
in this limit a shift in $S$ (in the Jordan frame) corresponds to an overall rescaling of the action.
We now have all the ingredients needed to solve the scalar equation of motion.
Using $-2F = G$, $F = \Psi$, $G = \Phi$, we adopt the notation of ref.~\cite{Cox:2012ee}. 
We introduce the KK decompositions 
\begin{equation}\label{eq:KKScalarDecomposition}
\Phi(x, y) = \sum_{n = 0}^{\infty}\Phi_n(y)\phi_n(x)~,~~~\varphi(x, y) = \sum_{n = 0}^{\infty}\varphi_n(y)\phi_n(x)~,
~~~v(x, y) = \sum_{n = 0}^{\infty}v_n(y)\phi_n(x)~.
\end{equation}
Notice that the 4D part of the three KK decompositions is equal. This is 
a direct consequence of the definition of $v(x,y)$ and the constraint in eq.~(\ref{eq:const}).
We define $\phi_n$ to be the solution of the equation of motion of a free 4D scalar field of mass $m_n$, namely $\Box \phi_n(x) - m_n^2\phi_n(x) = 0$.
At this level of the analysis we are interested in the mass spectrum $m_n$.
We can therefore solve the equation of motion focusing on a specific mode -- equivalently, $\Phi_n(y)$, $\varphi_n(y)$ or $v_n(y)$ -- and imposing the corresponding boundary conditions.
We consider the mode $\Phi_n$,  
for which the equation of motion is
\begin{equation}
\left(
\frac{d^2}{dy^2} +m_n^2 - k^2
\right)\left(
e^{ky}\Phi_n
\right) = 0~.
\end{equation}
We can write a generic solution in the form
\begin{equation}\label{eq:ScalarSolution}
\Phi_n(y) = N_n e^{-ky}\left[
\sin(\beta_n y) + w_n\cos(\beta_n y)
\right]~,
\end{equation}
with $\beta_n^2 = m_n^2 - k^2$, and $N_n$, $w_n$ constants.
The boundary conditions are given by eqs.~\eqref{eq:BC1}--\eqref{eq:BC2}.
Using the constraint in eq.~(\ref{eq:const}), we find
\begin{eqnarray}
\frac{9}{4k}\Phi_n^{\prime\prime} + 3\Phi_n^{\prime} - 2k\Phi_n
&=&  \frac{-4k + 9\mu_{\rm T}}{2} \left(
\Phi_n + \frac{3}{4k}\Phi_n^{\prime}
\right)~~~~~~{\rm at}~y = y_{\rm T}~,\\
\frac{9}{4k}\Phi_n^{\prime\prime} + 3\Phi_n^{\prime} - 2k\Phi_n
&=& 
-\frac{4k + 9\mu_{\rm P}}{2}
\left(
\Phi_n + \frac{3}{4k}\Phi_n^{\prime}
\right)~~~~~~{\rm at}~y = y_{\rm P}~.
\end{eqnarray}
We start considering the rigid limit $\mu_i \to \infty$.
The boundary condition at the TeV brane fixes $w_n = -3\beta_n/k$.
The boundary condition at the Planck brane can be solved analytically. 
We find
\begin{equation}\label{eq:ScalarSpectrum}
m_{\rm rad}^2 = \frac{8k^2}{9}~,~~~~~~~~~~~~m_{n}^2 = k^2 + \frac{n^2}{R^2}~,
\end{equation}
with $n = 1,2,3,\dots$. 
The lowest state in the KK tower corresponds to the radion. Notice that the mass of the radion vanishes in the absence of warping, $k = 0$.

The identification of the radion with the lowest state in the KK tower can be understood by means of the following argument. In the rigid limit $\mu_i \to \infty$,
if $k = 0$ we have from eq.~(\ref{eq:ScalarSpectrum}) a massless state and an ordinary KK tower with spectrum $m_n^2 = n^2/R^2$, $n = 1,2,3,\dots$. Note that there is no zero mode in the KK tower.
 The point is that in this setup it is straightforward to identify the states $m_n^2$ 
 with the KK decomposition of the scalar $S$. 
 Indeed, in this situation we just have a free scalar field in the bulk of a flat extra dimension (since $k=0$) with an infinite mass term on the two branes.
 After KK reduction, the equation of motion for the $y$-dependent variable reduces to an ordinary
 Schr\"odinger equation for a free particle in a one-dimensional well $0\leqslant y \leqslant \pi R$
 with infinite potential on the two walls. As well known from Quantum Mechanics, 
 the quantization condition precisely implies $m_n^2 = n^2/R^2$, $n = 1,2,3,\dots$, thus 
 making possible the identification of these states with the $k=0$ limit of 
 $m_{n}^2$
 in eq.~(\ref{eq:ScalarSpectrum}).
 Consequently, the remaining scalar degree of freedom can be identified with the radion field.
The second state in the KK tower corresponds to the dilaton, with mass $m_{\rm dil}^2 = k^2 + 1/R^2$.

For a generic value of $\mu_{\rm T,P}$, the boundary condition at the TeV brane gives 
\begin{equation}
w_n = -\frac{3\beta_n\mu_{\rm T}}{2(k^2 + \beta_n^2) + k\mu_{\rm T}} \,.
\end{equation} 
The boundary condition at the Planck brane can be solved numerically.
We show our result in figure~\ref{scalars_vs_BC}, with $\epsilon_{\rm T,P} \equiv 2k/\mu_{\rm T,P}$, and 
taking for simplicity $\epsilon_{\rm T} = \epsilon_{\rm P} \equiv \epsilon$.

At the first order in $\epsilon_{\rm T}$ -- and for generic value of $n$ -- we find
\begin{eqnarray}
m_{\rm rad}^2 &=& \frac{8k^2}{9}\left\{
1 - \frac{1}{9}
\left[
\epsilon_{\rm P} - \epsilon_{\rm T}-\left(
\epsilon_{\rm P} + \epsilon_{\rm T}
\right)\coth\frac{\pi kR}{3}
\right]
\right\}\simeq 
\frac{8k^2}{9}\left(
1- \frac{2\epsilon_{\rm T}}{9}
\right)
~,\label{eq:MassRadion}\\
m_{n}^2 &=&
 k^2 + \frac{n^2}{R^2}
\left[
1 - \frac{6\left(
n^2 + k^2 R^2
\right)(\epsilon_{\rm T} + \epsilon_{\rm P})}{9n^2\pi kR + \pi k^3 R^3}
\right]~.\label{eq:MassKK}
\end{eqnarray}

We now move to discuss the couplings of the scalar perturbations with SM fields.
To that end, we need to determine the proper normalization factor in eq.~(\ref{eq:ScalarSolution}).
By expanding the 5D action up to the quadratic order in the scalar fluctuations, 
and using the KK decompositions in eq.~(\ref{eq:KKScalarDecomposition}), 
we find the following normalization condition in 4D
\begin{eqnarray}\label{eq:ScalarNormalization}
S_n^{\rm 4D\,kin} &=& \mathcal{C}_n \int d^4x \phi_n(x)\left[
\Box - m_n^2
\right]\phi_n(x)~,\\
\mathcal{C}_n &=& \frac{27M_5^3}{16 k^2}\int_0^{\pi R}
dy e^{2ky}\left[
\left(\Phi_n^{\prime}\right)^2  +\frac{8 k^2}{3}\Phi_n^2
 +  \frac{8k}{3}\Phi_n^{\prime}\Phi_n \right] \equiv \frac{1}{2}~.
\end{eqnarray}
We can now extract the value of $N_n$ by plugging in $\mathcal{C}_n$ 
the solution found in eq.~(\ref{eq:ScalarSolution}). We find
\begin{eqnarray}\label{eq:Normalization}
N_n^{-2} &=& \frac{18 M_5^3}{16 k^2\left(2k^2 + 2\beta_n^2 + k\mu_{\rm T} \right)^2}\times \\
&&\left\{
2\left[2k\left(
k^2 + \beta_n^2
\right)^2 + \left(
-7k^4 + 2k^2\beta_n^2 + 9 \beta_n^4
\right)\mu_{\rm T} - 4k^3\mu_{\rm T}^2
\right]\sin^2\pi R\beta_n
\right.\nonumber\\
&&+\frac{3}{4\beta_n}\left[
4k^2\left(
k^2 + \beta_n^2
\right)^2 + 4k\left(
k^2 + \beta_n^2
\right)^2\mu_{\rm T} + \left(k^4 + 2k^2\beta_n^2 + 9\beta_n^4\right)\mu_{\rm T}^2
\right]\times\nonumber \\ &&\left(
2\pi R\beta_n - \sin2\pi R\beta_n
\right)+ \left.
3\beta_n\left[
k^4 + \beta_n^4 + 2k^2\left(
\beta_n^2 + \mu_{\rm T}^2
\right)
\right]\left(
2\pi R\beta_n + \sin2\pi R\beta_n
\right)
\right\}~.\nonumber
\end{eqnarray}
The coupling of the radion and the scalar KK modes with the SM Lagrangian at the TeV brane are described by the action
\begin{equation}
S_{\rm int}  = \int d^4x \sqrt{-\gamma_{\rm T}}e^{-\varphi/3}\mathcal{L}_{\rm SM}~.
\end{equation}
Expanding at the first order in the scalar fluctuation, and using the definition of the energy-momentum tensor (here for a generic metric $g$)
\begin{equation}
T^{\mu\nu} = \frac{2}{\sqrt{-g}}\frac{\delta \mathcal{S}}{\delta g_{\mu\nu}}~~~\Longrightarrow~~~\delta\mathcal{S} = \frac{1}{2}\int d^4x\sqrt{-g}T^{\mu\nu}\delta g_{\mu\nu}~,
\end{equation}
we find
\begin{equation}
S_{\rm int}^{\rm lin} = -\frac{1}{2}\sum_n \Phi_n(0)\int d^4 x \phi_n(x) {\rm Tr}[T]
-\frac{1}{3}\sum_n \varphi_n(0)
\int d^4 x  \phi_n(x) \mathcal{L}_{\rm SM}~.
\end{equation}
Defining the coupling constants
\begin{equation}\label{eq:ScalarCouplings}
\frac{1}{\Lambda_{\Phi}^{(n)}} \equiv 
\frac{\Phi_n(0)}{2}~,~~~~~~~~~
\frac{1}{\Lambda_{\varphi}^{(n)}}\equiv
\frac{\varphi_n(0)}{3}
\end{equation}
we find -- at the first order in $\epsilon_{\rm T/P}$, and for generic $n$ -- the following couplings with the trace of the energy-momentum tensor
\begin{eqnarray} 
\frac{1}{\Lambda_{\Phi}^{(0)}}
 &=& \frac{1}{6}\sqrt{\frac{k\left(1 + \coth\frac{k\pi R}{3}\right)}{M_5^3}}\left(
1 + \frac{4\epsilon_{\rm T}}{9}
\right)\simeq 
\frac{1}{6}\sqrt{\frac{k}{M_5^3}}\left(
1 + \frac{4\epsilon_{\rm T}}{9}
\right)
~, \label{eq:CAppox1} \\  
\frac{1}{\Lambda_{\Phi}^{(n)}}
&=&  \frac{2knR}{\sqrt{3\pi M_5^3R\left(n^2 + k^2 R^2\right)\left(9n^2 + k^2 R^2\right)}}\left(
1-\epsilon_{\rm T}
\right)~. \label{eq:CAppox2}
\end{eqnarray}
\begin{figure}[!htb!]
\minipage{0.48\textwidth}
  \includegraphics[width=1\linewidth]{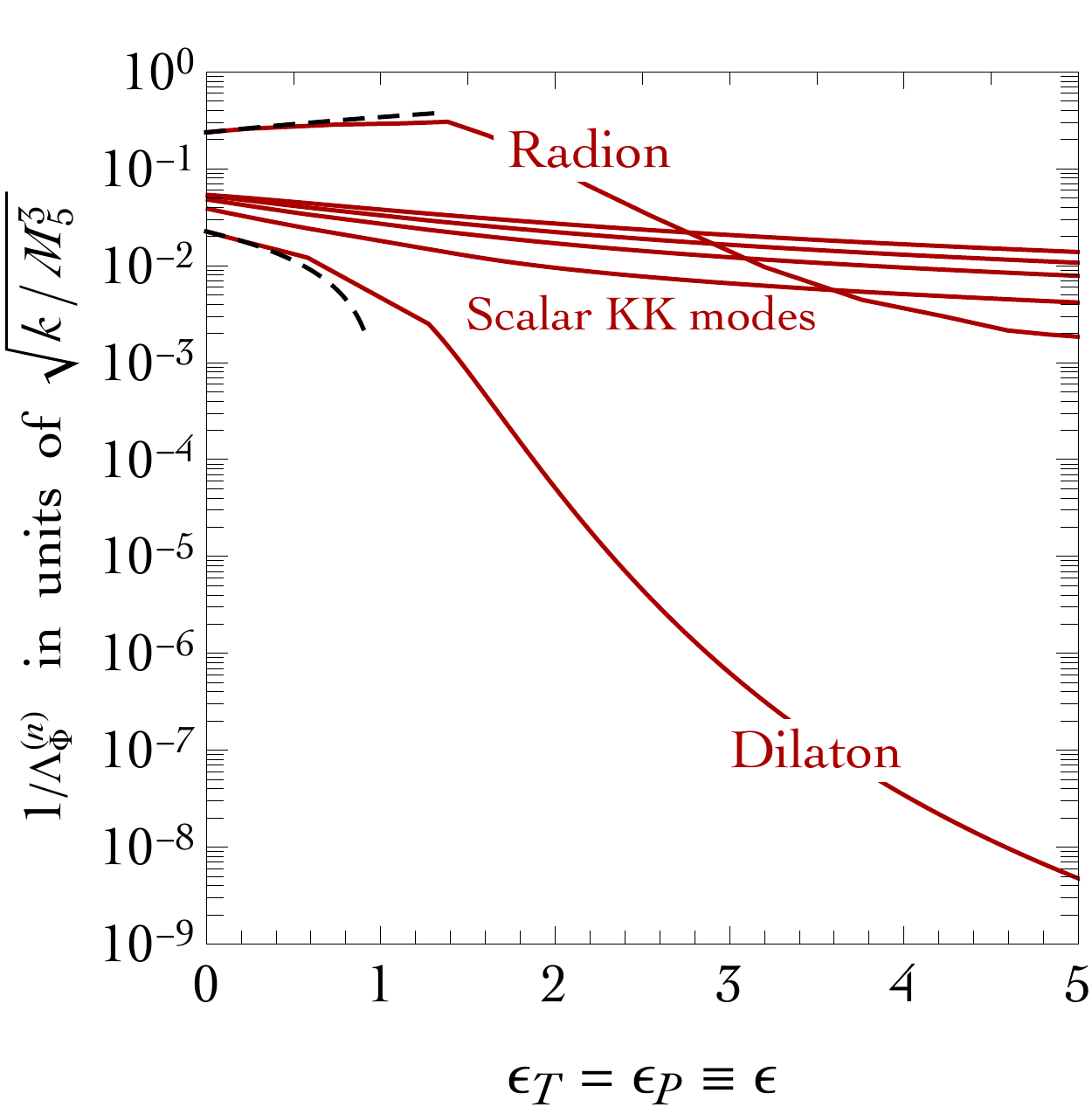}
\endminipage \quad
\minipage{0.48\textwidth}
  \includegraphics[width=1\linewidth]{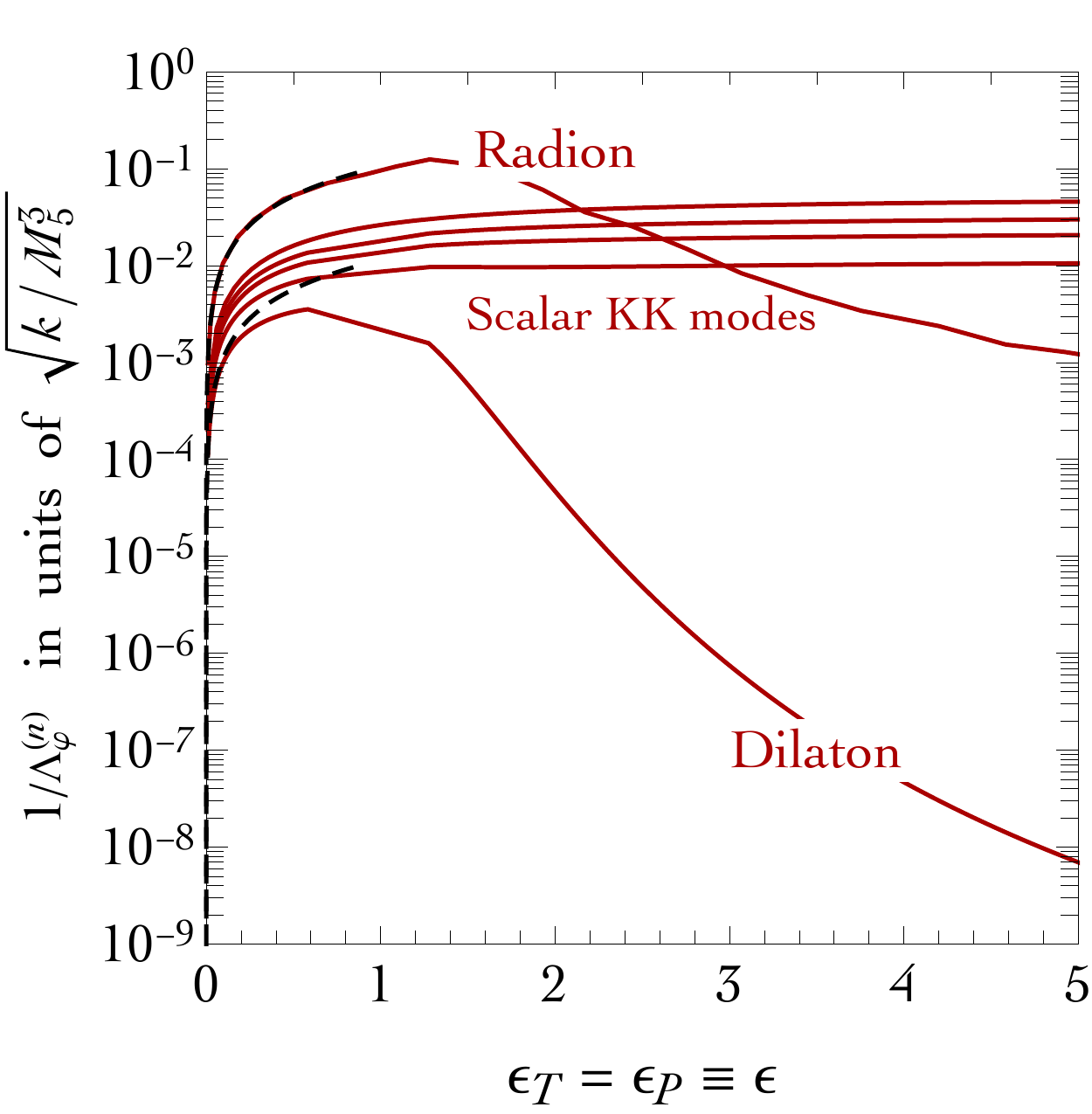}
\endminipage 
\vspace{-.2 cm}
\caption{\label{fig:Couplings} 
Scalar couplings with the trace of the energy-momentum tensor (left panel) and the 
SM Lagrangian (right panel). The dashed black lines (for simplicity, only for radion and dilaton) represent the analytical approximations in eqs.~\eqref{eq:CAppox1}--\eqref{eq:CAppox4}.
We only show the first few modes of the KK tower.
}
\end{figure}
while for the direct coupling with the SM Lagrangian we have
\begin{eqnarray}
\frac{1}{\Lambda_{\varphi}^{(0)}}
 &=& \frac{2\epsilon_{\rm T}}{27}\sqrt{\frac{k\left(1 + \coth\frac{k\pi R}{3}\right)}{M_5^3}}\simeq 
\frac{2\epsilon_{\rm T}}{27}\sqrt{\frac{k}{M_5^3}}
~,  \label{eq:CAppox3}  \\  
\frac{1}{\Lambda_{\varphi}^{(n)}}
  &=& 
\frac{n\epsilon_{\rm T}}{k\sqrt{3\pi}R^{3/2}}\sqrt{\frac{n^2 + k^2 R^2}{M_5^3(9n^2 + k^2 R^2)}}~. \label{eq:CAppox4}
\end{eqnarray}
In the rigid limit the direct coupling  with the SM Lagrangian vanishes.
This is a consequence of the junction conditions in eqs.~\eqref{eq:BC1}--\eqref{eq:BC2} in which 
a large value of $\lambda^{\prime\prime}_{{\rm T,P}}$ forces the scalar field fluctuation 
$\varphi$ on the two branes to vanish.
We note that 
the analytical approximations in eqs.~\eqref{eq:CAppox1}--\eqref{eq:CAppox4} do not feature an explicit $\epsilon_{\rm P}$-dependence. However, at higher orders, an explicit  
$\epsilon_{\rm P}$-dependence appears since it unavoidably enters via the scalar masses.
For this reason, in fig.~\ref{fig:Couplings}  we show the full numerical results for the couplings in eq.~(\ref{eq:ScalarCouplings}) under the assumption $\epsilon_{\rm T} = \epsilon_{\rm P} \equiv \epsilon$ already adopted in fig.~\ref{scalars_vs_BC}. 

In light of the discussion we put forward in section~\ref{sec:KK-masses-couplings} about the KK mode production cross sections, it is also interesting to present a useful analytical limit for the 
couplings in eq.~(\ref{eq:ScalarCouplings}) for a generic value of $\epsilon_{\rm T}$ and large $n$. 
By approximating the KK scalar spectrum
with $m_n^2 \approx n^2/R^2$, we find, at the leading order in a $kR/n$ expansion
\begin{equation}
\frac{1}{\Lambda_{\varphi}^{(n)\,2}} = \frac{1}{3\pi M_5^3 R}~,~~~~~~
\frac{1}{\Lambda_{\Phi}^{(n)\,2}} =  \frac{4}{3\pi M_5^3 R \epsilon_{\rm T}^2}\left(\frac{kR}{n}\right)^4~.
\end{equation}

Equipped with these results, we can now move to compute the dilaton/radion decays to SM particle pairs.

\section{Dilaton/radion decays to SM particle pairs}
\label{app:scalar-SM-decays}

Using the results of ref.~\cite{Cox:2012ee} (accounting for the difference in conventions described in footnote~\ref{KKdilaton-factors-2}), the partial decay widths of a KK scalar of mass $m_n$ are
\begin{align}
&\Gamma_{ZZ} = \left[\left(1 - x_Z + \frac{3}{4}x_Z^2\right)f_h(m_n)^2 - \kappa_n \left(1 + \frac{x_Z}{2}\right)f_h(m_n) + \frac34 \kappa_n^2\right](1-x_Z)^{1/2}\,\frac{\Gamma_0}{4}\,,\\
&\Gamma_{W^+W^-} = \left[\left(1 - x_W + \frac{3}{4}x_W^2\right)f_h(m_n)^2 - \kappa_n \left(1 + \frac{x_W}{2}\right)f_h(m_n) + \frac34 \kappa_n^2\right](1-x_W)^{1/2}\,\frac{\Gamma_0}{2}\,,\\
&\Gamma_{hh} = \left[1 + x_h + \frac{x_h^2}{4} - 12\xi - 6\xi x_h + 36\xi^2 - \frac{36\xi m_h^2}{m_n^2-m_h^2} - \frac{72\xi m_h^4}{m_n^2(m_n^2-m_h^2)} + \frac{216\xi^2m_h^2}{m_n^2-m_h^2} \right. \nonumber\\
&\qquad\qquad \left. +\, \frac{324\xi^2m_h^4}{(m_n^2-m_h^2)^2}
- \kappa_n \left(1 + \frac{x_h}{2} - 6\xi - \frac{18\xi m_h^2}{m_n^2-m_h^2}\right)
+ \frac{1}{4}\kappa_n^2\right] (1 - x_h)^{1/2}\,\frac{\Gamma_0}{4}\,, \\
&\Gamma_{f\bar f} = N_f\left(\frac{m_f}{m_n}\right)^2\, f_h(m_n)^2 (1-x_f)^{3/2}\, \Gamma_0\,, \\
&\Gamma_{gg} = \left|\frac{\alpha_s}{2\pi} b_{\rm QCD} + \kappa_n\right|^2 \Gamma_0 \,, \\
&\Gamma_{\gamma\gamma} = \frac18\left|\frac{\alpha}{2\pi}b_{\rm QED} + \kappa_n\right|^2 \Gamma_0 \,,
\end{align}
where
\beq
\Gamma_0 \equiv \frac{m_n^3}{8\pi\Lambda_\Phi^{(n)2}} \,,\qquad
x_i \equiv \frac{4m_i^2}{m_n^2} \,,\qquad
f_h(m_n) = 1 - \frac{6\xi m_n^2}{m_n^2 - m_h^2} \,,
\eeq
$\Lambda_\Phi^{(n)}$ is given in eqs.~\eqref{KKdilaton-couplings-0}--\eqref{KKdilaton-couplings-n}, $\kappa_n \equiv \Lambda_\Phi^{(n)}/\Lambda_\varphi^{(n)}$, $\xi$ is the Higgs-curvature coupling as defined in eq.~\eqref{xi}, $N_f = 3$ for each quark, $1$ for each charged lepton and $1/2$ for each (Majorana) neutrino, $b_{\rm QCD}=11-2n_Q/3$, where $n_Q$ is the number of quarks lighter than $m_n$, and $b_{\rm QED}=-(4/3)\sum_f N_f Q_f^2\,$, where the summation is over fermions lighter than $m_n$ and $Q_f$ is the electric charge. In the region of the $W$ and top masses, threshold effects are accounted for by taking
\beq
b_{\rm QED} = b_2 + b_Y - f_h(m_n)(2 + 3x_W + 3x_W(2-x_W)f(x_W)) + \frac83 f_h(m_n)x_t(1 + (1-x_t)f(x_t))
\eeq
with $b_2 = 19/6$ and $b_Y = -41/6$, and (relevant also for production)
\beq
b_{\rm QCD} = 7 + f_h(m_n)\, x_t\, (1 + (1-x_t)f(x_t)) \,,
\eeq
where
\beq
f(x) = \left\{\begin{array}{lcc}
\displaystyle\left[\sin^{-1}\left(\frac{1}{\sqrt x}\right)\right]^2 & \quad & x \geq 1 \vspace{3mm} \\
\displaystyle -\frac14\left[\ln\frac{1+\sqrt{1-x}}{1-\sqrt{1-x}} - i\pi\right]^2 & \quad & x < 1
\end{array}\right. \;.
\eeq

\section{Trilinear graviton/dilaton/radion decays}
\label{app:trilinear-decays}

In this section we discuss trilinear decays involving scalars and gravitons.
The interactions responsible for these processes are encoded in the action obtained by 
expanding the metric including both tensor and scalar perturbations
\begin{eqnarray}
ds^2 &=& e^{2\sigma(y)}
\left[
\left(
1+ 2F(x,y)\right)
(\eta_{\mu\nu} + 2h_{\mu\nu}(x,y)/M_5^{3/2}) dx^{\mu}dx^{\nu} + \left(
1 - 4F(x,y)
\right)dy^2
\right]~,	\nonumber\\
S(x,y) &=& S_0(y) + \varphi(x,y)~.
\end{eqnarray}
Although technically involved, the computation does not pose particular complications. 
After expanding the 5D action up to the cubic order in the fluctuations, 
one just needs to apply 
the KK reduction -- already introduced in appendix~\ref{app:grav-grav-decays} and 
appendix~\ref{app:ScalarSpectrum} for, respectively, tensor and scalar perturbations -- in order to extract the 4D Lagrangian interactions. 
In this process a non-trivial check is the exact cancellation of the Gibbons-Hawking-York terms against 
the total derivatives coming from the Ricci scalar. 

In addition to the trilinear graviton vertex already studied in appendix~\ref{app:grav-grav-decays}, 
the inclusion of scalar perturbations gives rise to three additional
 structures that we shall now discuss in detail. Let us start from the trilinear 5D action involving one graviton and two scalar fields (labelled `Tensor-Scalar-Scalar', TSS in the following).
We find
\begin{equation}
\mathcal{S}_{\rm TSS} = M_5^{3/2}\int_{0}^{\pi R}d^4x dy e^{3\sigma}h_{\mu\nu}
\left[
12\left(
\partial^{\mu}F
\right)\left(
\partial^{\nu}F
\right) + \frac{2}{3}\left(
\partial^{\mu}\varphi
\right)\left(
\partial^{\nu}\varphi
\right)
\right]~.
\end{equation}
This trilinear action is responsible for KK graviton decay into two scalars, $G_l \to \phi_m + \phi_n$, and 
KK scalar decay into a graviton and a scalar, $\phi_l \to G_m + \phi_n$. 
Next, we consider the trilinear 5D action involving two graviton and one scalar fields (labelled  TTS in the following).
We find
\begin{eqnarray}
\mathcal{S}_{\rm TTS} &=& \int_0^{\pi R} d^4x dy 6e^{3\sigma}\left\{
h^{\mu\nu}h_{\mu\nu}\left[
-4\sigma^{\prime}\partial_y F - 12F \sigma^{\prime\,2}
+S_0^{\prime}(\partial_y\varphi)/9 + 4S_0^{\prime\,2}F/3
\right] \right. \nonumber \\
&-& \left.
\partial_yh^{\mu\nu}\left[
F\left(
\partial_y h_{\mu\nu}\right) + 2h_{\mu\nu}\left(
\partial_yF + 6F\sigma^{\prime}
\right)
\right] +e^{5\sigma}\left[
\varphi V^{\prime}(S_0)/6 + FV(S_0)/3
\right]
\right\}\nonumber \\
&-& \sum_{i={\rm T}, {\rm P}}\int d^4x \frac{e^{4\sigma}}{M_5^3}
h_{\mu\nu}h^{\mu\nu}\left\{
\varphi\lambda_i^{\prime}[S_0(y_i)] + \frac{F}{4}\lambda[S_0(y_i)]
\right\}~.
\end{eqnarray}
This trilinear action is responsible for KK graviton decay into one scalar and one graviton, 
$G_l \to G_m + \phi_n$, and 
KK scalar decay into two gravitons, $\phi_l \to G_m + G_n$. Finally, we have 
the trilinear 5D action involving three scalar fields (labelled  SSS in the following).
We find
\begin{eqnarray}
\mathcal{S}_{\rm SSS} &=& \int_0^{\pi R} d^4x dy M_5^3\left\{e^{3\sigma}\left[
-48 F(\partial_{\mu}F)(\partial^{\mu}F) + 24F(\partial_y F)^2 + 128F^3S_{0}^{\prime\,2}/3 - 1536F^3\sigma^{\prime\,2}
\right.\right.\nonumber\\
&-&\left.\left. 12F^2S_0^{\prime}\varphi^{\prime}  - 2F\varphi^{\prime\,2} + 240(\partial_y F)F^2\sigma^{\prime} + 
12F^2\varphi S_0^{\prime}\sigma^{\prime} + 4F^2\varphi S_0^{\prime\prime}
\right]
\right.\nonumber\\
 &+&\left. e^{5\sigma}\left[
-F\varphi^2V^{\prime\prime}(S_0) - \varphi^3 V^{\prime\prime\prime}(S_0)/6
\right]\right\}\nonumber\\
&-& \sum_{i={\rm T}, {\rm P}}\int d^4x e^{4\sigma} \left\{
\frac{\varphi^3}{6}\lambda^{\prime\prime\prime}[S_0(y_i)] + 2F\varphi^2\lambda^{\prime\prime}[S_0(y_i)] + 
4F^2\varphi\lambda^{\prime}[S_0(y_i)]
\right\}~.
\end{eqnarray}
This trilinear action is responsible for KK scalar decay into two scalars, 
$\phi_l \to \phi_m + \phi_n$. In the following, we shall compute and discuss each one of these processes.
We describe the computations in full generality -- that is for an arbitrary choice of $\epsilon$ -- but we specialize all our numerical results as well as analytical approximations in the rigid limit $\epsilon = 0$.  

\subsection{KK graviton decays into two scalars}

The 4D Lagrangian density for a fixed triad $(n,m,l)$ of KK states is
\begin{equation}\label{eq:LTSS}
\mathcal{L}^{\rm TSS}_{(n,m,l)} = \frac{1}{M_5^{3/2} \sqrt{\pi R}}\,\mathcal{C}^{\rm TSS}_{(n,m,l)}(k,R)\,\tilde{h}_{\mu\nu}^{(n)}\left(\partial^{\mu}
\phi_{m}\right)\left(
\partial^{\nu}\phi_{l}
\right)~,
\end{equation}
where the dimensionless function $\mathcal{C}^{\rm TSS}_{(n,m,l)}(k,R)$ encodes the integral over the extra dimension.
We find
\begin{equation}\label{eq:ExtraDimIntegralTSS}
\mathcal{C}^{\rm TSS}_{(n,m,l)}(k,R) \equiv \int_0^{\pi R}dy e^{3\sigma}k\psi_n(y)\left[
3\tilde{\Phi}_m(y)\tilde{\Phi}_l(y) + \frac{2}{3}\tilde{\varphi}_m(y)\tilde{\varphi}_l(y)
\right]~.
\end{equation}
Notice that, compared with eq.~(\ref{eq:ScalarSolution}), we defined here the dimensionless scalar wavefunction 
$\tilde{\Phi}_n \equiv \sqrt{M_5^3/k}\Phi_n$. Furthermore, from the constraint in 
eq.~(\ref{eq:const}) we get
\begin{equation}
\tilde{\varphi}_n(y) = \frac{3}{2k}\left[
2k\tilde{\Phi}_n(y) + \frac{3}{2}\tilde{\Phi}^{\prime}_n(y)
\right]~.
\end{equation}
From eq.~(\ref{eq:LTSS}) we find the decay width for the process $G_l \to \phi_m + \phi_n$
\begin{equation}\label{eq:ScatteringTSS}
\Gamma_{G_l \to \phi_{m}\phi_n}(k,R) = \frac{|\mathcal{C}^{\rm TSS}_{(l,m,n)}(k,R)|^2}
{480\pi^2(1+\delta_{mn})m_l^7 RM_5^{3}}
\lambda^{5/2}(m_l^2, m_m^2, m_n^2)~,
\end{equation}
where $\lambda(x,y,z) \equiv x^2 + y^2 + z^2 - 2xy -2xz -2yz$.
\begin{figure}[!htb!]
\centering
\minipage{0.48\textwidth}
  \includegraphics[width=1\linewidth]{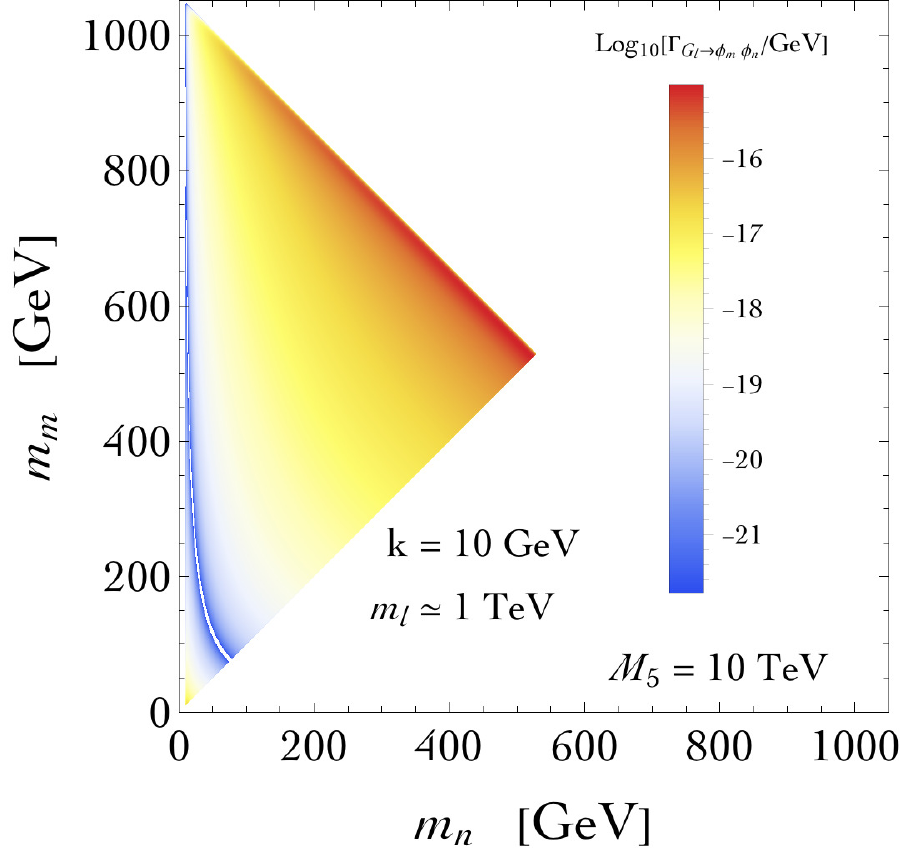}
\endminipage \quad
\minipage{0.48\textwidth}
  \includegraphics[width=1\linewidth]{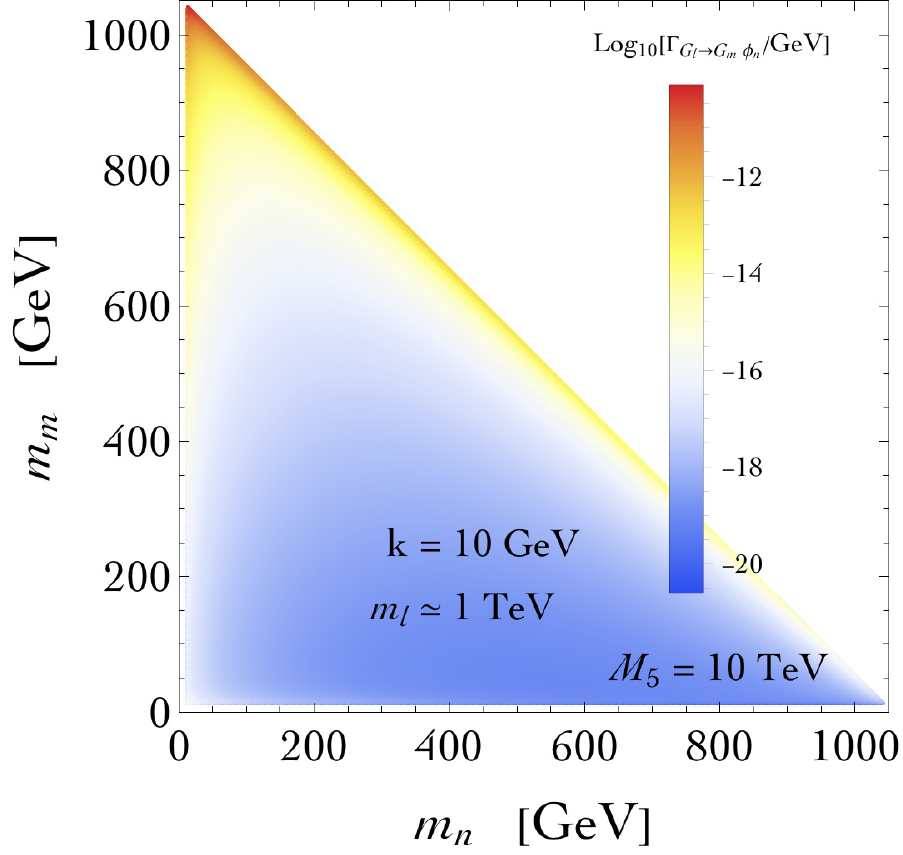}
\endminipage 
\vspace{-.2 cm}
\caption{\label{fig:GravitonDecayDensity}
Density plot of the decay widths 
 $\Gamma_{G_l \to \phi_{m}\phi_n}$ (left panel) and 
 $\Gamma_{G_l \to G_{m}\phi_n}$ (right panel) as a function of the final state masses. The example 
 shown is $M_5 = 10$~TeV, $k = 10$~GeV, for mode $l = 1000$. The corresponding mass of the 
 decaying particle is $m_{1000} \simeq 1$~TeV. 
}
\end{figure}
In the left panel of fig.~\ref{fig:GravitonDecayDensity} we show the density plot of the decay width 
 $\Gamma_{G_l \to  \phi_{m}\phi_n}$ as a function of the final state masses.
 From this plot it is evident that the decay width is dominated by states close to
 the phase space boundary. 
If compared with the trilinear graviton decay computed in appendix~\ref{app:grav-grav-decays}, 
it is easy to see that the KK graviton decay into two scalars is largely sub-dominant. 
The main reason is that 
in the process $G_l \to \phi_m + \phi_n$ conservation of angular momentum requires a final state in d-wave, $L = 2$. By direct computation, it is possible to show that the 
decay width $\Gamma_{G_l \to \phi_{m}\phi_n}$ is indeed proportional to the fourth power of the relative velocity
between the two final state particles, defined as $v_{\rm rel} = |\vec{p}_m/E_m - \vec{p}_n/E_n|$, in agreement with the usual correspondence between 
total orbital angular momentum $L$ and velocity $v_{\rm rel}$, which implies the 
scaling $v_{\rm rel}^{2L}$ in the non-relativistic limit.
Consequently, close to
 the phase space boundary
the decay width receives a further suppression due to the fact that 
final state particles are produced almost at rest with very small relative velocity.

Finally, we provide an analytical approximation for the sum over the kinematically allowed final state particles, in parallel with the discussion presented for the graviton decay into gravitons. 
We find
\begin{equation}
\Gamma_{G_l \to \sum\phi_{m}\phi_n} = \Gamma_{G_l \to \sum G_mG_n} \times
\frac{6 k^2}{595 m_l^2}~.
\end{equation}
where $\Gamma_{G_l \to \sum G_mG_n}$ is the inclusive width computed in eq.~(\ref{eq:TTTsum}). 
This result clearly shows the relative suppression between the two channels.

\subsection{KK graviton decays into a KK graviton and a scalar}

The 4D Lagrangian density for a fixed triad $(n,m,l)$ of KK states is
\begin{equation}\label{eq:LTTS}
\mathcal{L}^{\rm TTS}_{(n,m,l)} = \frac{1}{M_5^{3/2}(\pi R)}\,\left[
\mathcal{C}^{\rm TTS}_{(n,m,l)}(k,R) + \mathcal{T}_{(n,m,l)}^{\rm TTS}(k,R) - \mathcal{P}_{(n,m,l)}^{\rm TTS}(k,R)
\right]\phi_n \tilde{h}_{\mu\nu}^{(m)}\tilde{h}^{(l)\,\mu\nu}
\end{equation}
where $\mathcal{C}^{\rm TTS}_{(n,m,l)}(k,R)$ encodes the integration over the  
extra dimension, and we find
\begin{eqnarray}
\mathcal{C}^{\rm TTS}_{(n,m,l)}(k,R) &=& \int_0^{\pi R}dy e^{3\sigma}
\sqrt{k}
\left[
48k^2\tilde{\Phi}_n\psi_m \psi_l + 3\tilde{\Phi}_n(\partial_y\psi_n)(\partial_y\psi_l) 
+ 6(\partial_y \tilde{\Phi}_n)(\partial_y\psi_m)\psi_l 
\right.\nonumber \\
&+&\left.
24k\tilde{\Phi}_n\psi_m(\partial_y \psi_l) + 8k(\partial_y\tilde{\Phi}_n)\psi_m \psi_l 
+\frac{8k^2}{3}\tilde{\varphi}_n\psi_m\psi_l + \frac{4k}{3}(\partial_y\tilde{\varphi}_n)\psi_m\psi_l
\right] .\nonumber \\
\end{eqnarray}
The two functions $\mathcal{T}_{(n,m,l)}^{\rm TTS}(k,R)$ and $\mathcal{P}_{(n,m,l)}^{\rm TTS}(k,R)$ arise, respectively, 
from interactions localized on the TeV and Planck brane. We find
\begin{eqnarray}
\mathcal{T}_{(n,m,l)}^{\rm TTS}(k,R) &=& 4k^{3/2}\left[
\tilde{\varphi}_n(0)\psi_m(0)\psi_l(0) + 2 \tilde{\Phi}_n(0)\psi_m(0)\psi_l(0)
\right],\\
\mathcal{P}_{(n,m,l)}^{\rm TTS}(k,R) &=& 4k^{3/2}e^{2k\pi R}\left[
\tilde{\varphi}_n(\pi R)\psi_m(\pi R)\psi_l(\pi R) + 2 \tilde{\Phi}_n(\pi R)\psi_m(\pi R)\psi_l(\pi R)
\right].
\end{eqnarray}
We can now compute the decay width for the process 
$G_l \to G_m + \phi_n$.
We find
\begin{equation}\label{eq:TTSdecaywidth}
\Gamma_{G_l \to G_m \phi_n} = \frac{|\mathcal{S}^{\rm TTS}_{(n,m,l)} + \mathcal{S}^{\rm TTS}_{(n,l,m)}|^2}{
2880 M_5^3 m_l^7 m_m^4\pi^3 R^2}\lambda^{1/2}(m_l^2, m_m^2, m_n^2)\,\mathcal{F}(m_l^2,m_m^2,m_n^2)~,
\end{equation}
where $\mathcal{F}(x,y,z)\equiv 
x^4 + x^3(26y - 4z)+ 2x^2(63y^2 - 28y z + 3z^2) + 2x(13y - 2z)(y - z)^2 + (y - z)^4$, and we introduced the short-hand notation 
$\mathcal{S}^{\rm TTS}_{(n,m,l)} \equiv \mathcal{C}^{\rm TTS}_{(n,m,l)}(k,R) + 
\mathcal{T}_{(n,m,l)}^{\rm TTS}(k,R) - \mathcal{P}_{(n,m,l)}^{\rm TTS}(k,R)$.

In the right panel of fig.~\ref{fig:GravitonDecayDensity} we show the density plot of the decay width 
 $\Gamma_{G_l \to G_{m}\phi_n}$ as a function of the final state masses.
 From this plot we see that the decay width is dominated by heavy gravitons and light scalars in the final state.
 
We provide an analytical approximation for the sum over the kinematically allowed final state particles, in parallel with the discussion presented for the graviton decay into gravitons. 
We find
\begin{equation}
\Gamma_{G_l \to \sum G_{m}\phi_n} =\Gamma_{G_l \to \sum G_mG_n} \times
\frac{1524 k^{3/2}}{595 m_l^{3/2}}~.
\end{equation}

\subsection{KK graviton decays: branching ratios}

We can now summarize our results.

\begin{figure}[!htb!]
\begin{center}
\includegraphics[width=0.55\textwidth]{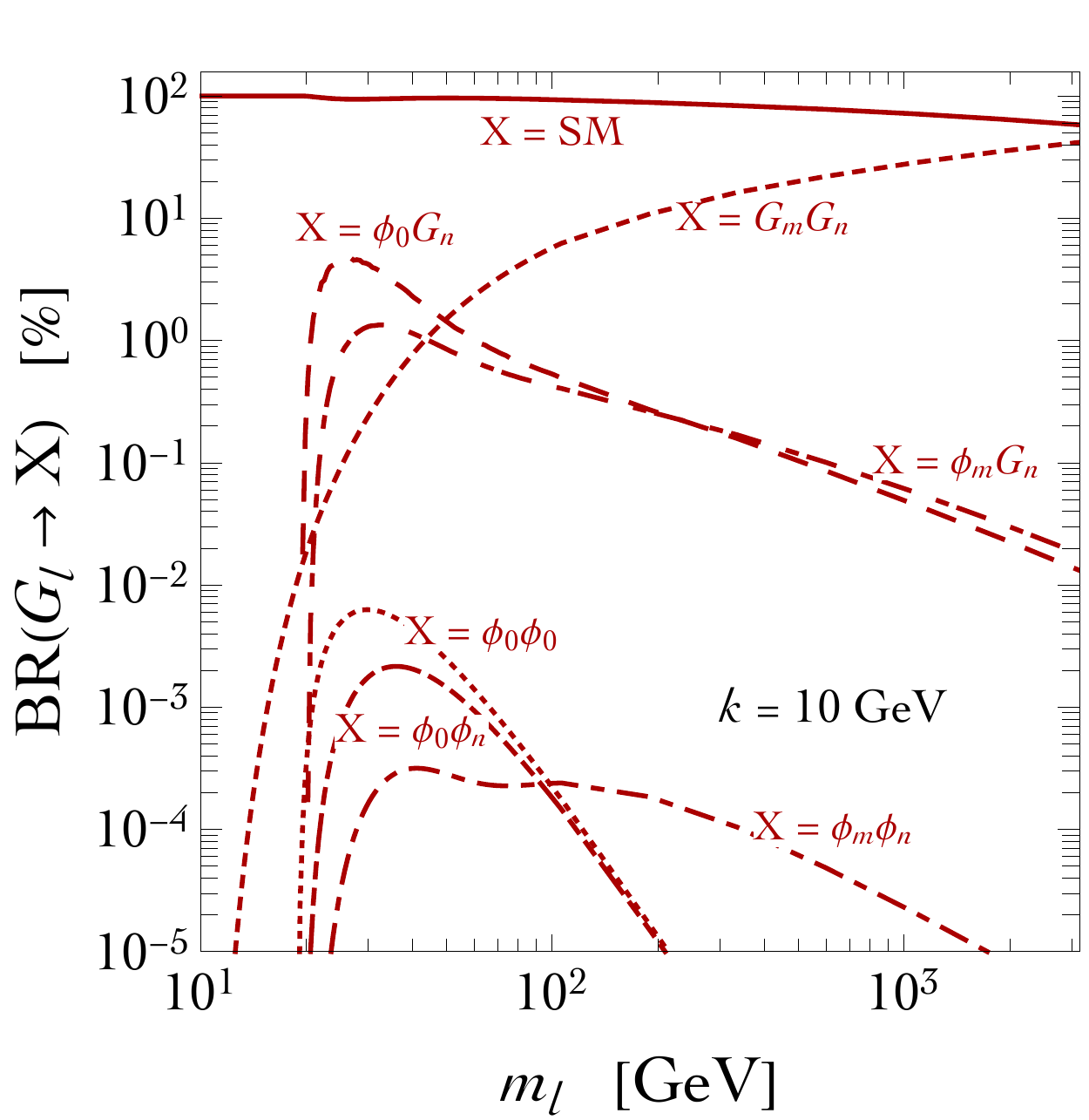}
\end{center}
\caption{{ Branching ratios of the graviton KK modes for $k = 10$~GeV and rigid boundary conditions. We take $M_5 = 10$~TeV.}}
\label{graviton_BRs_10}
\end{figure}

In fig.~\ref{graviton_BRs_10} we show the branching ratios of the graviton KK modes.
For clarity, in the final states involving KK scalars we plotted separately 
the case with a radion in the final state. 
All the decay widths are inclusive, meaning that we consider the sum over all kinematically allowed final state particles. For the inclusive processes, 
in the limit $l \gg kR\gg 1$ we find the simple scaling relations
\begin{equation}\label{eq:graviton_scaling}
\Gamma_{G_l\to \sum G_m G_n} \sim \frac{m_l^{7/2}\sqrt{k}}{M_5^3}~,~~~~~
\Gamma_{G_l\to \sum G_m\phi_n}  \sim \frac{m_l^2 k^2}{M_5^3}~,~~~~~
 \Gamma_{G_l\to \sum \phi_m \phi_n} \sim \frac{m_l^{3/2} k^{5/2}}{M_5^3}~,
\end{equation}
that should be compared with the decay width into SM states
\begin{equation}
\Gamma_{G_l\to {\rm SM}} \sim \frac{m_l^3 k}{M_5^3}~.
\end{equation}
As shown in fig.~\ref{graviton_BRs_10}, and already discussed in the main text, 
graviton decays into gravitons competes with the SM decay channels at small $k$ since 
the ratio of their decay widths parametrically grows as $\sim (m_l/k)^{1/2}$. 
On the contrary, KK graviton decays into, respectively, a KK graviton and a scalar and two scalars feature the 
suppressions $\sim (k/m_l)^{3/2}$ and $(k/m_l)^{2}$ if compared to the graviton into SM decay.

Let us consider these cases in more detail.  It is useful to remember that in the graviton case the wavefunction integrals prefer small mass splittings, which is physically understood from the view of extra-dimensional momentum conservation in the $k\to0$ limit.

For the decay of graviton into a graviton and a scalar there is no wavefunction suppression.  This is because in the stiff limit the boundary conditions for the dilaton are Dirichlet-Dirichlet and, when integrated over the wavefunction for two gravitons, one no longer requires that $m_l \sim m_m + m_n$ as the integrand is $y$-parity odd in the limit $k\to 0$.  Naively one would expect the total width should then scale as $\Gamma_{G_l\to \sum G_m\phi_n}  \sim m_l^4 / M_5^3$.  However, in the limit $k\to0$ the dilaton acquires an additional $\mathcal{Z}_2$ symmetry that would completely forbid such decays.  Thus one expects the amplitude should still scale proportional to $k$ due to the vertex factor itself.  Thus, when squared, this explains the form of the decay width in eq.~(\ref{eq:graviton_scaling}), and the extra suppression relative to the $G_l\to \sum G_m G_n$ decays.

The suppression for $G_l\to \sum \phi_m \phi_n$ decays can be understood from momentum conservation.  The wavefunction integral suppression is still in action as in the $k\to0$ limit the integrand is $y$-parity even.  In addition to this the decay of a spin-2 graviton to two scalars will be d-wave suppressed, hence the matrix element is itself proportional to the final state momenta, which are proportional to the mass splitting.  When squared this leads to an additional factor of $(k/m_l)^{2}$ relative to the $G_l\to \sum G_m G_n$ decays.

\subsection{KK scalar decays into two scalars}

The 4D Lagrangian density for a fixed triad $(n,m,l)$ of KK states is
\begin{equation}\label{eq:LSSS}
\mathcal{L}_{\rm SSS}^{(n,m,l)} =  \frac{1}{M_5^{3/2}}\left[
\mathcal{C}_{(n,m,l)}^{\rm SSS\,(1)}(k,R)\phi_n(\partial_{\mu}\phi_m)(\partial^{\mu}\phi_l) +
\mathcal{B}_{(n,m,l)}^{\rm SSS}(k,R) \phi_n \phi_m \phi_l
\right]~,
\end{equation}
where the coefficient of the derivative interaction comes from the extra-dimensional integral
\begin{equation}
\mathcal{C}_{(n,m,l)}^{\rm SSS\,(1)}(k,R) = 6\int_0^{\pi R}dy e^{3\sigma}k^{3/2}\tilde{\Phi}_n(y) \tilde{\Phi}_m(y) \tilde{\Phi}_l(y)~,
\end{equation}
while the coefficient of the trilinear interaction has the structure $\mathcal{B}_{(n,m,l)}^{\rm SSS}(k,R) \equiv 
\mathcal{C}_{(n,m,l)}^{\rm SSS\,(2)}(k,R) + \mathcal{T}^{\,\rm SSS}_{(n,m,l)}(k,R) + \mathcal{P}^{\,\rm SSS}_{(n,m,l)}(k,R)$. 
The first term arises from the following extra-dimensional integral
\begin{eqnarray}\label{eq:I2}
\mathcal{C}_{(n,m,l)}^{\rm SSS\,(2)}(k,R) &=& \int_0^{\pi R}dy e^{3\sigma}k^{3/2}\left(
-3\tilde{\Phi}_n\tilde{\Phi}^{\prime}_m\tilde{\Phi}^{\prime}_l + 64 k^2 \tilde{\Phi}_n \tilde{\Phi}_m \tilde{\Phi}_l -
6k\tilde{\Phi}_n\tilde{\Phi}_m \tilde{\varphi}^{\prime}_l 
+ \tilde{\Phi}_n\tilde{\varphi}_m^{\prime}\tilde{\varphi}_l^{\prime}  \right.\nonumber \\
&-&\left. 20 k\tilde{\Phi}_n^{\prime}\tilde{\Phi}_m\tilde{\Phi}_l  + 4k^2\tilde{\Phi}_n \tilde{\varphi}_m \tilde{\varphi}_l
-\frac{16}{81}k^2 \tilde{\varphi}_n \tilde{\varphi}_m \tilde{\varphi}_l
\right)~,
\end{eqnarray}
while $\mathcal{T}^{\,\rm SSS}_{(n,m,l)}(k,R)$ and $\mathcal{P}^{\,\rm SSS}_{(n,m,l)}(k,R)$ describe interactions localized, respectively, 
on the TeV and Planck branes. We find
\begin{eqnarray}
\mathcal{T}^{\,\rm SSS}_{(n,m,l)}(k,R) &\equiv& k^{3/2}
\left[-\frac{1}{6}\left(\frac{4k}{27} - \mu_{\rm T}\right)\tilde{\varphi}_n(0)\tilde{\varphi}_m(0)\tilde{\varphi}_l(0) +
\left(-\frac{4k}{9} + \mu_{\rm T}\right)\tilde{\varphi}_n(0)\tilde{\varphi}_m(0)\tilde{\Phi}_l(0)\right. \nonumber \\ &-&\left. 
\frac{4k}{3}\tilde{\varphi}_n(0)\tilde{\Phi}_m(0)\tilde{\Phi}_l(0)\right]~,\\
\mathcal{P}^{\,\rm SSS}_{(n,m,l)}(k,R) &\equiv&  \frac{e^{2k\pi R}k^{3/2}}{162}\left[
(27\mu_{\rm P} + 4k)\tilde{\varphi}_n(\pi R)\tilde{\varphi}_m(\pi R)\tilde{\varphi}_l(\pi R) \right. \\
 &+& \left. (72k + 162\mu_{\rm P})\tilde{\varphi}_n(\pi R)\tilde{\varphi}_m(\pi R)\tilde{\Phi}_l(\pi R) + 
 216k\tilde{\varphi}_n(\pi R)\tilde{\Phi}_m(\pi R)\tilde{\Phi}_l(\pi R)
 \right]~. \nonumber
\end{eqnarray}
Notice that both TeV and Planck brane contributions vanish in the rigid limit. 
Equipped with this result, we can compute the decay width for the process 
$\phi_l \to \phi_m + \phi_n$. 
We find
\begin{eqnarray}
&&\Gamma_{\phi_l \to \phi_m \phi_n}(k,R) = \frac{\lambda^{1/2}(m_l^2,m_m^2, m_n^2)}{16\pi (1+\delta_{mn}) m_l^3M_5^3} \times \\
&& \left|
\left(m_l^2 + m_m^2 + m_n^2\right)\mathcal{C}_{(n,m,l)}^{\rm SSS\,(1)}(k,R) + \sum_{\rm perm.}\left[
\mathcal{C}_{(n,m,l)}^{\rm SSS\,2}(k,R) + \mathcal{T}^{\,\rm SSS}_{(n,m,l)}(k,R) + \mathcal{P}^{\,\rm SSS}_{(n,m,l)}(k,R)
\right]
\right|^2~.\nonumber
\end{eqnarray}

\begin{figure}[!htb!]
\centering
\minipage{0.33\textwidth}
  \includegraphics[width=1\linewidth]{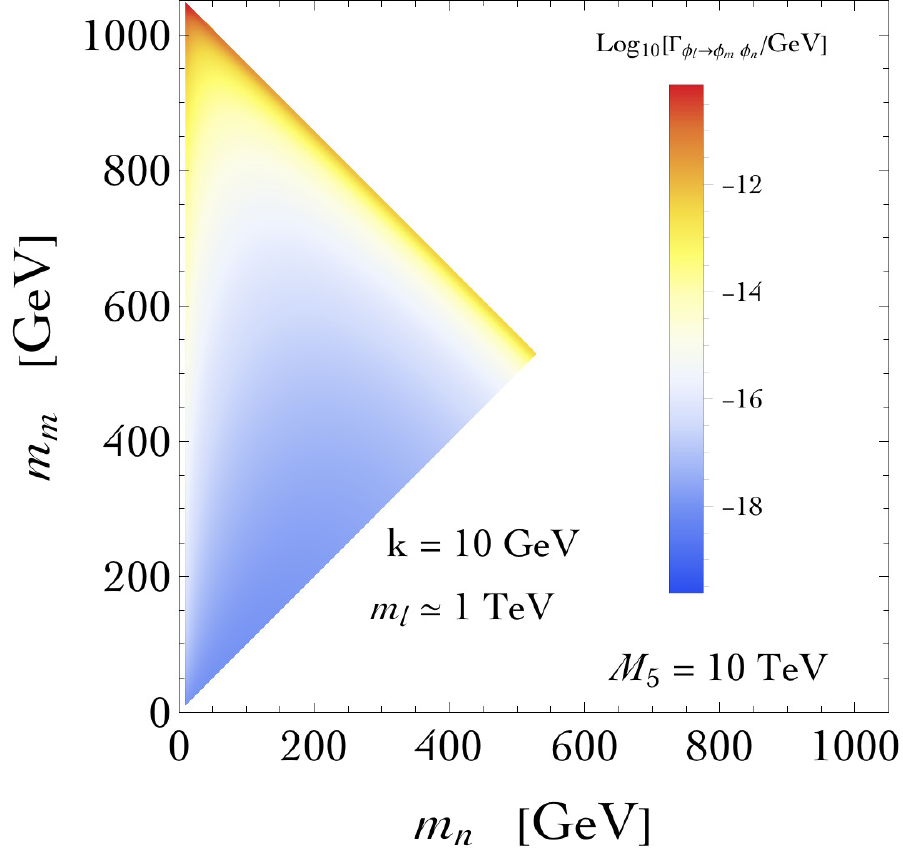}
\endminipage 
\minipage{0.33\textwidth}
  \includegraphics[width=1\linewidth]{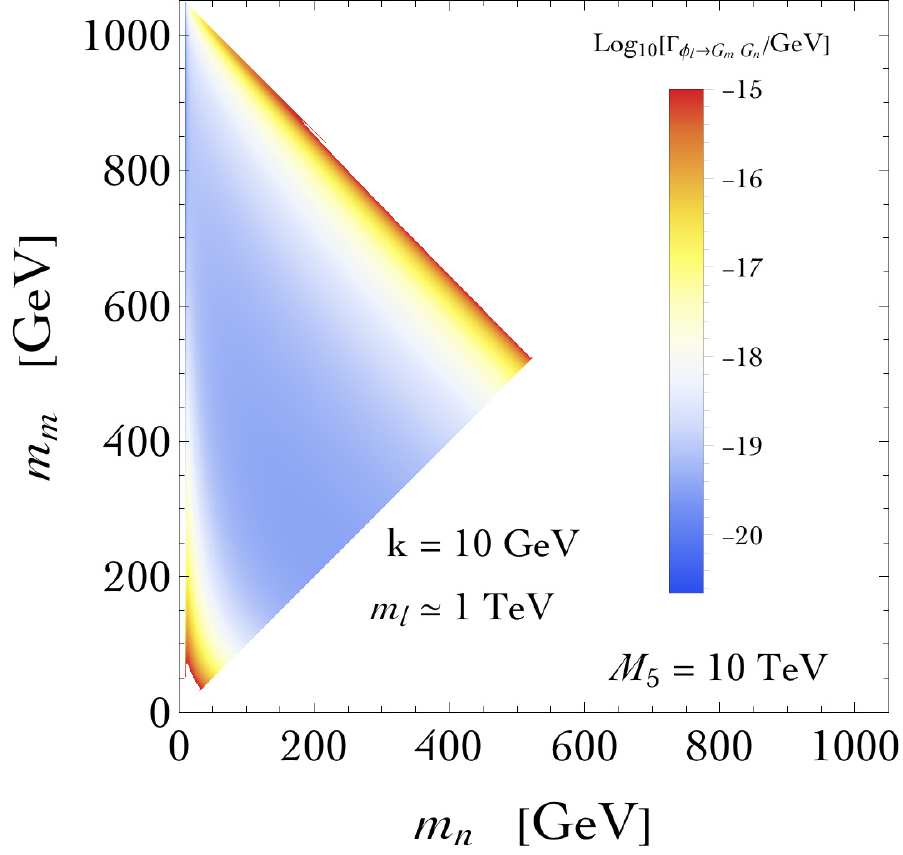}
\endminipage 
\minipage{0.33\textwidth}
  \includegraphics[width=1\linewidth]{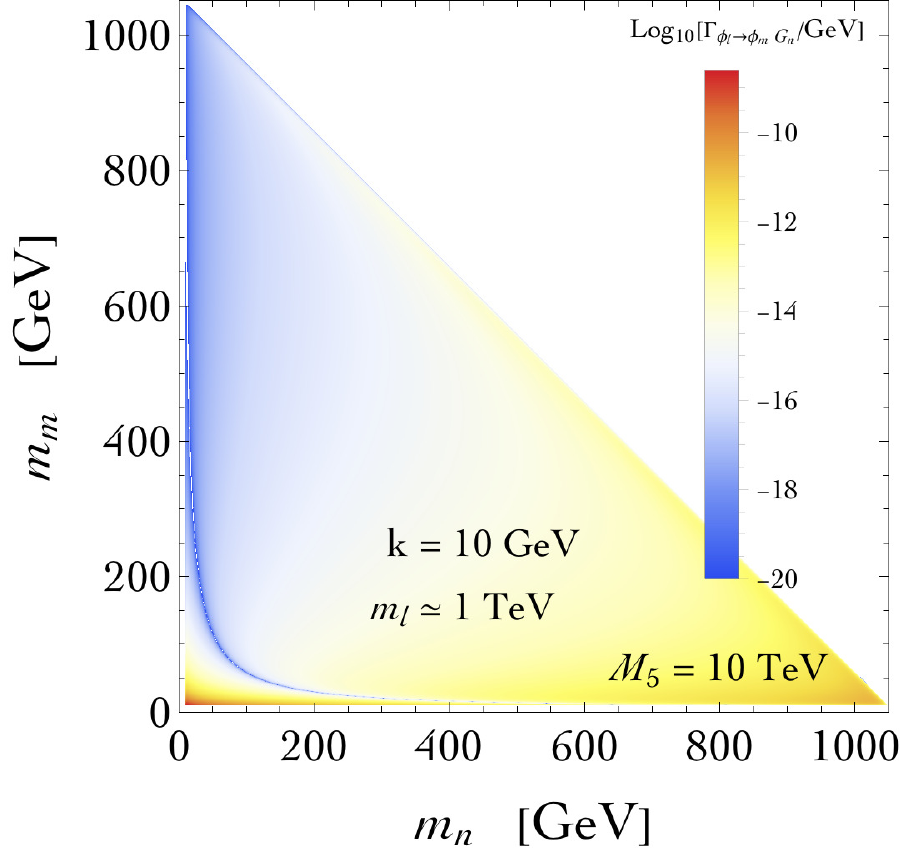}
\endminipage 
\vspace{-.2 cm}
\caption{\label{fig:ScalarDecayDensity}
Density plot of the decay widths 
 $\Gamma_{\phi_l \to \phi_{m}\phi_n}$ (left panel), 
 $\Gamma_{\phi_l \to G_{m}G_n}$ (central panel) and $\Gamma_{\phi_l \to \phi_{m}G_n}$ (right panel)
 as a function of the final state masses. The example 
 shown is $M_5 = 10$~TeV, $k = 10$~GeV, for mode $l = 1000$. The corresponding mass of the 
 decaying particle is $m_{1000} \simeq 1$~TeV. 
}
\end{figure}
In the left panel of fig.~\ref{fig:ScalarDecayDensity} 
we show the density plot of the decay width 
 $\Gamma_{\phi_l \to \phi_{m}\phi_n}$ as a function of the final state masses.
  From this plot we see that the decay width is dominated by the region close to the phase space boundary.

\subsection{KK scalar decays into two KK gravitons}

The decay process $\phi_l \to G_m + G_n$ is controlled by the Lagrangian density in eq.~(\ref{eq:LTTS}).
We find the width
\begin{equation}\label{eq:STTdecaywidth}
\Gamma_{\phi_l \to G_m G_n} = \frac{|\mathcal{S}^{\rm TTS}_{(l,n,m)} + \mathcal{S}^{\rm TTS}_{(l,m,n)}|^2}{
2880(1+\delta_{mn})M_5^3 m_l^3 m_n^4 m_m^4\pi^3 R^2}\,
\lambda^{1/2}(m_l^2,m_m^2, m_n^2)\,
\mathcal{F}(M_m^2,M_n^2,M_l^2)~.
\end{equation}
In the central panel of fig.~\ref{fig:ScalarDecayDensity} 
we show the density plot of the decay width 
 $\Gamma_{\phi_l \to G_m G_n}$ as a function of the final state masses.
  From this plot we see that the decay width is dominated by the region close to the phase space boundary with an additional sizable contribution coming from the decay into two light gravitons.

\subsection{KK scalar decays into a KK graviton and a scalar}

The decay process $\phi_l \to \phi_m + G_n$ is controlled by the Lagrangian density in 
eq.~(\ref{eq:LTSS}). We find the width
\begin{equation}
\Gamma_{\phi_l \to \phi_m G_n}(k,R) = 
\frac{|\mathcal{C}^{\rm TSS}_{(n,m,l)}(k,R)|^2}{96M_5^{3}m_l^3 m_n^4\pi^2 R}
\lambda^{5/2}(m_l^2,m_m^2, m_n^2)~.
\end{equation}
In the right panel of fig.~\ref{fig:ScalarDecayDensity} 
we show the density plot of the decay width 
 $\Gamma_{\phi_l \to \phi_m G_n}$ as a function of the final state masses.
  From this plot we see that the decay width prefers the presence of a light scalar in the final state.
  
  \subsection{KK scalar decays: branching ratios}
  
  We can now summarize our results.
  We already discussed in fig.~\ref{scalar_BRs_10} the 
 branching ratios of the scalar KK modes for $k = 10$~GeV, rigid boundary conditions and $\xi = 0$.
 For the inclusive processes, 
in the limit $l \gg kR\gg 1$ we find the simple scaling relations
\begin{equation}\label{eq:scalar_scaling}
 \Gamma_{\phi_l\to \sum\phi_m \phi_n} \sim \frac{m_l^{2} k^2}{M_5^3}~,~~~~~
 \Gamma_{\phi_l\to \sum\phi_m G_n}  \sim \frac{m_l^2 k^2}{M_5^3}~,~~~~~
 \Gamma_{\phi_l\to \sum G_m G_n} \sim \frac{m_l k^{3}}{M_5^3}~,
\end{equation}
that should be compared with the decay width into SM states
\begin{equation}
\Gamma_{\phi_l\to {\rm SM}} \sim \frac{m_l k^3}{M_5^3}~.
\end{equation}
As already noticed and discussed in the main text, the suppression of the SM decay modes 
makes all decay channels in eq.~(\ref{eq:scalar_scaling}) important for the phenomenology.

\section{Dijet angular distributions}
\label{app:jj-ang}

Analyses of the dijet angular distributions (such as refs.~\cite{Aaboud:2017yvp,Sirunyan:2017ygf,CMS-PAS-EXO-16-046}) use the variable $\chi$, defined in terms of the rapidities of the two jets as
\beq
\chi \equiv \exp(|y_1 - y_2|) \, ,
\eeq
which can be expressed in terms of Mandelstam variables as
\beq
-t = \left\{
\begin{array}{cc}
\displaystyle\frac{s}{\chi + 1} & \quad\mbox{for}\quad \displaystyle 0 < -t < \frac{s}{2} \vspace{3mm}\\
\displaystyle\frac{s}{1/\chi + 1} & \quad\mbox{for}\quad \displaystyle \frac{s}{2} < -t < s
\end{array}\right. \; .
\eeq
Therefore, the angular distributions of the various dijet processes can be obtained from the expressions for
\beq
\frac{d\sigma}{dt}\left(s,t,u\right)
\eeq
available in appendix~A of ref.~\cite{Giudice:2004mg} as
\beq
\frac{d\sigma}{d\chi}\left(s,\chi\right) = \frac{s}{(\chi+1)^2}\left[\frac{d\sigma}{dt}\left(s,\,-\frac{s}{\chi+1},\,-\frac{s}{1/\chi+1}\right) + \frac{d\sigma}{dt}\left(s,\,-\frac{s}{1/\chi+1},\,-\frac{s}{\chi+1}\right)\right] .
\eeq
The effective graviton propagators in our model are different from those of ref.~\cite{Giudice:2004mg} and are derived in appendix~\ref{app:Seff}.

\section{Representing the KK graviton tower by an effective propagator}
\label{app:Seff}

For computing matrix elements involving KK gravitons, it is sometimes useful to sum the propagators (times couplings) of the whole KK graviton tower and work in terms of the effective propagator
\begin{align}
{\cal S}_{\rm eff}(s) &\equiv 
\sum_{n=1}^\infty \frac{1}{\Lambda_G^{(n)2}}\,\frac{1}{s - m_n^2 + i\epsilon} \nonumber\\
&\simeq \frac{1}{\pi M_5^3}\int_k^\infty dm\, \sqrt{1 - \frac{k^2}{m^2}}\;\frac{1}{s - m^2 + i\epsilon} \label{Seff}\\
&= -\frac{1}{2M_5^3}\,\frac{1}{\sqrt{-s + k^2} + k} \;. \nonumber
\end{align}
This expression reduces to
\beq
{\cal S}_{\rm eff}(s) \;\to\;
\left\{
\begin{array}{cl}
\displaystyle -\frac{1}{2M_5^3}\,\frac{i}{\sqrt{s}} & \quad\mbox{for}\; s \gg k^2
\vspace{3mm}\\
\displaystyle -\frac{1}{4 k M_5^3} & \quad\mbox{for}\; |s| \ll k^2
\end{array}\right. \quad.
\label{S-lim}
\eeq

Another useful quantity, relevant for on-shell KK graviton production (at $\sqrt s > k$), is the sum of squared propagators,
\begin{align}
{|{\cal S}(s)|^2}_{\rm eff} &\equiv 
\sum_{n=1}^\infty \frac{1}{\Lambda_G^{(n)4}}\,\frac{1}{(s - m_n^2)^2 + s\,\Gamma_n^2(s)} \nonumber\\
&\simeq \frac{1}{\pi^2 M_5^6 R}\int_k^\infty dm\, \left(1 - \frac{k^2}{m^2}\right)^{3/2}\frac{1}{(s - m^2)^2 + s\,\Gamma^2(s)} \label{S2eff}\\
&\simeq \frac{1}{4 M_5^6 s}\times\frac{2}{\pi R\,\Gamma(s)} \quad\mbox{for $s\gg k^2$}\nonumber \,,
\end{align}
where $\Gamma(s)$ is the width of KK gravitons with mass near $\sqrt s$. For $s \approx k^2$, there is an extra suppression factor of approximately
\beq
\left(1 - \frac{k^2}{s}\right)^{3/2} \,,
\label{S2eff-corr}
\eeq
as can be seen by using $m^2 \simeq s$ in the $k$-dependent prefactor in the integrand.

Note the important difference between the result of eq.~\eqref{S2eff} and the na\"{i}ve square of eq.~\eqref{S-lim}, which has been also discussed in refs.~\cite{Kisselev:2008xv,Franceschini:2011wr} in the context of the low-curvature RS model. When using the formulas of ref.~\cite{Giudice:2004mg} as envisioned in appendix~\ref{app:jj-ang}, one should use ${|{\cal S}(s)|^2}_{\rm eff}$ whenever $|{\cal S}(s)|^2$ with $\sqrt s > k$ appears, but ${\cal S}_{\rm eff}(s)$ when ${\cal S}(s)$ is not squared. For ${\cal S}(s)$ with $\sqrt s < k$, ${\cal S}(t)$, or ${\cal S}(u)$, one should use ${\cal S}_{\rm eff}(s)$, ${\cal S}_{\rm eff}(t)$, or ${\cal S}_{\rm eff}(u)$, respectively.

\section{An exact solution to Einstein's equations}
\label{app:exact}
Working in the comoving coordinate $ds^2 = e^{2 \Sigma(z)} \eta^{\mu\nu} dx_\mu dx_\nu+dz^2$ we find that a solution of Einstein's equation and the dilaton equation of motion from the action in eq.~(\ref{ancora}) is
\be
S(z) = \frac{3}{2} \log \left( \frac{4 k z}{\sqrt{3 \varepsilon}} +1 \right)  \, , ~~~~
\Sigma(z) = \frac{\sqrt{3 \varepsilon}}{6}  k z+ \frac{1}{4} \log \left( \frac{4 k z}{\sqrt{3 \varepsilon}} +1  \right)  \, ,
\ee
where we set the integration constants such that $\Sigma(0) = 0$ and $S(0) = 0$.  Note that there is no smooth limit to either AdS or CW/LD geometry, suggesting this is a branch of the general solution to the equations of motion that is rather unique.  If one reparameterises $k = \sqrt{3 \varepsilon} \tilde{k}$ then in the limit $\varepsilon \to 0$ one appears to recover an LD-like solution, however inspecting \Eq{ancora} one sees that in this limit the bulk potential is vanishing.  Thus this limit corresponds to having a massless bulk scalar with non-trivial boundary conditions on $S$ and $S'$ at the branes, due to a brane potential.

\bibliographystyle{utphys}
\bibliography{clockwork-pheno}

\end{document}